\documentclass[usenatbib,useAMS]{mn2e}

\usepackage{natbib}
\usepackage{graphicx,epsfig}
\usepackage{latexsym,amssymb}
\usepackage[fleqn]{amsmath}

\newcommand{\aj}{AJ}% 
\newcommand{\apj}{ApJ}% 
\newcommand{\apjl}{ApJ}% 
\newcommand{\apjs}{ApJS}% 
\newcommand{\aap}{A\&A}% 
\newcommand{\mnras}{MNRAS}% 
\newcommand{\pasp}{PASP}% 
\newcommand{\physrep}{Phys.~Rep.}% 
\newcommand{\dgr}{$^\circ$}
\newcommand{\Msun}{M$_\odot$}

\newcommand{\hkpc}{$h^{-1}$\,kpc}
\newcommand{\hMpc}{$h^{-1}$\,Mpc}
\newcommand{\hMsun}{$h^{-1}$\,M$_\odot$}
\newcommand{\sphq}{\sin^2\varphi}
\newcommand{\cphq}{\cos^2\varphi}
\newcommand{\sthq}{\sin^2\vartheta}
\newcommand{\cthq}{\cos^2\vartheta}
\newcommand{\gdm}{\ensuremath{\gamma_\mathrm{dm}}}
\newcommand{\ndm}{\ensuremath{n_\mathrm{dm}}}
\newcommand{\rsdm}{\ensuremath{r_{s,\mathrm{dm}}}}
\newcommand{\rvdm}{\ensuremath{r_{\mathrm{180},\mathrm{dm}}}}
\newcommand{\Mvdm}{\ensuremath{M_{\mathrm{180},\mathrm{dm}}}}
\newcommand{\Rhdm}{\ensuremath{R'_{h,\mathrm{dm}}}}
\newcommand{\cdm}{\ensuremath{c_\mathrm{dm}}}
\newcommand{\badm}{\ensuremath{(b/a)_\mathrm{dm}}}
\newcommand{\cadm}{\ensuremath{(c/a)_\mathrm{dm}}}
\newcommand{\gs}{\ensuremath{\gamma_\star}}
\newcommand{\ns}{\ensuremath{n_\star}}
\newcommand{\rss}{\ensuremath{r_{s,\star}}}

\newcommand{\Mts}{\ensuremath{M_{\mathrm{tot},\star}}}
\newcommand{\Rhs}{\ensuremath{R'_{h,\star}}}

\newcommand{\bas}{\ensuremath{(b/a)_\star}}
\newcommand{\cas}{\ensuremath{(c/a)_\star}}
\newcommand{\rs}{\ensuremath{r_s}}
\newcommand{\rv}{\ensuremath{r_\mathrm{180}}}
\newcommand{\Mv}{\ensuremath{M_\mathrm{180}}}
\newcommand{\mv}{\ensuremath{m_\mathrm{180}}}
\newcommand{\Rh}{\ensuremath{R'_h}}
\newcommand{\st}{\ensuremath{\sigma_\mathrm{tot}}}
\newcommand{\sig}[1]{\ensuremath{\sigma_{#1}}}
\newcommand{\Rein}{\ensuremath{R_\mathrm{ein}}}
\newcommand{\Rrad}{\ensuremath{R_\mathrm{rad}}}
\newcommand{\Mein}{\ensuremath{M_\mathrm{ein}}}

\newcommand{\fpdm}{\ensuremath{f'_\mathrm{dm}}}
\newcommand{\rmd}{\mathrm{d}} % upright d
\newcommand{\mathpc}{\mathrm{pc}}
\newcommand{\gravlens}{\textsc{gravlens}}
\newcommand{\idl}{\textsc{idl}}
\newcommand{\Ie}{\ensuremath{\langle I_e \rangle}}

\title[Strong lensing simulations]{Galaxy density profiles and shapes
  -- I. simulation pipeline for lensing by realistic galaxy models}

\author[van de Ven, Mandelbaum \& Keeton]{%
  Glenn van de Ven$^1$\thanks{\texttt{glenn@ias.edu}, Hubble Fellow},
  Rachel Mandelbaum$^1$\thanks{\texttt{rmandelb@ias.edu}, Hubble Fellow},
  Charles R.\  Keeton$^2$\thanks{\texttt{keeton@physics.rutgers.edu}}
  \\
  $^1$Institute for Advanced Study, Einstein Drive, Princeton NJ
  08540, USA 
  \\
  $^2$Department of Physics and Astronomy, Rutgers University, 136
  Frelinghuysen Road, Piscataway NJ, 08854, USA
}

\begin{document}

%\date{Draft \today \hfill\fbox{\textbf{\emph{DO NOT DISTRIBUTE}}}}

\maketitle 

\begin{abstract}
  Studies of strong gravitational lensing in current and upcoming wide
  and deep photometric surveys, and of stellar kinematics from
  (integral-field) spectroscopy at increasing redshifts, promise to
  provide valuable constraints on galaxy density profiles and shapes.
  However, both methods are affected by various selection and
  modelling biases, whch we aim to investigate in a consistent way.
  In this first paper in a series we develop a flexible but efficient
  pipeline to simulate lensing by realistic galaxy models.  These
  galaxy models have separate stellar and dark matter components, each
  with a range of density profiles and shapes representative of
  early-type, central galaxies without significant contributions from
  other nearby galaxies. We use Fourier methods to calculate the
  lensing properties of galaxies with arbitrary surface density
  distributions, and Monte Carlo methods to compute lensing statistics
  such as point-source lensing cross-sections.  Incorporating a
  variety of magnification bias modes lets us examine different survey
  limitations in image resolution and flux. We rigorously test the
  numerical methods for systematic errors and sensitivity to basic
  assumptions.  We also determine the minimum number of viewing angles
  that must be sampled in order to recover accurate
  orientation-averaged lensing quantities. We find that for a range of
  non-isothermal stellar and dark matter density profiles typical of
  elliptical galaxies, the combined density profile and corresponding
  lensing properties are surprisingly close to isothermal around the
  Einstein radius.  The converse implication is that constraints from
  strong lensing and/or stellar kinematics, which are indeed
  consistent with isothermal models near the Einstein radius, cannot
  trivially be extrapolated to smaller and larger radii.
\end{abstract}

\begin{keywords}
  gravitational lensing -- stellar dynamics -- galaxies: photometry --
  galaxies: kinematics and dynamics -- galaxies: structure -- methods:
  numerical 
\end{keywords}

%=====================================================================
\section{Motivation}
\label{S:motivation}
%=====================================================================

%---------------------------------------------------------------------
\subsection{Learning from galaxy density profiles and shapes}
\label{SS:intro1}
%---------------------------------------------------------------------

Observational constraints on galaxy density profiles and shapes can be
used to study a great variety of problems, from basic cosmology to the
connections between dark matter and baryons.  For example, the
spherically-averaged form of the dark matter halo density profile, and
the distribution of halo shapes, both depend on cosmological
parameters to the extent that those parameters affect the process of
structure formation \citep[e.g.,][]{2001MNRAS.321..559B,
  2006MNRAS.367.1781A}, and on the physics of galaxy formation to the
extent that baryons modify the dark matter distribution
\citep[e.g.,][]{1986ApJ...301...27B, 2004ApJ...616...16G,
  2004ApJ...611L..73K, 2007ApJ...658..710N, 2008ApJ...672...19R}.
While there is still some disagreement in the literature about the
theoretical predictions, even among those papers cited above, one key
result is that there is considerable scatter in both the density
profiles and shapes of dark matter halos, which presumably reflects
different formation histories \citep[e.g.,][]{2002ApJ...568...52W}.

Several observational tools have been used to constrain the density
profiles and shapes of galaxies, at different length and mass scales,
and redshifts. The main tools we consider are gravitational lensing
and stellar kinematics. X-ray data are also useful for understanding
the density profiles and shapes of massive galaxies and galaxy
clusters, but are beyond the scope of our investigation.

Gravitational lensing is the deflection of light from distant sources
by the gravitational fields of intervening lens galaxies.  Strong
lensing occurs in the central regions of galaxies (and clusters) where
the light bending is extreme enough to produce multiple images of a
background source.  It can be used to study the mass distributions of
individual galaxies on projected scales of typically several kpc
\citep[for a review, see ][]{Saas-Fee}.  Weak lensing is a
complementary phenomenon in which the deflection of light slightly
distorts the shapes of background sources without creating multiple
images \citep[for a review, see][]{2001PhR...340..291B}.  Weak lensing
probes galaxy mass distributions on scales from tens of kpc out to
about 10 Mpc, but only yields constraints on ensemble averages, since
currently weak lensing by galaxies can only be detected by stacking
many lens galaxies. While galaxy clusters can by studied on an
individual basis using weak lensing, our focus is on galaxies.

High-quality (two-dimensional) data on the kinematics of stars as well
as gas in the inner parts (a few to ten kpc) of galaxies are now
readily available, thanks in particular to strong progress in
integral-field spectroscopy in the last decade. Kinematics in the
outer parts of late-type galaxies can often be observed from the
presence of neutral hydrogen \citep[e.g.,][]{1981AJ.....86.1825B,
  1985ApJ...295..305V, 1996MNRAS.281...27P, 2007MNRAS.376.1513N}. In
the outer parts of early-type galaxies, however, cold gas is scarce
\citep[but see e.g.][]{1994ApJ...436..642F, 1997AJ....113..937M,
  2008MNRAS.383.1343W}, so we are left with discrete kinematic tracers
such as planetary nebulae and globular cluster
\citep[e.g.,][]{2003ApJ...591..850C, 2007ApJ...664..257D} out to tens
of kpc. 

Current and upcoming wide and deep photometric surveys will reveal a
vast number of strong lensing events \citep[e.g.,][]{
  2004AAS...20510827F, 2004NewAR..48.1085K, 2004ApJ...601..104K,
  2005NewAR..49..387M}, and will also allow for extensive,
complementary weak lensing analysis
\citep[e.g.][]{2002SPIE.4836...10T, 2004SPIE.5489...11K}. At the same
time, there is a rapid increase in the availability of two-dimensional
kinematics of galaxies nearby \citep[e.g.,][]{2004MNRAS.352..721E,
  2006MNRAS.373..906M}, and even at high(er) redshift
\citep[e.g.,][]{2007ApJ...668..738V, 2007ApJ...671..303B}. These data
sets, individually as well as combined, may provide strong constraints
on galaxy density profiles and shapes, but the analyses involved are
subject to selection and modelling biases.

%---------------------------------------------------------------------
\subsection{Selection and modelling biases}
\label{SS:intro2}
%---------------------------------------------------------------------

If we want to use strong lensing and kinematics to make robust
tests of cosmological predictions, we need to answer two questions:
Can strong lensing and kinematic analyses yield accurate constraints
on galaxy density profiles?  Are the galaxies in which we can make
the measurements (especially in the case of strong lensing)
representative of all galaxies?  We refer to these two concerns
as \emph{modelling biases} and \emph{selection biases}, respectively.

The question of selection bias is particularly important given the
diversity in the galaxy population. It is well known that strong
lensing favours early-type over late-type galaxies, and massive
galaxies over dwarfs \citep[e.g.,][]{1984ApJ...284....1T,
  1991MNRAS.253...99F}.  But even within the population of massive
early-type galaxies, to what extent does strong lensing favour galaxies
whose dark matter halos have inner slopes that are steeper than
average, or concentrations that are higher than average?  Also, to
what extent does strong lensing favour galaxies with particular shapes
and/or orientations with respect to the line-of-sight? To phrase these
questions formally, consider some parameter $x$ describing the galaxy
density profile or shape (e.g., the inner slope of the dark matter
density profile). We need to understand how the distribution
$p_\mathrm{SL}(x)$ among strong lens galaxies compares to the
underlying distribution $p(x)$ for all galaxies.
Any analysis that includes strong lensing with some other technique
used to study the same systems will suffer from strong lensing-related
selection biases, since we never get to choose which systems will be a
strong lens.  Selection biases are critical when we want to interpret
constraints on lens galaxy density profiles and/or shapes from strong
lensing, possibly in combination with kinematics
\citep[e.g.][]{2005ApJ...623..666R, 2006ApJ...649..599K} in comparison
with predictions from cosmological models.

On the other hand, modelling biases may occur because of (often
unavoidable) assumptions made when analysing gravitational lensing
and/or kinematic data because of the finite number of constraints
available from the data. Since galaxies are in general non-spherical,
we can only correctly interpret the observations if we know the
viewing direction, and even then the deprojection might not be unique
\citep[e.g.][]{1987IAUS..127..397R}. Moreover, strong lensing is
subject to the so-called mass-sheet degeneracy: part of the deflection
and magnification of the light from the background source can be due
to mass along the line-of-sight that is not associated with the lens
galaxy itself\footnote{Part of this mass-sheet degeneracy may be
  overcome in future surveys through the measurements of time delays
  for an adopted Hubble constant.}. In addition, the constraints from
strong lensing on the galaxy density are typically limited in radius
to around the Einstein radius, and even then, without secure,
unaffected measurements of the flux of the lensing images, the density
profile is difficult to recover.  Kinematics can be obtained over a
larger radial extent, but in particular stellar kinematics suffer from
the so-called mass-anisotropy degeneracy: a change in the measured
line-of-sight velocity dispersion can be due to a change in total
mass, but may also be the result of velocity anisotropy.

% Some notes by Chuck: Selection bias is important when we interpret
% constraints on lens galaxy density profiles (e.g., Rusin et al.;
% Treu/Koopmans et al.)  in comparison with predictions from
% cosmological models.  Modeling bias may also be important there,
% because most of that work has been done with single-component models.
% Another issues to mention in relation to modeling bias is the
% quad/double ratio: that ratio is sensitive to the distribution of
% galaxy shapes, but real galaxy shapes are (presumably) more
% complicated than the simple ellipsoids usually used in the
% calculations. We could mention selection bias in relation to dark
% matter substructure, because substructure studies are usually done
% with quad lenses which are biased even among lens galaxies; does that
% mean, for example, that quad lenses will tend to have
% higher-than-average amounts of projected substructure?  I suppose
% modeling bias may be relevant for substructure, too, because the
% macromodel flux ratios (which provide the benchmark when interpreting
% observed flux ratios) may be different for realistic two-component
% models than for usual SIE+shear models.

In a series of papers, we are developing a pipeline for using
realistic galaxy models to simulate strong lensing and kinematic data,
and assess how both selection and modelling biases affect typical
strong lensing and kinematic analyses. In this first paper, we present
the simulation pipeline for point-source lensing. We focus on strong
lensing of quasars by early-type, central galaxies at different mass
scales, using two-component mass profiles (dark matter plus stellar
component) that are consistent with existing photometry and stacked
weak lensing data from SDSS
\citep{2003MNRAS.341...33K,2006MNRAS.368..715M}, $N$-body, and
hydrodynamic simulations. In \citeauthor{2008paperII}
(\citeyear{2008paperII}, hereafter Paper~II), we use this pipeline to
study selection biases in strong lensing surveys. In future work we
will address modelling biases in strong lensing and kinematics (both
when studied separately and when combined). Our overall goal is to
determine how strong lensing, kinematics, and possibly other probes
(weak lensing, and X-ray data for cluster mass scales) can be used to
obtain robust, unbiased constraints on galaxy density profiles and
shapes.

To make our presentation coherent, and to clarify our notation and
terminology, we first review the analytic treatment of ellipsoidal
mass distributions with different profiles and shapes
(Section~\ref{S:profiles-shapes}), and the basic theory of strong
lensing (Section~\ref{S:stronglensing}).  We then describe our
simulation pipeline for point-source lensing in depth.  In
Section~\ref{S:galmodels} we present our choices for the masses,
profiles, and shapes of the galaxy models. In
Section~\ref{S:lensingcalc} we discuss the numerical methods we use
for lensing calculations, including numerous tests. A summary of the
simulation pipeline, and some implications for strong lensing analyses
are given in Section~\ref{S:conclusions}.

%=====================================================================
\section{Galaxy density profiles and shapes}
\label{S:profiles-shapes}
%=====================================================================

In this section we describe our treatment of galaxy density profiles
and shapes.  While our approach is fairly conventional, it is important
to present it carefully to clarify our notation and terminology,
collect useful technical results, and provide a firm foundation for
the work in this series of papers.  This section focuses on the
formal framework; the specific galaxy models used in our simulations
are discussed in Section~\ref{S:galmodels}.

%---------------------------------------------------------------------
\subsection{Notation}
\label{SS:notation}
%---------------------------------------------------------------------

We compute distances using a flat $\Lambda$CDM cosmology with
$\Omega_m=0.27$ and $h=0.72$.  We use $x$, $y$, and $z$ to denote
intrinsic, three-dimensional (3d) coordinates, and $x'$ and $y'$ for
the projected, two-dimensional (2d) coordinates. Similarly, $r$ is the
intrinsic radius ($r^2=x^2+y^2+z^2$) and $R'$ is the projected radius
($R'^2=x'^2+y'^2$). For non-spherical galaxies, we use $a$, $b$, and
$c$ for the major, intermediate, and minor semi-axis lengths of the
intrinsic, 3d density profiles; and we use $a'$ and $b'$ for the major
and minor semi-axis lengths of the projected, 2d surface densities.
Subscripts ``dm'' and ``$\star$'' are used to indicate whether a
quantity describes the dark matter or stellar component of the galaxy
model.  Any gas component that has not cooled to form stars --- which
is expected to be subdominant relative to the stellar component for
early type galaxies \citep[e.g.][]{2005RSPTA.363.2693R,
  2006MNRAS.371..157M} --- is implicitly included in the component
that we label dark matter.

%---------------------------------------------------------------------
\subsection{Ellipsoidal shapes}
\label{SS:ellipsoidalshapes}
%---------------------------------------------------------------------

We consider mass density distributions $\rho(x,y,z) = \rho(m)$ that
are constant on ellipsoids
\begin{equation}
  \label{eq:ellipsoid}
  m^2 = \frac{x^2}{a^2} + \frac{y^2}{b^2} + \frac{z^2}{c^2},
\end{equation}
with $a \ge b \ge c$. The major semi-axis length $a$ is a scale
parameter, whereas the intermediate-over-major ($b/a$) and
minor-over-major ($c/a$) axis ratios determine the shape. In the
oblate or prolate axisymmetric limit we have $a=b>c$ (pancake-shaped)
or $a>b=c$ (cigar-shaped), respectively, while in the spherical
limit $a=b=c$ (so then $m=r/a$).  Note that $m$ is a dimensionless
ellipsoidal radius.

Under the thin-lens approximation, the gravitational lensing
properties are fully characterised by the mass density projected along
the line-of-sight. We introduce a new Cartesian coordinate system
$(x'',y'',z'')$, with $x''$ and $y''$ in the plane of the sky and the
$z''$-axis along the line-of-sight. Choosing the $x''$-axis in the
$(x,y)$-plane of the intrinsic coordinate system (cf.\ 
\citealt{1989ApJ...343..617D} and their Fig.~2), the transformation
between both coordinate systems is known once two viewing angles, the
polar angle $\vartheta$ and azimuthal angle $\varphi$, are specified.
The intrinsic $z$-axis projects onto the $y''$-axis; for an
axisymmetric galaxy model the $y''$-axis aligns with the short axis
of the projected mass density,\footnote{For an oblate galaxy, the
  alignment ($\psi=0$) follows directly from
  equation~\eqref{eq:misalignment_psi} since $T=0$. For a prolate
  galaxy it is most easily seen by exchanging $a$ and $c$ ($c>b=a$),
  so that the $z$-axis is again the symmetry axis (instead of the
  $x$-axis).}
but for a triaxial galaxy model the $y''$-axis is misaligned by an
angle $\psi \in [-\pi/2,\pi/2]$ such that (cf.\ equation~B9 of
\citealt{1988MNRAS.231..285F})
\begin{equation}
  \label{eq:misalignment_psi}
    \tan2\psi = \frac{T\sin2\varphi\cos\vartheta}
    {\sthq + T\left(\sphq\cthq-\cphq\right)}\ ,
\end{equation}
where $T$ is the triaxiality parameter defined as $T =
(a^2-b^2)/(a^2-c^2)$. A rotation through $\psi$ transforms the
coordinate system $(x'',y'',z'')$ to $(x',y',z')$ such that the $x'$
and $y'$ axes are aligned with the major and minor axes of the
projected mass density (respectively), while $z'=z''$ is along the
line-of-sight (see also Section~\ref{SSS:misalignment} below).

Projecting $\rho(m)$ along the line-of-sight yields a surface mass
density $\Sigma(x',y') = \Sigma(m')$ that is constant on ellipses
in the sky-plane,
\begin{eqnarray}
  \label{eq:surfmassell}
  \Sigma(m')  
  = \int_{-\infty}^{\infty} \rho(m) \, \rmd z'
  = \frac{abc}{a'b'} \; 2 \int_{0}^{\infty} 
  \rho(m) \, m \; \rmd u,
\end{eqnarray}
where we have used $z' = abc \, \sinh(u)/(a'b') + \mathrm{constant}$,
and $m = m' \cosh(u)$. The sky-plane ellipse is given by
\begin{equation}
  \label{eq:ellipse}
  m'^2 = \frac{x'^2}{a'^2} + \frac{y'^2}{b'^2}.
\end{equation}
The projected major and minor semi-axis lengths, $a'$ and $b'$,
depend on the intrinsic semi-axis lengths $a$, $b$, and $c$ and
the viewing angles $\vartheta$ and $\varphi$ as follows:
\begin{equation}
  \label{eq:apbpellipse}
  a'^2 = \frac{2\,A^2}{B - \sqrt{B^2 - 4 A^2}},
  \qquad
  b'^2 = \frac{2\,A^2}{B + \sqrt{B^2 - 4 A^2}},
\end{equation}
where $A$ and $B$ are defined as
\begin{eqnarray}
  \label{eq:defA}
  A^2 & = & 
  a^2 b^2 \cthq + ( a^2 \sphq + b^2 \cphq ) c^2 \sthq,
  \\ \label{eq:defB}
  B & = &  
  a^2 (\cphq\cthq+\sphq)
  \nonumber \\  
  && + b^2 (\sphq\cthq+\cphq) + c^2 \sthq.
\end{eqnarray}
It follows that $A=a'b'$ is proportional to the area of the ellipse.

The flattening $b'/a'$ of the projected ellipses actually depends
on the viewing angles $(\vartheta,\varphi)$ and the intrinsic axis
ratios $(b/a,c/a)$, and is independent of the scale length.  Since
the intrinsic and projected semi-major axis lengths are directly
related via the left equation of \eqref{eq:apbpellipse}, the scale
length can be set by choosing either $a$ or $a'$.

Finally, we define the negative logarithmic slope of the mass
density $\gamma(m) = - \rmd\ln\rho(m)/\rmd\ln m$ and of the surface mass
density $\gamma'(m') = - \rmd\ln\Sigma(m')/\rmd\ln m'$. Substituting
equation~\eqref{eq:surfmassell}, the latter follows as
\begin{equation}
  \label{eq:logslopesurfell}
  \gamma'(m') = 
  \frac{\int_{0}^{\infty} [\gamma(m) - 1] \rho(m)\,m'\; \rmd u}
  {\int_{0}^{\infty} \rho(m)\,m\; \rmd u},
\end{equation}
which in general has to be evaluated numerically.

%---------------------------------------------------------------------
\subsection{Choice of density profiles}
\label{SS:densprof}
%---------------------------------------------------------------------

We consider two families of density profiles, motivated by both
observations and simulations of galaxies.  Historically, there has
been considerable interest in cusped density profiles that have a
shallower power-law at small radii and a steeper power-law at large
radii, with a smooth transition in between.  This family includes the
\citet{1990ApJ...356..359H} profile that is commonly used to model the
stellar components of early-type galaxies and spiral galaxy bulges,
along with (generalised) NFW \citep{1997ApJ...490..493N} profiles
often used to describe the dark matter profiles of simulated galaxies.

An alternative, observationally motivated family of models is obtained
by deprojecting the \cite{1968adga.book.....S} profile that is known
to give a good fit to the surface brightness profiles of early-type
galaxies and spiral galaxy bulges. There have been claims that
deprojected S\'ersic profiles actually provide a better fit to
simulated dark matter halos than (generalised) NFW profiles
\citep[e.g.,][]{2004MNRAS.349.1039N, 2005ApJ...624L..85M}.  Several
recent studies argue that a S\'ersic profile that is not deprojected
---also referred to as an \citet{Einasto65} profile--- provides an
even better fit \citep{2006AJ....132.2685M, 2008MNRAS.387..536G},
although when compared against a generalised NFW or S\'ersic profile
the improvement is marginal at best.

%% In support of our claim that a cusped profile is a reasonable
%% representation of reality, we note that 
X-ray \citep[e.g., ][]{2001MNRAS.324..877A, 2002MNRAS.335..256A,
  2005ApJ...628..655V, 2006ApJ...640..691V} and weak lensing
\citep[e.g., ][]{2003ApJ...598..804K, 2007ApJ...668..643L}
observations of individual galaxy clusters suggest that outside of
$\sim 50$ kpc, the density profiles are consistent with the NFW
profiles seen in N-body simulations (no attempts were made to compare
with the S\'ersic model, however). The mass range for these
observations is about a factor two above our higher mass model.

Since there is still no consensus opinion about which models are best,
we consider how the choice of density profile affects our conclusions
by using both cusped density profiles and deprojected S\'ersic density
profiles

\subsection{Cusped density profiles}
\label{SS:cuspdens}
%---------------------------------------------------------------------

The cusped density distribution with inner slope $\gamma$ and outer
slope $n$,
\begin{equation}
  \label{eq:denscusp}
  \rho(m) = \frac{\rho_0}{m^\gamma (1+m)^{n-\gamma}},
\end{equation}
includes the following well-known spherical ($m=r/a$) density profiles:
$(\gamma,n)=(1,4)$, the Hernquist profile \citep{1990ApJ...356..359H};
$(2,4)$, the Jaffe profile \citep{1983MNRAS.202..995J}; and $(1,3)$,
the NFW profile \citep{1997ApJ...490..493N}.
% and Moore $(3/2,3)$

The mass enclosed within the ellipsoidal radius $m$ (for inner slopes
$\gamma<3$) is
\begin{eqnarray}
  \label{eq:masscusp}
  M(m) & = & 4 \pi a b c \rho_0 \; 
  \left(\frac{m}{1+m}\right)^{3-\gamma}
  \frac{1}{3-\gamma}
  \nonumber \\ && \times \; \;
  {}_2F_1\left[n-\gamma,3-\gamma;4-\gamma;\frac{m}{1+m}\right], 
  %\left[m/(1+m)\right]^{3-\gamma}
\end{eqnarray}
where ${}_2F_1[\alpha,\beta,\gamma;x]$ is the hypergeometric function.
For outer slopes $n>3$, equation~\eqref{eq:masscusp} reduces to $M(m) = 4
\pi a b c \rho_0 \, \beta[n-3,3-\gamma;m/(1+m)]$. When $m \to \infty$,
the latter incomplete beta function becomes the complete beta function
$\beta[n-3,3-\gamma]$, and we obtain a finite total mass.

Substituting $n=4$ into equation~\ref{eq:masscusp} yields $M(m) = 4 \pi a b
c \rho_0 [m/(1+m)]^{3-\gamma}/(3-\gamma)$, so that for $\gamma=1$, we
find the total mass $M=2 \pi a b c \rho_0$ of the Hernquist profile,
which we adopt for the stellar component when using cusped density
profiles. For outer slopes $n \le 3$, the total mass is infinite, but
for certain half-integer values of $\gamma$ and $n$ the expression for
the enclosed mass simplifies significantly. For example, for $n=3$,
the value which we adopt for the outer slope of the dark matter
profile, we find
\begin{equation}
  \label{eq:masscuspn3}
  \frac{M(m)}{4 \pi a b c \rho_0} = 
  \begin{cases}
    %2\sinh^{-1}(\sqrt{m}) - 2\sqrt{m}(23m^2+35m+15)/[15(1+m)^{5/2}] \\
    \ln(1+m) - m(2+3m)/[2(1+m)^2], \\
    2\sinh^{-1}(\sqrt{m}) - 2\sqrt{m}(4m+3)/[3(1+m)^{3/2}], \\
    \ln(1+m) - m/(1+m), \\
    2\sinh^{-1}(\sqrt{m}) - 2\sqrt{m/(1+m)}, \\
    \ln(1+m), \\
    %2\sinh^{-1}(\sqrt{m}) & \text{$\gamma=5/2$}, \\
  \end{cases}
\end{equation}
for inner slope values of $\gamma=\{0,1/2,1,3/2,2\}$, respectively.
%Similar expressions for $n=5/2$ and $n=2$ are given in Appendix~XX.

Although for certain values of $\gamma$ and $n$, lengthy analytic
expressions for the surface mass density $\Sigma(m')$ can be derived,
we evaluate the integral in equation~\eqref{eq:surfmassell}
numerically. 
%Analytic expressions for $\Sigma(m')$ for $\gamma=\{1/2,1,3/2\}$ and
%$n=\{3,4\}$ are given in Appendix~XX.
The logarithmic slope $\gamma(m)$ of the cusped density profile is
\begin{equation}
  \label{eq:logslopedenscusp}
  \gamma(m) = (\gamma + m \, n)/(1 + m),
\end{equation}
while $\gamma'(m')$ of the surface mass density follows from 
numerical evaluation of equation~\eqref{eq:logslopesurfell}.

%---------------------------------------------------------------------
\subsection{Deprojected S\'ersic density profiles}
\label{SS:sersicdens}
%---------------------------------------------------------------------

It has long been known that the surface brightness profiles of
early-type galaxies and of spiral galaxy bulges are well fit by a
\cite{1968adga.book.....S} profile $I(R) \propto
\exp[-(R/R_e)^{1/n}]$, with the effective radius $R_e$ enclosing half
of the total light.  A key conceptual difference from the cusped
models is that the profile does not converge to a particular inner
slope on small scales.  The deprojection of the S\'ersic profile has
to be done numerically.  However, the analytic density profile of
\cite{1997A&A...321..111P},
\begin{eqnarray}
  \label{eq:rhosersic}
  \rho(m) & = & \frac{\rho_0}{m^{p_n}} \; 
  \exp\left[-b_n\,m^{1/n} \right],
\end{eqnarray}
provides a good match to the deprojected S\'ersic profile when the
inner negative slope is given by 
\begin{equation}
  \label{eq:psersic}
  p_n = 1 - \frac{0.6097}{n} + \frac{0.05563}{n^2}.
\end{equation}
The enclosed mass for the Prugniel-Simien model is
\begin{equation}
  \label{eq:masssersic}
  M(m) = 4 \pi a b c \rho_0 \; 
  n \, b_n^{(p_n-3)n} \, \gamma[(3-p_n)n,b_n\,m^{1/n}],
\end{equation}
where $\gamma[p;x]$ is the incomplete gamma function, which in the
case of the total mass reduces to the complete gamma function
$\Gamma[p] = \gamma[p;\infty]$.

The expression for the surface mass density is, to high accuracy, the
S\'ersic profile
\begin{equation}
  \label{eq:surfsersic}
  \Sigma(m') = \Sigma_0 \;
  \exp\left[-b_n\,(m')^{1/n} \right].
\end{equation}
Given the enclosed projected mass
\begin{equation}
  \label{eq:2dmasssersic}
  M'(m') = 2 \pi a' b' \Sigma_0 \; 
  n \, b_n^{-2n} \, \gamma[2n,b_n\,(m')^{1/n}],
\end{equation}
the requirement that the total intrinsic and projected mass have to be
equal yields a normalisation 
\begin{equation}
  \label{eq:normalizationsersic}
  \Sigma_0 = \frac{abc}{a'b'} \, \rho_0 \, 
  \frac{2\,\Gamma[(3-p_n)n]}
  {b_n^{(1-p_n)n}\,\Gamma[2n]}.
\end{equation}
The value of $b_n$ depends on the index $n$ and the choice for the
scale length. The latter is commonly chosen to be the effective radius
$R_e$ in the surface brightness profile, which contains half of the
total light. We adopt a similar convention requiring that the ellipse
$m'=1$ contains half of the projected mass. This choice results in the
relation $\Gamma[2n] = 2\,\gamma[2n,b_n]$, which to high precision can
be approximated by \citep{1999A&A...352..447C}
\begin{equation}
  \label{eq:approxb}
  b_n = 2\,n 
  - \frac{1}{3} 
  + \frac{4}{405} \, \frac{1}{n} 
  + \frac{46}{25515} \, \frac{1}{n^2}. 
\end{equation}

The logarithmic slope of the Prugniel-Simien and S\'ersic profile are
related by
\begin{equation}
  \label{eq:logslopedenssersic}
  \gamma(m) = p_n + \gamma'(m),
\end{equation}
where the logarithmic slope of the surface mass density is
\begin{equation}
  \label{eq:logslopesurfsersic}
  \gamma'(m') = (b_n/n) \, (m')^{1/n}.
\end{equation}
%

%=====================================================================
\section{Strong lensing}
\label{S:stronglensing}
%=====================================================================

In this section, we review the strong lensing concepts that are most
important for our work.  See \citet{1992grle.book.....S} and
\citet{Saas-Fee} for more discussion of strong lensing theory.
The specific lensing calculations used in our simulations are
discussed in Section~\ref{S:lensingcalc}.

%---------------------------------------------------------------------
\subsection{Basic theory}
\label{SS:lenstheory}
%---------------------------------------------------------------------

The gravitational lensing properties of a galaxy with surface mass
density $\Sigma(R')$ are characterised by the lens potential $\phi$
(in units of length squared) that satisfies the two-dimensional
Poisson equation
\begin{equation} \label{eq:Poisson}
  \nabla^2 \phi = 2 \kappa\,,
  \quad\mbox{where}\quad
  \kappa \equiv \Sigma/\Sigma_c\,.
\end{equation}
Here $\kappa$ is the surface mass density scaled by the critical
density for lensing ($\Sigma_c$, see Section~\ref{SS:redshifts}), and
is referred to as the ``convergence.''  The positions and
magnifications of lensed images depend on the first and second
derivatives of the lens potential, respectively.  The image
positions are the solutions of the lens equation,
\begin{equation}\label{E:lensequation}
  \vec{\beta} = \vec{\theta} - \vec{\alpha}(\vec{\theta})\,,
  %\vec{\beta} = \vec{\theta} - \vec{\alpha}(\vec{\theta})/D_L\,,
\end{equation}
where $\vec{\alpha} = \vec{\nabla}\phi$ is the deflection angle (in
units of length), while $\vec{\theta} = (\theta_1,\theta_2)$ and
$\vec{\beta} = (\beta_1,\beta_2)$ are two-dimensional angular
positions on the sky in the lens and source planes, respectively.  The
angular coordinates are related to the projected physical coordinates
by $\theta_1 = x'/D_L$ and $\theta_2 = y'/D_L$, where $D_L$ is the
angular diameter distance to the lens; and $\beta_1 = x'/D_S$ and
$\beta_2 = y'/D_S$, where $D_S$ is the angular diameter distance to
the source.\footnote{Note that because of this trivial conversion
  between angular and physical units, we do not explicitly distinguish
  between angles and lengths in the lens plane.  For example in the
  text we refer to the deflection angle in physical units (e.g., by
  stating that $\alpha(\Rein)=\Rein$) whereas our plots may show
  $\alpha$ in angular units.}  The magnification of an image at
position $(x',y')$ is given by
\begin{equation} \label{eq:mu}
  \mu = \left[ \left(1-\frac{\partial^2\phi}{\partial x'^2}\right)
    \left(1-\frac{\partial^2\phi}{\partial y'^2}\right)
    - \left(\frac{\partial^2\phi}{\partial x' \partial y'}\right)^2
  \right]^{-1} .
\end{equation}
A typical lens galaxy has two ``critical curves'' along which the
magnification is formally infinite, which map to ``caustics'' in the
source plane.  The caustics bound regions with different numbers of
images (see Section~\ref{SS:resolution} for examples).

If the lens is circularly symmetric, the deflection is radial and
has amplitude
\begin{equation} \label{eq:circdef}
  \alpha(R') = \frac{2}{R'} \int_{0}^{R'}
    \kappa(\tilde{R}')\,\tilde{R}'\,\rmd\tilde{R}'\,.
\end{equation}
The outer or ``tangential'' critical curve corresponds to the ring
image that would be produced by a source at the origin.  The radius of
this ring is the Einstein radius $\Rein$, which is given
mathematically by the solution of the equation $\alpha(\Rein) =
\Rein$.  This definition is equivalent to saying that the Einstein
radius bounds the region within which the average convergence is
unity, such that the enclosed mass $\Mein = \pi \Rein^2 \Sigma_c$.  By
contrast, the inner or ``radial'' critical curve has radius $\Rrad$
given by the solution of $\rmd\alpha/\rmd R' = 1$.

If the lens is non-circular, the conditions for the critical curves
and caustics are more complicated, but we can still retain some
concepts from the circular case. In particular, we can say that the
outer/tangential critical curve is principally determined by the
enclosed mass, whereas the inner/radial critical curve is determined
by the slope of the deflection curve. When we map the critical curves
in the lens plane to the caustics in the source plane, the
arrangement of curves is inverted: the outer critical curve
corresponds to the inner caustic; while the inner critical curve
corresponds to the outer caustic.\footnote{If the lens is highly
  flattened, the tangential caustic can actually pierce the radial
  caustic; some specific examples are discussed below (e.g.,
  Section~\ref{SS:resolution}).  However, this situation is unusual, and it
  does not substantially alter the present discussion.}  Since the
outer caustic bounds the multiply-imaged region, it determines the
total lensing cross-section. This means the cross-section is sensitive
to the surface mass density (in particular its logarithmic slope) at
small radii in the lens plane, which will be important to bear in
mind when interpreting our results.

While a non-circular lens has non-circular critical curves, it is
still occasionally useful to characterise the scale for strong lensing
with a single Einstein radius.  We generalise the definition of the
Einstein radius to a non-circular lens by using the monopole
deflection, or the deflection angle produced by the monopole moment of
the lens galaxy,
\begin{equation} \label{eq:monodef}
  %\alpha_0(R') = \frac{1}{\pi R'} \int_{0}^{R'} \tilde{R} \rmd \tilde{R}
  %\int_{0}^{2\pi} \rmd \tilde{\theta} \kappa(\tilde{R},\tilde{\theta})
  \alpha_0(R') = \frac{1}{\pi R'} 
  \int_{0}^{2\pi} \int_{0}^{R'}
  \kappa(\tilde{R},\tilde{\theta}) \;
  \tilde{R} \, \rmd \tilde{R} \; \rmd \tilde{\theta}
  = R'\,\overline{\kappa}(R')\,,
\end{equation}
where $\overline{\kappa}(R')$ is the average convergence within radius
$R'$, and $(\tilde{R},\tilde{\theta})$ are polar coordinates on the
sky.
% \footnote{Strictly speaking it would be better to write the integral
%   in terms of angular variables, but we do not distinguish between
%   angular and physical lengths because they are trivially related by
%   the angular distance to the lens ($D_L$).}
The Einstein radius is then given by $\alpha_0(\Rein) = \Rein$, or
equivalently $\overline{\kappa}(\Rein) = 1$.  This is the definition that
emerges naturally from models of non-spherical lenses
\citep[e.g.,][]{2001ApJ...554.1216C}, and it matches the conventional
definition in the circular case.

It is instructive to think of lensing in terms of Fermat's principle
and say that lensed images form at stationary points of the arrival
time surface \citep[e.g.,][]{1986ApJ...310..568B}.  Since the arrival
time is a 2d function of angles on the sky, there are three types of
stationary points: minima, maxima, and saddle points.  We can classify
them using the eigenvalues of the inverse magnification tensor,
\begin{equation} \label{eq:parities}
  \lambda_{\pm} = (1-\kappa) \pm \gamma\,,
  \quad\mbox{where}\quad
  \gamma = \left[{(1-\kappa)^2 - \mu^{-1}}\right]^{1/2} .
\end{equation}
Here, $\gamma$ is the shear\footnote{In general the shear has two
  components conveniently expressed in complex notation
  \citep[e.g.][]{1992grle.book.....S}, but as usual in strong lensing
  analyses we only use the amplitude (or norm) for which $\gamma$ is
  the standard symbol. The context should make clear whether we are using
  $\gamma$ to refer to a lensing shear or to
  the negative logarithmic slope of the density profile.}, which
quantifies how much a resolved image is distorted. The eigenvalues are
both positive for a minimum, both 
negative for a maximum, and mixed for a saddle point.  Since our
galaxies have central surface densities shallower than $\Sigma \propto
R'^{-1}$, the lensing ``odd image theorem'' applies: the total number
of images must be odd, and the number of minimum, maximum, and
saddle point images must satisfy the following relation
\citep{1981ApJ...244L...1B,1992grle.book.....S}:
\begin{equation} \label{eq:imgnum}
  N_\mathrm{min} + N_\mathrm{max} = N_\mathrm{sad} + 1\,.
\end{equation}
In practice, images at maxima are rarely observed because they form
near the centres of lens galaxies and are highly demagnified; the only
secure observation of a maximum image in a galaxy-scale lens required
a deep and dedicated search \citep{2004Natur.427..613W}.  We do
compute maxima because they are useful in checking our numerical
methods (see Section~\ref{SS:imgconfig}), but we focus our analysis on
minimum and saddle point images because they constitute the bulk of
lensing astrophysics.

%---------------------------------------------------------------------
\subsection{Redshift dependence}
\label{SS:redshifts}
%---------------------------------------------------------------------

The redshifts of the lens ($z_L)$ and source ($z_S$) play an important
role in determining what region of a galaxy is relevant for strong
lensing.  Roughly speaking, strong lensing occurs in the region where
the surface mass density exceeds the critical density for lensing,
\begin{equation}
  \label{eq:sigmacrit}
  \Sigma_c = \frac{c^2}{4\pi G}\frac{D_S}{D_L D_{LS}}\ ,
\end{equation}
where the $D$ values are the angular diameter distances to the
lens ($D_L$), to the source ($D_S$), and from the lens to the source
($D_{LS}$).  Note that $D_{LS}$ can be computed for a flat universe
using the relation
\begin{equation}
  D_{LS} = D_S - \left(\frac{1+z_L}{1+z_S}\right)D_L ,
\end{equation}
and for more general cosmologies, see \cite{1999astro.ph..5116H}.
% Fig.~\ref{F:sigmacrit} shows how the critical density depends on the
% lens and source redshifts.
For a fixed source redshift, $\Sigma_c$ is
lowest when the lens is roughly halfway between the observer and
source, and it increases as the lens moves toward the observer or
source.  For a fixed lens redshift, the critical density is formally
infinite for $z_S \le z_L$, and it decreases monotonically as the
source redshift increases past $z_L$.

% %%%FIG
% \begin{figure}
%   \begin{center}
%     \includegraphics[width=0.75\columnwidth,trim=0.5in 0 0 0]{scinvcontour.fiducial.ps}
%     \caption{\label{F:sigmacrit}Critical surface density, plotted as
%       $\log[\Sigma_c/(\mathrm{M}_\odot\,\mathrm{pc}^{-2})]$, as a
%       function of lens and source redshift.  For $z_S<z_L$, where
%       $\Sigma_c$ is formally infinite, we have set it to the maximum
%       value shown, $10^4$\,\Msunpcsq. Asterisks indicate the curve in
%       the ($z_L,z_S$) plane with our fiducial value of $\Sigma_c$.}
%   \end{center}
% \end{figure} 
% %%%FIG

We adopt fiducial redshifts of $z_L = 0.3$ for the lens and $z_S = 2$
for the source. The resulting critical density, $\Sigma_c =
2389\,M_\odot\,\mathpc^{-2}$, is such that quasar images generally
appear at about one effective radius from the centre of the lens
galaxies. At this radius, the stellar component and dark matter halo
may both play significant roles in the lensing signal, which is
important to keep in mind when interpreting our results in this and
subsequent papers.

Our results apply equally well to any other combination of lens and
source redshifts that yield $\Sigma_c = 2389\,M_\odot\,\mathpc^{-2}$.
% (see Fig.~\ref{F:sigmacrit}). 
The reason is that if we fix the galaxy density profile and vary $z_L$
and $z_S$ so as to keep $\Sigma_c$ fixed, the strong lensing region
will always have the same physical scale and hence the same relation
to the density profile.  If we instead vary $z_L$ and $z_S$ in a way
that increases $\Sigma_c$, that will tend to push the lensed images to
smaller radii and hence make the stellar component somewhat more
dominant.  Conversely, varying the redshifts in a way that lowers
$\Sigma_c$ will tend to make the dark matter component more
significant. We examine the effect of varying $z_L$ and $z_S$ on our
results explicitly in subsequent papers.

We note that the lens redshift also affects strong lensing in the
conversion from physical to angular scales.  Even if the physical size
of the lensing region stays fixed as we vary the source and lens
redshifts, the angular separation between the lensed images will vary
with $z_L$.  This may lead to \emph{observational} selection effects:
for example, resolution-limited surveys may miss lenses for which
$z_L$ is high and the images cannot be resolved, while spectroscopic
surveys may miss lenses for which $z_L$ is low and the images fall
outside the spectroscopic slit or fiber.  However, our focus in this
work is on \emph{physical} selection biases in strong lensing, which
means that we consider the lens and source redshifts only in
terms of how they affect $\Sigma_c$. 

Our choice $z_S = 2$ for the source redshift is typical of quasar
lenses in the CASTLES\footnote{CASTLES (see
  http://cfa-www.harvard.edu/castles/) is a collection of uniform HST
  observations of mostly point-source lenses from several samples with
  differing selection criteria, rather than a single,
  uniformly-selected survey.} sample, but our choice $z_L = 0.3$
places the lens galaxy on average at a smaller distance. The latter
is, however, closer to the typical lens redshift in the SLACS sample
of galaxy/galaxy strong lenses \citep{2006ApJ...638..703B,
  2008ApJ...682..964B}, but those lenses are explicitly selected to
have source redshifts $z_S \lesssim 0.8$ \citep{2004AJ....127.1860B}.
As a result, our fiducial value of $\Sigma_c$ is typically between
that of CASTLES and SLACS lenses. The Einstein radii at $\sim 1 \,
R_e$ in our case appear at larger radii in the CASTLES sample
\citep[$2 - 3 \, R_e$;][]{2003ApJ...587..143R}, and at smaller radii
in the SLACS sample \citep[$0.3 - 0.9 \,
R_e$;][]{2006ApJ...649..599K}.  The statistics of the SLACS survey are
much more involved than the statistics of quasar lens surveys, because
SLACS lenses have extended rather than point-like sources, and the
survey relies on fiber spectra from SDSS.  SLACS lens statistics have
recently been examined by \cite{2008ApJ...685...57D}, and we do not
consider them explicitly in the current work.

%=====================================================================
\section{Construction of Galaxy Models}
\label{S:galmodels}
%=====================================================================

In this section we construct the galaxy models used in our
simulations. We seek models with a variety of density profiles and
shapes that not only are realistic but also sample the range of
systematic effects in strong lensing. To that end, we must carefully
consider the mass and length scales
(Sections~\ref{SS:massscales}--\ref{SS:lengthscales}), the profiles
(Section~\ref{SS:densityprofiles}), and the shapes
(Section~\ref{SS:densityshapes}) of both the stellar and dark matter
components.  In addition to a main set of simulations with cusped
density profiles, we also create a subset of simulations based on
deprojected S\'ersic density profiles (Section~\ref{SS:sersicsim}).
After discussing the choice of the model parameters, we briefly
describe the part of our pipeline that computes surface mass density
maps for these models (Section~\ref{SS:compsurfmaps}), which can then
be used for lensing calculations (Section~\ref{S:lensingcalc}).

%---------------------------------------------------------------------
\subsection{Mass scales} 
\label{SS:massscales}
%---------------------------------------------------------------------

We use stellar and dark matter halo masses from SDSS for early-type
galaxies, where analyses of the spectra give an estimate of the
stellar mass \citep{2003MNRAS.341...33K}, and weak lensing analyses
from $0.02$--$2$\hMpc\ yield the dark matter mass
\citep{2006MNRAS.368..715M}.  The former are subject to systematic
uncertainties due to the initial mass function (IMF) at the $\sim
20$--$30$ per cent level, whereas the latter is subject to $\sim 10$
per cent uncertainty in the modelling in addition to $\sim 10$--$15$
per cent statistical error.

The selection of mass scales begins with the ansatz that we would like
to approximately bracket the typical luminosity and mass range of
observed lensing systems. Given the range of lens galaxy luminosities
in the CASTLES and SLACS strong lens samples, we choose models with
total luminosities of $\sim 2 L_*$ and $\sim 7 L_*$ in the (SDSS)
$r$-band. We refer to the lower-luminosity model interchangeably as
``galaxy scale'' or ``lower mass scale'' model.  The higher-luminosity
model is essentially a massive galaxy with convergence that is
significantly boosted by a group dark matter halo, but we often refer
to it as a ``group scale'' or ``higher mass scale'' model for brevity.
In both cases, the model includes concentric stellar and dark matter
components; thus, our current results cannot be applied to satellites
in groups/clusters, which will be the subject of future work. We also
neglect contributions of nearby (satellite) galaxies to the surface
density, focusing only on the host galaxy, and we neglect
contributions from other structures along the line of sight
\citep[e.g.][]{2004ApJ...612..660K, 2006ApJ...641..169M,
  2006ApJ...646...85W}.  The masses are selected from the
\cite{2006MNRAS.368..715M} SDSS weak lensing analysis using results
for early-type galaxies as a function of luminosity, as shown in
Table~3 or Fig.~4 of that paper.

The model parameters for these mass scales are summarised in
Table~\ref{T:massscales}. We define the virial radius entirely using
comoving quantities, with $\rv$ satisfying the relation
\begin{equation} 
  \label{eq:defvirialradius} 
  180\overline{\rho} =
  \frac{3\Mv}{4\pi \rv^3}, 
\end{equation} 
where $\Mv$ is the virial mass and $\overline{\rho}$ is the mean
density of the universe (using $\Omega_m=0.27$). The definition is in
principle arbitrary, but this choice was used to analyse the weak
lensing results. In the weak lensing results, the profiles were found
to be consistent with NFW for early-type galaxies on all scales
studied (see also \citealt{2006MNRAS.372..758M}, which focuses on
$z\sim 0.25$ Luminous Red Galaxies). Thus,
not only our use of the masses, but also the form of the density
profiles, is observationally motivated outside of $\sim 20$\hkpc.  

%%%TAB
\begin{table*} 
  \caption{\label{T:massscales}Summary of the spherical
    density profile parameters for galaxy and group scale models.
    With the exception of \Rhs, all numbers given are intrinsic
    (3d).}  
  \begin{tabular}{l|l|r|r|r} \hline \hline
    Description & Symbol & Galaxy-scale & Group-scale & Unit \\
    \hline
    \multicolumn{5}{c}{Dark matter component} \\
    \hline 
    Virial mass
    & $\Mvdm$ & $3.4 \times 10^{12}$ & $6.7 \times 10^{13}$ & \hMsun \\
    & $\Mvdm$ & $4.7 \times 10^{12}$ & $9.3 \times 10^{13}$ & \Msun \\
    Virial radius
    & $\rvdm$ & $390.7$ & $1055$ & comoving \hkpc \\
    & $\rvdm$ & $417$ & $1127$ & physical kpc \\
    & $\rvdm$ & $97.1$ & $262$ & arcsec \\
    Concentration
    & $\cdm$ ($\gdm=1$) & $8.4$ & $5.6$ & -- \\
    Scale radius
    & $\rsdm$ ($\gdm=1$) & $49.7$ & $201$ & physical kpc \\
    & $\rsdm$ ($\gdm=1$) & $11.56$ & $46.8$ & arcsec \\
    \hline
    \multicolumn{5}{c}{Stellar component} \\
    \hline 
    % Virial mass & $\Mvs$ & $11.2 \times 10^{10}$ & $5.3 \times
    % 10^{11}$ & \Msun \\
    % Virial radius
    % & $\rvs$ & $110.8$ & $186.5$ & comoving \hkpc \\
    % & $\rvs$ & $118$ & $199.3$ & physical kpc \\
    % & $\rvs$ & $27.54$ & $46.3$ & arcsec \\
    Total mass
    & $\Mts$ & $1.16 \times 10^{11}$ & $5.6 \times 10^{11}$ & \Msun \\
    Half-mass radius
    & $\Rhs$ & $4.07$ & $11.7$ & kpc \\
    (projected)
    & $\Rhs$ & $0.95$ & $2.7$ & arcsec \\
    Scale radius
    & $\rss$ ($\gs=1$) & $2.24$ & $6.4$ & physical kpc \\
    & $\rss$ ($\gs=1$) & $0.52$ & $1.5$ & arcsec \\
    % Concentration
    % & $\cs$ ($\gs=1$) & $52.3$ & $31.1$ & \\
    \hline 
  \end{tabular} 
\end{table*} 
%%%TAB

%---------------------------------------------------------------------
\subsection{Length scales} 
\label{SS:lengthscales}
%---------------------------------------------------------------------

We choose the dark matter halo concentration from the
\cite{2001MNRAS.321..559B} result that $\cdm \approx 10
(M/M_\mathrm{nl})^{-0.13}$, after converting to our mass definition
and using inner slope $\gdm=1$, $z_L = 0.3$, $\sigma_8=0.75$ and
$\Omega_m=0.27$. The nonlinear mass $M_\mathrm{nl}$ is the mass in a
sphere within which the rms linear density fluctuation is equal to
$\delta_c$, the overdensity threshold for spherical collapse. We
obtain $\cdm \simeq 8.4$ for the lower mass scale and $\cdm \simeq
5.6$ for the higher mass scale (see also Table~\ref{T:massscales}).
The scale radius of the dark matter component then follows as $\rsdm =
\rvdm/\cdm$.

We determine the scale length of the stellar components based on SDSS
photometry of galaxies of the appropriate luminosity. Fitting de
Vaucouleurs profiles, i.e., S\'ersic profiles with index $n=4$, to the
growth curve of these elliptical galaxies yields the half-light (or
effective) radius $R_e$. Assuming a constant stellar mass-to-light
ratio, $R_e$ is equal to the projected stellar half-mass radius $\Rhs$
as given in Table~\ref{T:massscales}. For a Hernquist cusped density
profile with $(\gs,\ns)=(1,4)$ as we adopt for the stellar component,
the scale radius follows as $\rss = \Rhs/1.8153$. 

Given the large scatter in dark matter halo concentration values in
simulations \citep[0.15 dex;][]{2001MNRAS.321..559B}, we will also
redo a subset of the analysis using concentrations lower and higher
than the fiducial values used for the main analysis.  This work will
allow us to quantify biases associated with halo concentration.
Realistically, we expect the changes in the concentration $\cdm$ to be
somewhat degenerate with changes in the inner slope $\gdm$, as both
change the amount of dark matter in the inner parts, although there
may be some differences because changing $\gdm$ affects the inner
logarithmic slope while changing $\cdm$ does not.  In Paper II we
quantify the approximate degeneracy between changes in $\cdm$ and
$\gdm$ for the dark matter component, while keeping the stellar
component fixed.

%---------------------------------------------------------------------
\subsection{Density profiles} \label{SS:densityprofiles}
%---------------------------------------------------------------------

For our main set of simulations, we use Hernquist and generalised NFW
profiles for stellar and dark matter components, respectively (cf.\
Sections~\ref{SS:densprof} and~\ref{SS:cuspdens}).  While there is some
uncertainty over the true dark matter profile in $N$-body simulations
(particularly in the inner regions; see Section~\ref{SS:sersicdens}), and
the initial dark matter profile may in any case be modified due to the
presence of baryons, we use the generalised NFW model because it
allows freedom in the form of the cusp.  As discussed in more detail
below (Section~\ref{SS:sersicsim}), we also examine how our results depend
on the form of the density profiles by running a set of simulations
that use deprojected S\'ersic profiles for the intrinsic (3d)
densities of both the stellar and dark matter components.  Here we
focus on the main set of simulations with cusped profiles.

\subsubsection{Inner and outer slope}
\label{SSS:innerandouterslope}

For our main set of simulations with cusped density profiles,
we fix the stellar component to have a Hernquist profile, with inner
slope $\gs=1$ and outer slope $\ns=4$.  Based on cosmological $N$-body
simulations, we fix the outer slope of the dark matter simulation to
$\ndm=3$ \citep[e.g.,][]{2004MNRAS.349.1039N}.  

The inner slope of the dark matter profile is more problematic.  While
$N$-body simulations can suggest forms for dark matter halo profiles
\citep[e.g.,][]{2004MNRAS.349.1039N}, there is some question whether
those profiles are affected by the presence of a stellar component,
and if so, in what way.  One formalism for analysing how a stellar
component affects a dark matter profile is called adiabatic
contraction \citep[AC: ][]{1980ApJ...242.1232Y, 1986ApJ...301...27B,
  2004ApJ...616...16G, 2005ApJ...634...70S}.  The physical picture is
that as the baryons cool and condense into the centre of the system,
they draw some of the dark matter in as well, causing the dark matter
halo profile to steepen.  Persuasive observational evidence for or
against AC in the galaxies we are modelling does not yet exist, though
there are claims that the theoretical assumptions behind AC are not
valid for these galaxies.  \cite{2007ApJ...658..710N} suggest that the
opposing effect, growth through dissipationless infall of matter onto
the galaxy (which tends to push the dark matter outwards from the
centre) may cancel out the steepening due to baryonic cooling.  We
expect the effects of infalling matter to be even more significant for
our higher mass model, which is equivalent to a fairly massive group.
Furthermore, hydrodynamic simulations that demonstrate that adiabatic
contraction dominates may suffer from over-cooling of the baryonic
component, so we cannot use them to determine the extent of the effect
either.  Thus, we cannot be certain how much the dark matter profiles
may be modified due to the different processes involved in galaxy
formation.  

We consider the range of possibilities in two ways.  First, for the
main set of simulations with cusped density profiles, we examine three
values of the dark matter inner slope: $\gdm = \{0.5,1.0,1.5\}$.  As
we shall see, this allows for a broad range of lensing properties.
Second, for a subset of simulations with $\gdm=1$ NFW dark matter
profiles, we explicitly include the effects of AC, using the
formalisms in \cite{1986ApJ...301...27B} and
\cite{2004ApJ...616...16G}.

\subsubsection{Normalisation}
\label{SSS:normalization}

When the inner slope of the dark matter $\gdm$ is changed, the virial
mass $\Mv$ also changes. To preserve $\Mv$ we must change either the
density amplitude $\rho_0$ or the scale radius $r_s$ of the dark
matter (or both at the same time). Preserving in addition the
(reduced) shear at the virial radius $\rv$, would break this
``degeneracy'' in the normalisation. However, for individual strong
lens galaxies, weak lensing and other constraints on the density are
very weak if present at all. Nevertheless, the different choices for
the normalisation might not result in significantly different lensing
cross sections.

%%%FIG
\begin{figure}
\begin{center}
\includegraphics[width=1.0\columnwidth]{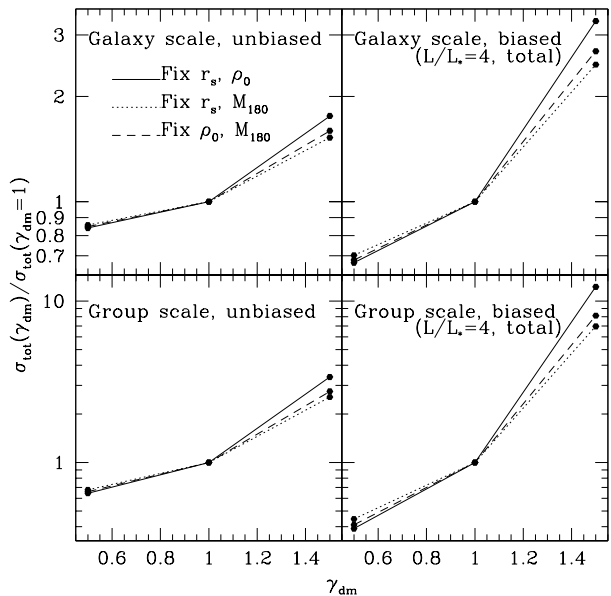}
\caption{\label{F:normalization} Variation in total lensing
  cross-section for different normalisations for the lower, galaxy
  (top) and higher, group (bottom) mass scale. The change in
  cross-section is shown relative to that of the fiducial case with
  dark matter inner slope of $\gdm=1$. The panels on the left show the
  relative difference in unbiased cross-sections, while the panels on
  the right are for biased cross-sections with a limiting luminosity
  of $4L_*$ (see Section~\ref{SS:magbias} for details). The solid
  curves show the relative difference in cross-section when both $r_s$
  and $\rho_0$ are kept fixed, while the dotted and dashed curve show
  the difference when preserving $\Mv$ by varying $\rho_0$ (fixing
  $r_s$) or $r_s$ (fixing $\rho_0$), respectively.}
\end{center}
\end{figure}
%%%FIG

To investigate this, we compute the change in total lensing
cross-section with respect to the fiducial case with $\gdm=1$ when
varying $\rho_0$ and/or $r_s$ to preserve $\Mv$. The results are
summarised in Fig.~\ref{F:normalization} for both the lower (top) and
higher (bottom) mass scale. The curves in the left panels show the
relative difference in unbiased cross-sections, while the curves in
the right panels are for biased cross-sections with a limiting
luminosity of $4L_*$ (see Section~\ref{SS:magbias} below). The solid
curves show the relative difference in cross-section when both $r_s$
and $\rho_0$ are kept fixed when changing the inner slope away from
the fiducial value $\gdm=1$, i.e., not preserving the virial mass
$\Mv$. The dotted and dashed curve show the difference when preserving
$\Mv$ by varying either $\rho_0$ (fixing $r_s$) or $r_s$ (fixing
$\rho_0$), respectively.

The results for the unbiased and biased cross-section are similar.  At
$\gdm=0.5$, the offsets between the solid, dotted and dashed curves
corresponding to the three different normalisations are marginal,
which is expected since for a shallower slope the contribution of the
dark matter component decreases with respect to the stellar component
which we keep fixed. However, the dark matter contribution becomes
important when $\gdm=1.5$, in particular for the high-mass scale as
can be seen from the different scaling of the vertical axes. We see
that in all cases the solid curve leads to higher cross-sections than
the dotted and dashed curve when we preserve $\Mv$, which is expected
given the increase in mass of tens of per cent. 

The latter two normalisations result in changes in the cross-section
of which the relative difference is well within the combined
statistical and systematic uncertainty in the cross-sections (i.e.,
while we have attempted to make these models fairly realistic, there
are some systematic uncertainties, so we do not trust them to per cent
level precision). This implies that varying $\rho_0$ and/or $r_s$ to
preserve $\Mv$ does not change the lensing properties in a way that is
significantly different. In what follows we choose to fix $r_s$ (and
hence the concentration $\rv/r_s$), because it is simpler and in line
with preserving the (weak) concentration-mass relation for dark matter
halos \citep[e.g.][]{2001MNRAS.321..559B} and the size-luminosity
relation for early-type galaxies \citep[e.g.][]{2003MNRAS.343..978S,
  2007AJ....133.1741B}, as discussed further in
Section~\ref{SSS:sersicnormalization} below.

\subsubsection{Intrinsic profiles}
\label{SSS:intrinsicprofiles}

%%%FIG
\begin{figure*}
\begin{center}
\includegraphics[width=0.9\textwidth]{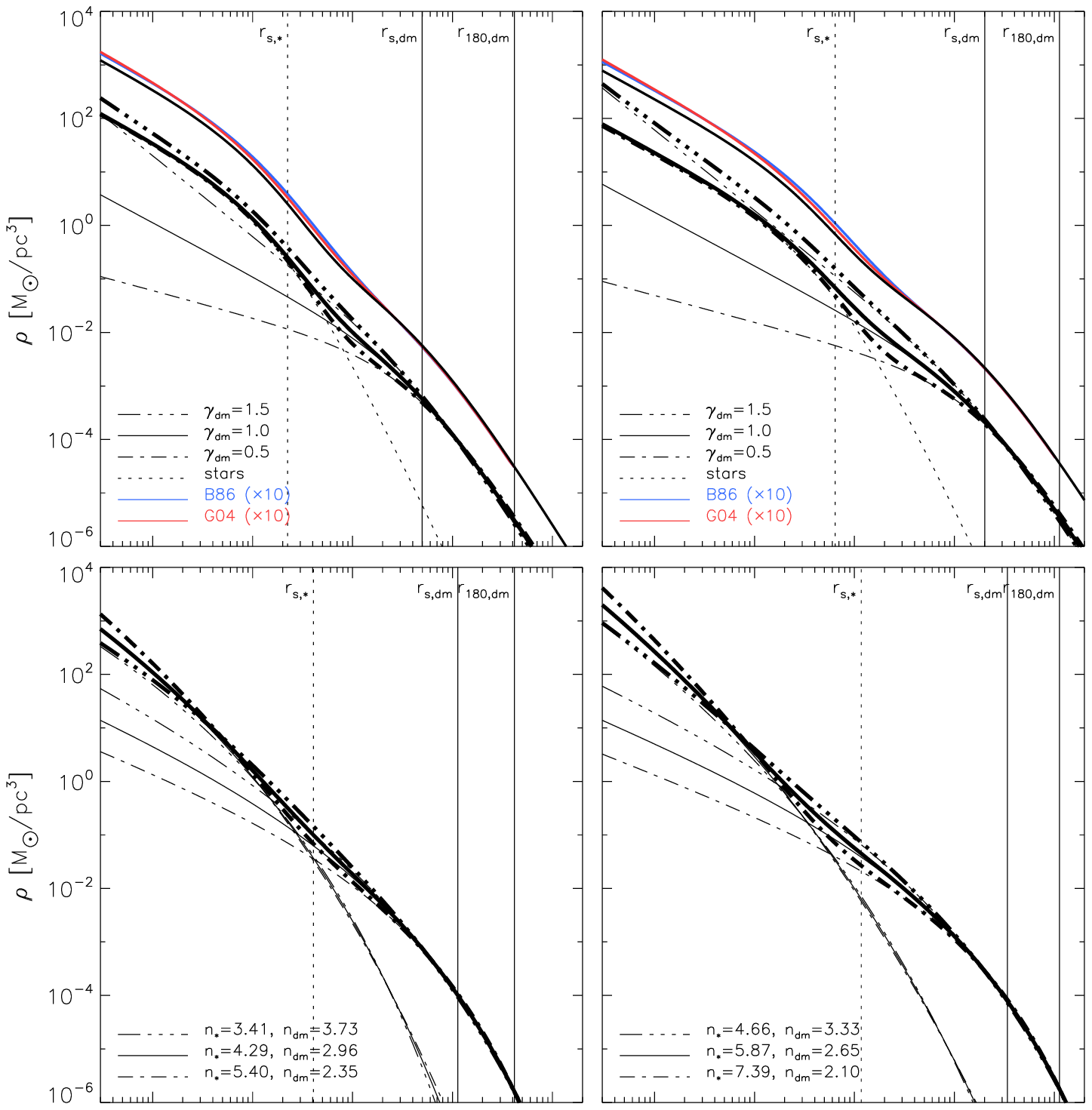}
\vfill
\includegraphics[width=0.9\textwidth]{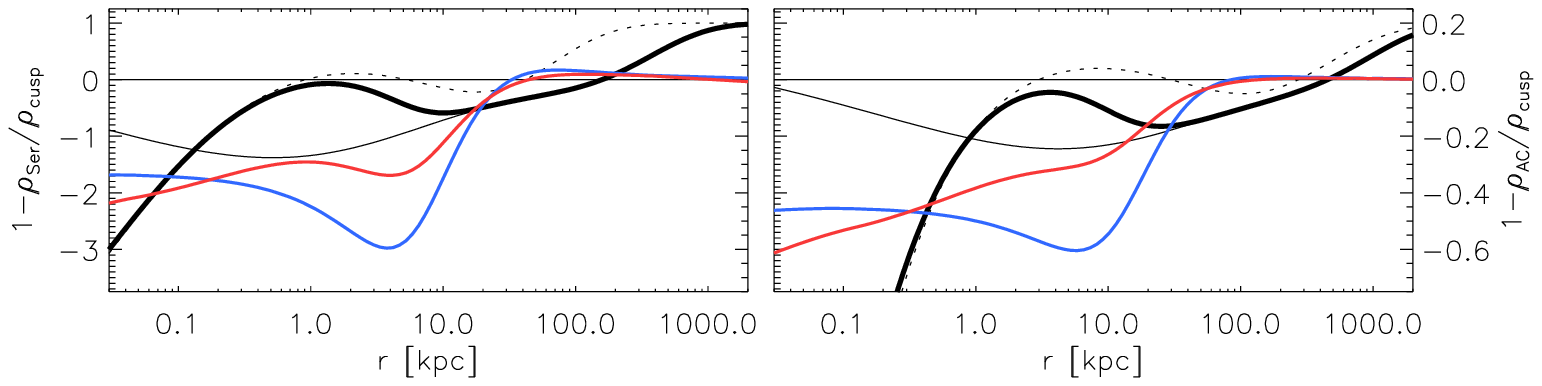}
\caption{\label{F:dens} 
  Intrinsic mass density versus intrinsic radius for cusped (top) and
  deprojected S\'ersic (middle) density profiles, for the lower (left)
  and higher (right) mass scale.
  Top panels: The dotted curve is the Hernquist profile of the
  stars, while the single-dot-dashed, solid and triple-dot-dashed
  curves represent the generalised NFW dark matter profiles with
  inner slopes $\gdm$ of 0.5, 1.0 (NFW) and 1.5, respectively.  The
  corresponding thick curves show the total mass density profile by
  combining that of the stars and dark matter.
  The blue and red curves show the total mass density profile
  (increased by a factor ten for clarity) taking into account
  adiabatic contraction (AC) of the fiducial cusped (NFW) dark matter
  profile (shown as black curve for comparison) using the
  prescriptions of \protect\cite{1986ApJ...301...27B} and
  \protect\cite{2004ApJ...616...16G}, respectively.
  Middle panels: The thin curves indicate the stellar and dark matter
  profiles which combined yield the thick curves. The solid curves are
  based on the fiducial S\'ersic indices for both components, whereas
  the two other curves indicate the variation in the mass density when
  adopting the limits in $\ndm$ and $\ns$ (see Section~\ref{SS:sersicsim}).
  Bottom panels: The black curves show the relative difference in the
  cusped and deprojected S\'ersic mass densities for the stellar
  component (dotted), dark matter component (thin solid) and the
  combination (thick solid), using the fiducial density profile
  parameters. The blue and red curves show the relative difference
  between fiducial cusped dark matter profile and those after taking
  into account AC, with corresponding axis labels on the right side.
  The dotted vertical line indicates the scale radius $\rss$ of the
  stellar component, whereas the innermost and outermost solid
  vertical lines show respectively the scale radius $\rsdm$ and virial
  radius $\rvdm$ of the dark matter component, using the fiducial
  density profile parameters.
}
\end{center}
\end{figure*}
%%%FIG

For the spherical density profiles, we present a set of plots showing
various quantities for both mass scales. We begin with
Fig.~\ref{F:dens}, which shows the intrinsic, 3d mass density
profile $\rho(r)$ for the cusped density profiles (NFW plus
Hernquist), for both mass scales. It is apparent that for $\gdm=0.5$
and $\gdm=1$ density profiles (stellar plus dark matter), the dark
matter component is negligible for scales below $\sim 2$ kpc for both
mass scales, whereas the $\gdm=1.5$ models have significant
contributions from dark matter for all scales shown. The difference
between these cases is somewhat reduced when considering the profiles
in projection along the line-of-sight, due to the dominant dark matter
contributions at large intrinsic radius. Nonetheless, we expect a
significant break in the strong lensing properties going from $\gdm=1$
to $\gdm=1.5$.

The colour curves in Fig.~\ref{F:dens} show the effects of AC on the
cusped, $\gdm = 1$ density profiles. We can see (bottom-left panel)
that for the lower mass scale, which has a higher ratio of stellar to
halo mass, AC tends to increase the density on scales below $\sim 10$
kpc by $\sim 40$ per cent depending on the model used, at the expense
of a slightly decreased density ($\sim 3$ per cent) for larger scales.
For the higher mass scale (bottom-right panel), the density below 10
kpc tends to increase by $\sim 30$ per cent for the
\cite{2004ApJ...616...16G} AC prescription, again at the expense of a
percent-level decrease on much larger scales.

%%%FIG
\begin{figure*}
\begin{center}
\includegraphics[width=0.9\textwidth]{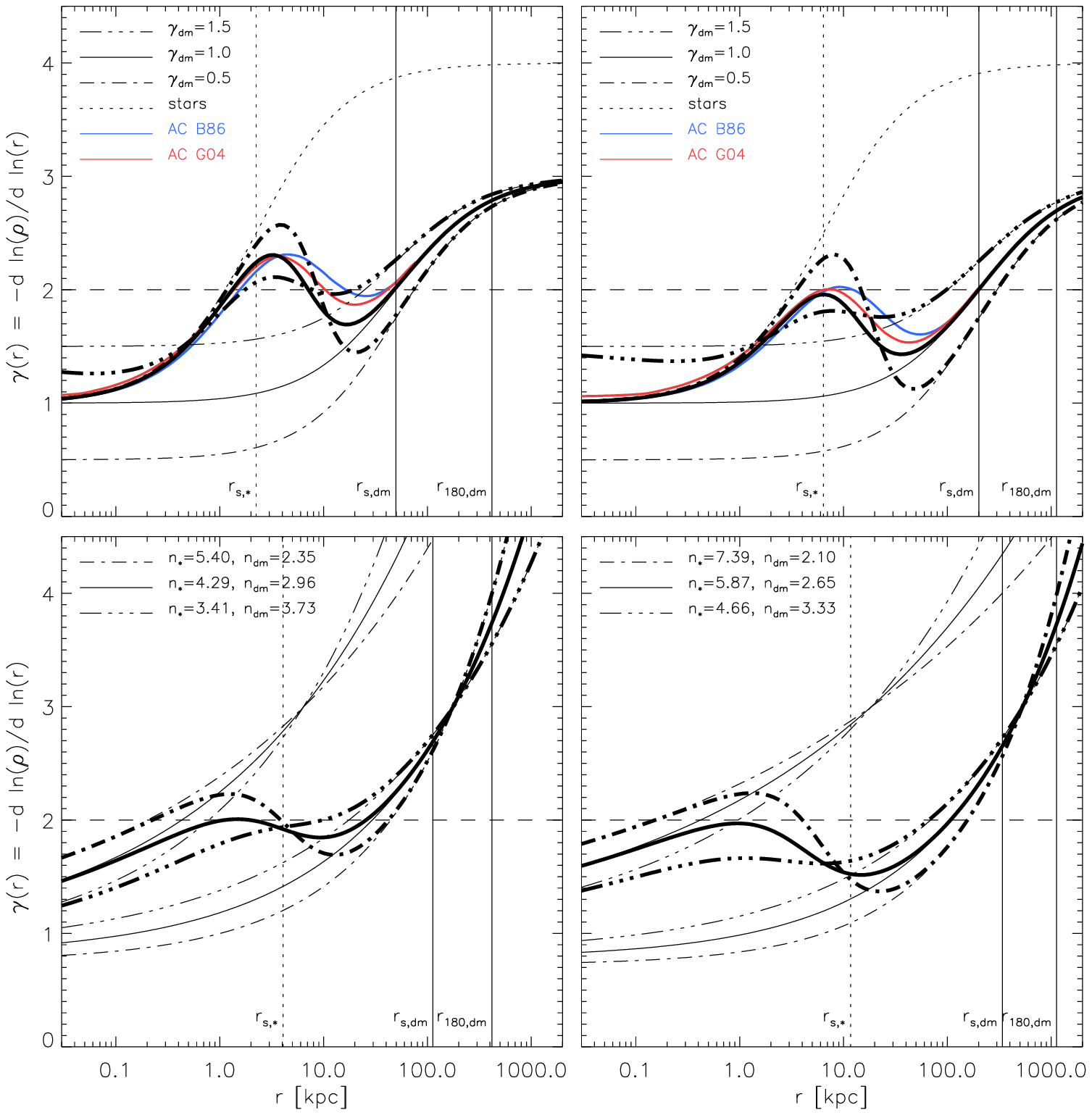}
%\vfill
%\includegraphics[width=0.9\textwidth]{slopedensrel.ps}
\caption{\label{F:3dlogslope} 
  The negative logarithmic slope of the intrinsic mass density as a
  function of the intrinsic radius, for cusped (top) and deprojected
  S\'ersic (middle) density profiles, and the relative difference
  (bottom), for the lower (left) and higher (right) mass scale. 
  The horizontal dashed line indicates the slope of an isothermal
  density profile.
  The meaning of the other curves is the same as in Fig.~\ref{F:dens}.
}
\end{center}
\end{figure*}
%%%FIG

Fig.~\ref{F:3dlogslope} shows the negative logarithmic slope
$\gamma(r)$ of the intrinsic mass density for both mass scales.
%for cusped and S\'ersic density profiles.
We can see that for the individual components of the models, the
logarithmic slopes behave smoothly according to
equation~\eqref{eq:logslopedenscusp}, but for the full model, there
can be complex, non-monotonic behaviour.
The latter ``wiggle'' in Fig.~\ref{F:3dlogslope} is stronger for the
cusped than the S\'ersic density profiles, reflecting the break in the
cusped density profile while the S\'ersic density profile turns over
smoothly.

The intrinsic profiles of our galaxy models are consistent with the
total mass density profile derived by \cite{2007ApJ...667..176G},
based on a joint strong-lensing and weak-lensing analysis of 22
early-type galaxies from the SLACS sample. The inferred intrinsic
density profile (their Fig.~8b) has an amplitude which falls in
between our lower-mass galaxy and higher-mass group scale (for an
adopted Hubble constant $h=0.72$).  Moreover, the corresponding slope
matches very well the wiggle in the logarithmic slope of our models in
Fig.~\ref{F:3dlogslope}.  Since this non-monotonic variation is around
a value of $\gamma=2$ (indicated by the dashed horizontal line in
Fig.~\ref{F:3dlogslope}), our models as well as the total mass density
inferred by \cite{2007ApJ...667..176G} are rather close to isothermal
over a large radial range around the Einstein radius.  However, the
density overall is \emph{not} consistent with isothermal, with clear
deviations at small and large radii, which happen well within the
two-decade radial range for which \cite{2007ApJ...667..176G} claim
consistency with isothermal. We discuss this ``isothermal conspiracy''
further when studying the profiles in projection next.

\subsubsection{Projected profiles}
\label{SSS:projectedprofiles}

%%%FIG
\begin{figure*}
\begin{center}
\includegraphics[width=0.9\textwidth]{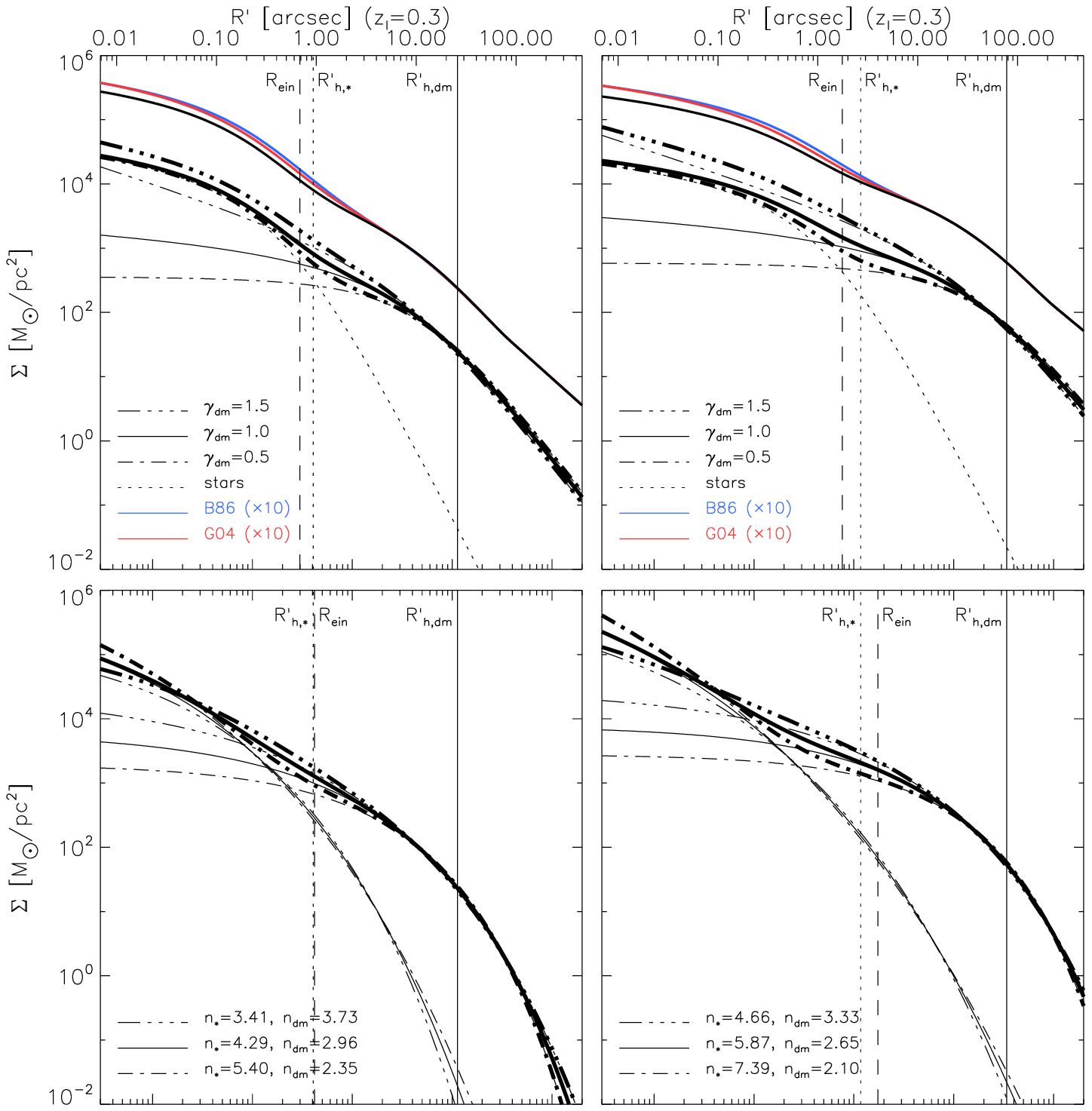}
\vfill
\includegraphics[width=0.9\textwidth]{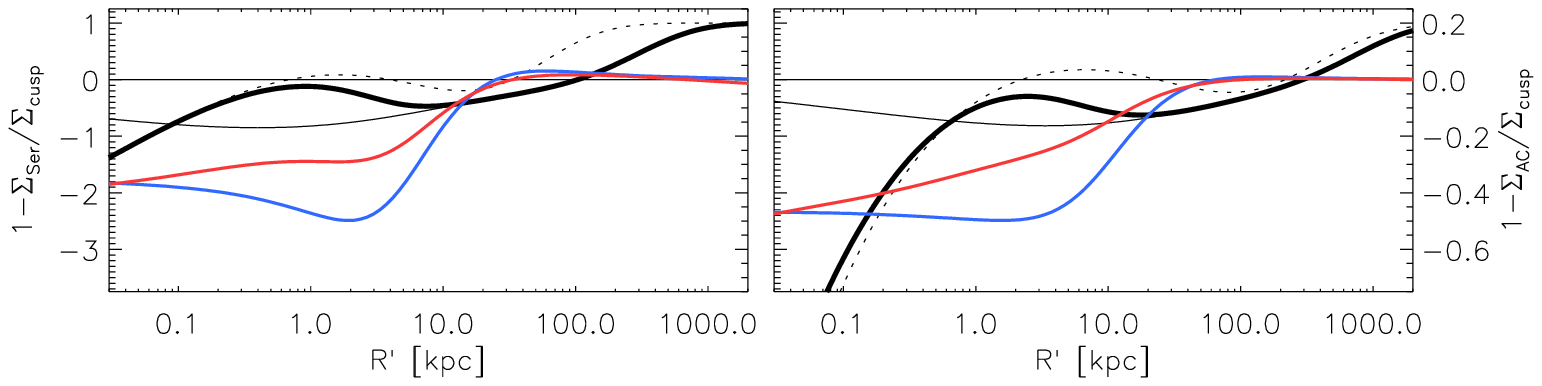}
\caption{\label{F:surf} Projected (or surface) mass density versus
  projected radius, for cusped (top) and deprojected S\'ersic (middle)
  density profiles, and the relative difference (bottom), for the
  lower (left) and higher (right) mass scale.  At the top, projected
  radius is given in angular units for a (lens) galaxy at the fiducial
  redshift of $z_L = 0.3$, corresponding to an angular diameter
  distance of $D_L \simeq 900$\,Mpc.  With a source at redshift $z_S =
  2.0$, the vertical dashed lines indicate the Einstein radii $\Rein$
  for the composite cusped and deprojected S\'ersic models, using the
  fiducial density profile parameters.  The dotted and solid vertical
  line indicate the half-mass radii $\Rhs$ and $\Rhdm$ containing half
  of the total mass of the stellar component and half of the virial
  mass of the dark matter component, respectively. In case of a
  constant mass-to-light ratio, $\Rhdm$ is equal to the effective
  radius $R_e$ of the surface brightness.  The meaning of the curves
  is the same as in Fig.~\ref{F:dens}.}
\end{center}
\end{figure*}
%%%FIG

In addition to the plots of the intrinsic, 3d profiles, we also show
projected, 2d quantities, which are observationally more relevant.
Fig.~\ref{F:surf} shows the surface mass density $\Sigma$
as a function of the projected radius $R'$, in physical units (kpc)
and at the top in angular units (arcsec) for a lens galaxy at the
fiducial redshift of $z_L = 0.3$.  The total (dark matter plus stars)
$\Sigma(R')$ for $\gdm = 0.5$ and $\gdm = 1.0$ are nearly identical
and star-dominated for $R \la 1$\,kpc. As anticipated, this is a
smaller transverse scale than that for which they are identical in 3d.

The colour curves in Fig.~\ref{F:surf} show the effect of AC on the
projected cusped, $\gdm = 1$ density profiles. As expected, the
projection effects have brought in larger scales where AC is less
significant, so the projected profiles with AC tend to be higher by
$\sim 30$ (lower mass scale) and $\sim 25$ (higher mass scale) per
cent on the scales of interest (a few kpc for the lower mass scale,
$\sim 10$ kpc for the higher mass scale).

%%%FIG
\begin{figure*}
\begin{center}
\includegraphics[width=0.9\textwidth]{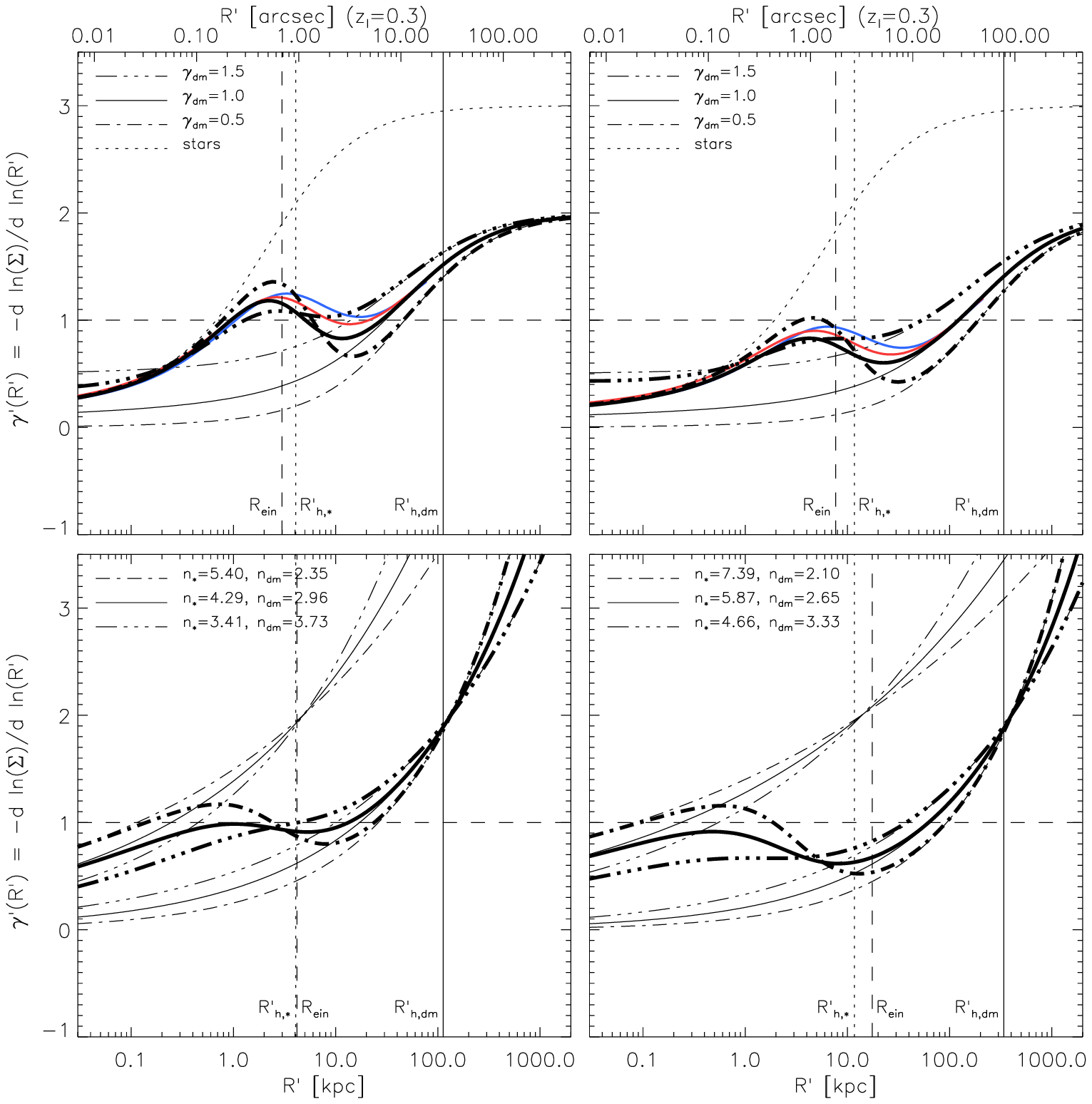}
%\vfill
%\includegraphics[width=0.9\textwidth]{slopesurfrel.ps}
\caption{\label{F:2dlogslope} The negative logarithmic slope of the
  surface mass density as a function of the projected radius, for
  cusped (top) and deprojected S\'ersic (middle) density profiles, and
  the relative difference (bottom), for the lower (left) and higher
  (right) mass scale. The horizontal dashed line indicates the slope
  of a (projected) isothermal density profile. The meaning of the
  other curves is the same as in Fig.~\ref{F:surf}.  }
\end{center}
\end{figure*}
%%%FIG

Fig.~\ref{F:2dlogslope} shows the negative logarithmic slope of
the surface mass density, $\gamma'(R')$.  If the density profile were
a pure power-law, the 2d and 3d negative logarithmic slopes would be
related by $\gamma' = \gamma - 1$.  Comparing
Fig.~\ref{F:2dlogslope} to Fig.~\ref{F:3dlogslope}, we see that
this simple relation does hold at small and large radii for our cusped
models, since these models are asymptotically power-laws. However, it
is not valid at intermediate radii, because our models --- especially
the full models containing both stellar and dark matter components ---
are far from being simple power-laws. This is particularly notable in
the vicinity of the Einstein radius, where the 3d slope varies much
stronger with radius than the 2d slope.

%%%FIG
\begin{figure*}
\begin{center}
\includegraphics[width=0.9\textwidth]{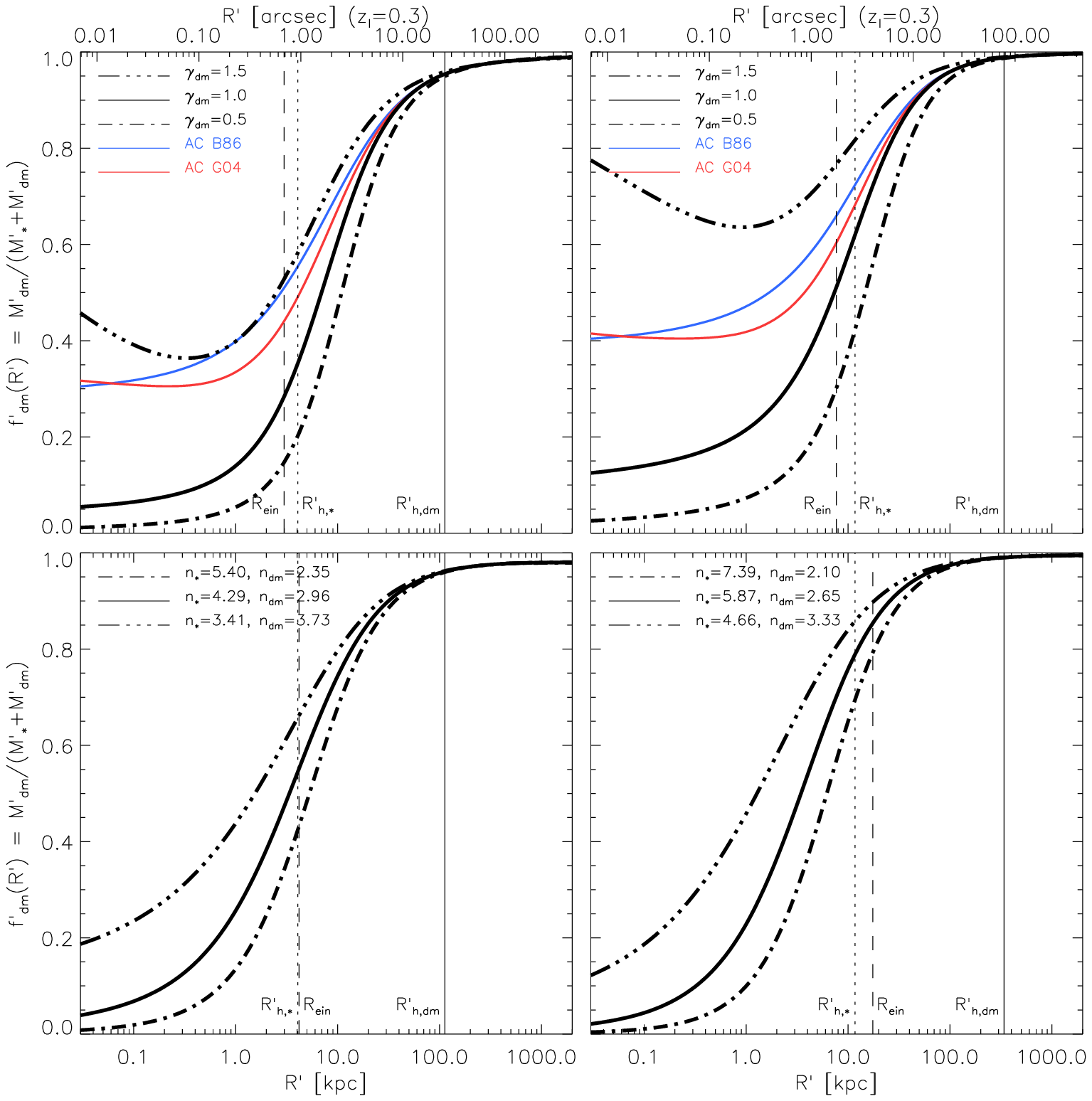}
%\vfill
%\includegraphics[width=0.9\textwidth]{projfdmrel.ps}
\caption{\label{F:2dfdm}
  The fraction of projected dark matter within a projected radius, for
  cusped (top) and deprojected S\'ersic (middle) density profiles, and
  the relative difference (bottom), for the lower (left) and higher
  (right) mass scale.
  The meaning of the curves is the same as in Fig.~\ref{F:surf}.
}
\end{center}
\end{figure*}
%%%FIG

Fig.~\ref{F:2dfdm} shows the projected dark matter fraction
$\fpdm$ within projected radius $R'$. As shown, by varying the dark
matter inner slope in the cusped models, we are able to explore models
with widely varying fractions of dark matter.  The dark matter
fractions in our models are consistent with the range of values
inferred from attempts to compare stellar and dark matter masses in
observed lens systems \citep{2003ApJ...595...29R, 2005ApJ...623..666R,
  2005ApJ...623L...5F, 2008ApJ...684..248B}.

%%%FIG
\begin{figure*}
\begin{center}
\includegraphics[width=0.9\textwidth]{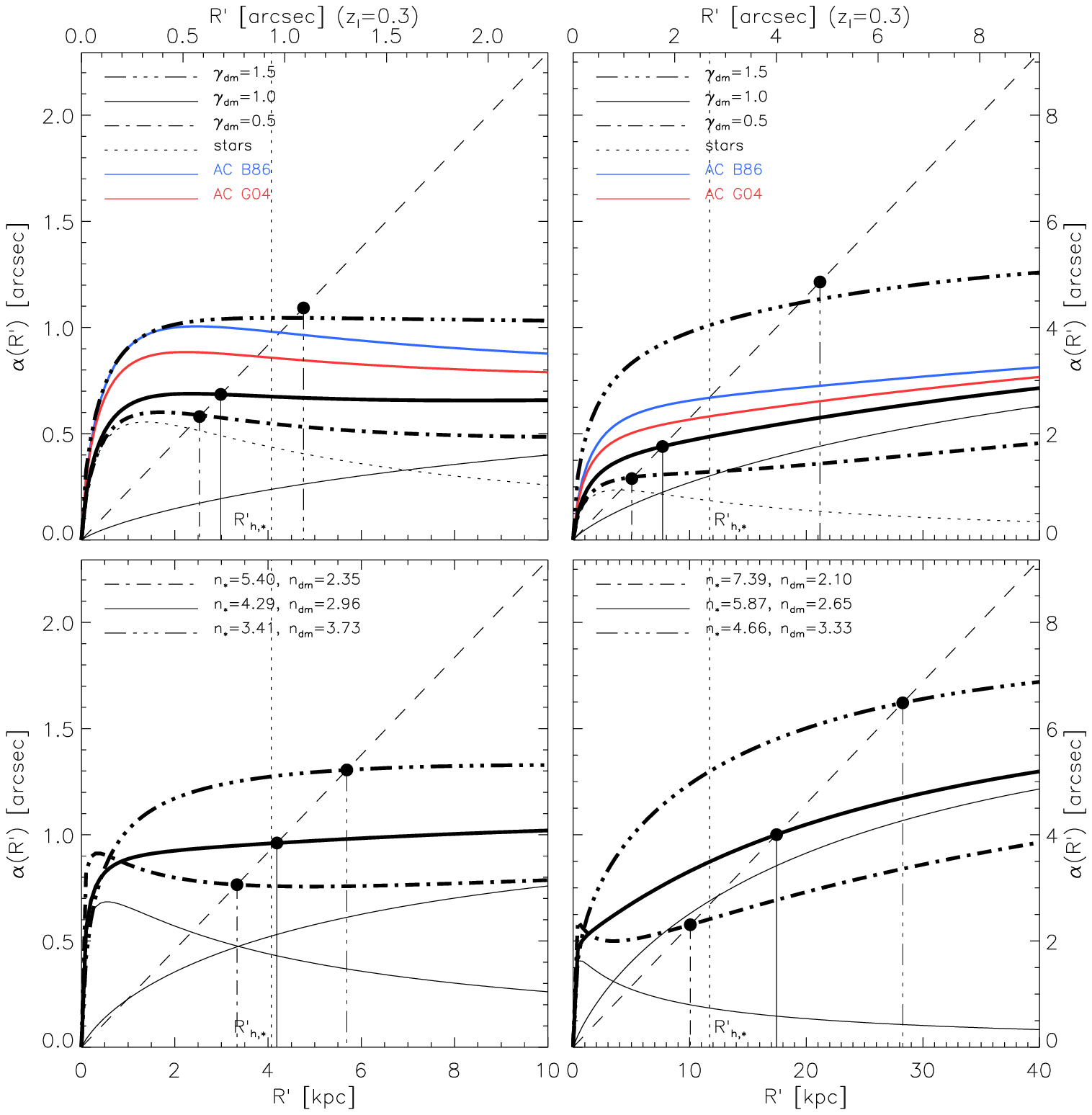}
%\vfill
%\includegraphics[width=0.9\textwidth]{alpharel.ps}
\caption{\label{F:alpha}
  Deflection angle (or derivative of the projected potential) as
  function of projected radius, for cusped (top) and deprojected
  S\'ersic (middle) density profiles, and the relative difference
  (bottom), for the lower (left) and higher (right) mass scale.
  The intersections of the diagonal one-to-one (dashed) line with the
  (thick) curves of the combined stellar and dark matter components
  yield the Einstein radii $\Rein$ as indicated by the filled circles
  and the corresponding vertical lines.
  The meaning of the other curves is the same as in Fig.~\ref{F:surf}.
  The deflection curves of the stellar and dark matter component
  separately show respectively a strong decrease and increase towards
  larger radii. However, the combination results for the lower mass
  scale in an almost flat deflection curve around the Einstein radius
  and beyond, similar to a constant deflection angle in case of an
  isothermal density profile.
}
\end{center}
\end{figure*}
%%%FIG

Finally, Fig.~\ref{F:alpha} shows the deflection angle $\alpha(R') =
\rmd \phi / \rmd R'$ as function of the projected radius $R'$. The
Einstein radii follow from the solution of $\alpha(\Rein) = \Rein$,
which corresponds to the intersections of the dashed one-to-one line
with the thick curves, indicated by the filled circles. The
corresponding Einstein radii appear at about the half-mass radius
$\Rhs$, which in case of a constant stellar mass-to-light ratio
corresponds to the effective radius $R_e$.

Figs.~\ref{F:2dlogslope} and \ref{F:alpha} contain an important
point related to attempts to combine strong lensing and stellar
dynamics to measure the density profile in the vicinity of the
Einstein radius \citep[e.g.,][]{2004ApJ...611..739T,
  2006ApJ...649..599K}. The slope of the composite surface mass
density for models with $\gdm$ of $0.5$, $1.0$ and $1.5$ are all close
to unity near the Einstein radius. For the lower-mass scale models
(top-left panel) there is a little more spread around a somewhat
higher value $\gamma' \simeq 1.2 \pm 0.2$ than for the higher-mass
scale (top-right panel) with $\gamma' \simeq 0.9 \pm 0.1$. However,
all are close to the slope $\gamma'=1$ of a projected isothermal
density as indicated by the horizontal dashed line in
Fig.~\ref{F:2dlogslope}. In Fig.~\ref{F:alpha}, the deflection
angle for the higher-mass scale (top-right panel) increases slowly
with radius around the Eintein radius, but for the lower-mass scale
the curves are nearly flat around (and beyond) $\Rein$, similar to a
constant deflection angle in case of an isothermal density. Even
though the stellar density and dark matter density with different
inner slopes are clearly non-isothermal, the combined density has a
projected slope and corresponding deflection angle consistent
with isothermal around the Einstein radius. This finding implies
that it will be challenging to use strong lensing to constrain the
dark matter inner slope $\gdm$ for a two-component model without
bringing in information from significantly different scales.
% \citep[see also][]{2008arXiv0807.4175V}.

We note that the apparent isothermality of our composite models is a
completely unintended consequence of the mass models we have chosen,
not something we have imposed by design.  However, it provides support
for the realism of our models, when compared against observations.
There is much evidence from strong lensing (especially in combination
with kinematics) that early-type galaxies have total density profiles
that are approximately isothermal in the vicinity of the Einstein
radius \citep[e.g.,][]{2003ApJ...595...29R, 2004ApJ...611..739T,
  2005ApJ...623..666R, 2006ApJ...649..599K}.  The fact that a
superposition of non-isothermal stellar and dark matter profiles can
produce a total profile that is approximately isothermal over a wide
range of scales --- known as the ``isothermal conspiracy'' --- is
related to the tendency towards flat rotation curves originally used
to argue for a dark matter component.  Finally, the same set of SLACS
lenses can be fit with a single power-law profile that is consistent
with isothermal \citep{2006ApJ...649..599K}, or by a more physically
intuitive combination of a stellar Hernquist profile and a dark matter
halo \citep{2007ApJ...671.1568J}.  All in all, the quasi-isothermal
nature of our models appears to be consistent with interpretations of
strong lensing by real galaxies.

%%%FIG
% \begin{figure*}
% \begin{center}
% \includegraphics[width=0.9\textwidth]{shear.ps}
% \vfill
% \includegraphics[width=0.9\textwidth]{shearrel.ps}
% \caption{\label{F:alpha} Shear curves \dots}
% \end{center}
% \end{figure*}
%%%FIG

%---------------------------------------------------------------------
\subsection{Density shape models}
\label{SS:densityshapes}
%---------------------------------------------------------------------

We consider seven different models for the shape of the galaxy
density: spherical, moderately oblate, very oblate, moderately
prolate, very prolate, triaxial, and a mixed model described below. In
all but the mixed model, the axis ratios used for the dark matter and
stellar components are related in a simple way (see the full
description in Section~\ref{SSS:axisratios}). In all cases the axes of the
two components are intrinsically aligned in three dimensions.

\subsubsection{Axis ratios}
\label{SSS:axisratios}

We adopt the mean values $\badm=0.71$ and $\cadm=0.50$ from
\cite{2002ApJ...574..538J} for the axis ratios of the triaxial density
of the dark matter component. These ratios were determined from
cosmological $N$-body simulations, without any stellar component.  We
set $\badm = 1$ for the oblate shapes, with the same $\cadm=0.50$ for
the moderately oblate case, and $\cadm = 0.71 \times 0.50 \simeq 0.36$
for the very oblate case. We set $\badm = \cadm$ for the prolate
shapes, with $\cadm = 0.71$ for the moderately prolate case and $\cadm
= \sqrt{0.71 \times 0.51} \simeq 0.60$ for the very prolate case. Note
that in the very oblate and very prolate case these choices preserve
the product $\badm \times \cadm$ of the triaxial case, and hence the
enclosed ellipsoidal mass.

Next, we use results from the cosmological gasdynamics simulations in
\cite{2004ApJ...611L..73K} to make a (crude) conversion from the dark
matter shape to the rounder shape of the stars: $\bas = 0.6 \, \badm +
0.4$, and similarly for $\cas$.  In practice, in that paper, the axis
ratios were a function of separation from the centre of the halo; we
neglect those effects in this work. Also, in that work the dark matter
halos were themselves made more round due to the presence of baryons,
with up to $\sim 15$ per cent effects out to the virial radius.
However, those simulations may suffer from baryonic overcooling, which
would tend to accentuate these effects, so we do not incorporate the
rounding of the dark matter component at this time.

The final model, the mixed shape model, is constructed using the very
oblate stellar shape combined with the triaxial dark matter halo
shape. Part of the motivation for doing so is based on the results
given by \cite{1992MNRAS.258..404L} for the axis ratio distributions
and 3d shapes of early-type galaxies as inferred from the
distributions of their projected shapes. In that paper, the elliptical
sample containing $2\,135$ galaxies was found to have a projected
shape distribution consistent with a (weakly) triaxial intrinsic shape
distribution. Fitting Gaussian distributions in both axis ratios, they
found a best-fitting mean and standard deviation of $0.95$ and $0.35$
in $\bas$, and $0.55$ and $0.20$ in $\cas$. The $4\,782$ S0 galaxies
in their sample had a projected shape distribution with best-fit mean
value $\bas=1$ consistent with oblate axisymmetry, and with
best-fitting mean and standard deviation of $0.59$ and $0.24$ in
$\cas$. 

Further motivation is provided by detailed dynamical models of nearby
elliptical and lenticular galaxies that accurately fit their observed
surface brightness and (two-dimensional) stellar kinematics. These
models show that the inner parts of lenticular as well as many
elliptical galaxies, collectively called fast-rotators
\citep{2007MNRAS.379..401E}, are consistent with an oblate
axisymmetric shape \citep{2007MNRAS.379..418C}. The slow-rotator
ellipticals, on the other hand, might be rather close to oblate
axisymmetric in the centre but rapidly become truly triaxial at
larger radii (\citeauthor{2008MNRAS.385..647V}
\citeyear{2008MNRAS.385..647V}, \citeyear{2008PhDT.........5V}).

We thus consider this mixed model, with a triaxial dark matter shape
as in cosmological $N$-body simulations plus an oblate stellar shape as
inferred from observations, as being potentially consistent with
reality for the lower mass scale. The triaxial shape for both dark
matter and stars may be more appropriate for the higher mass scale.

%%%TAB
\begin{table}
  \caption{Adopted axis ratios of the ellipsoidal intrinsic mass density for the dark
    matter and stellar component for seven different shape models.} 
\label{T:axisratios}
\begin{center}
  \begin{tabular}{*{5}{l}}
    \hline
    \hline
    Shape & $\badm$ & $\cadm$ & $\bas$ & $\cas$ \\
    \hline
    Spherical          & 1.00 & 1.00 & 1.00 & 1.00 \\
    Moderately oblate  & 1.00 & 0.50 & 1.00 & 0.70 \\
    Very oblate        & 1.00 & 0.36 & 1.00 & 0.62 \\
    Moderately prolate & 0.71 & 0.71 & 0.83 & 0.83 \\
    Very prolate       & 0.60 & 0.60 & 0.76 & 0.76 \\
    Triaxial           & 0.71 & 0.50 & 0.83 & 0.70 \\
    Mixed              & 0.71 & 0.50 & 1.00 & 0.62 \\
    \hline
  \end{tabular}
\end{center}
\end{table}
%%%TAB

The resulting axis ratios for each shape model, both for the dark matter and stellar
component, are given in Table~\ref{T:axisratios}.  
While we do not explore a broad range of axis ratios due to the
prohibitive numbers of simulations involved, we can still get some
sense of the trend with axis ratio by looking at the changes from
spherical to moderately and very oblate or prolate shapes.

\subsubsection{Scale lengths for non-spherical density profiles}
\label{SSS:scalelengthnonspherical}

When constructing the non-spherical density profiles, we fix the
concentration to be the same as in the spherical case, but set the
scale length $a$ such that the mass within the ellipsoidal virial
radius $\mv = \rv/a$ is equal to the virial mass $\Mv$. This is done
using
\begin{equation}
  \label{eq:changeshape}
  a^3 \, (b/a) \, (c/a) = r_s^3.
\end{equation}
In this case, the length scale of the halo is rescaled while
maintaining its concentration $\rv/\rs$.

\subsubsection{(Mis)alignment of projected axes}
\label{SSS:misalignment}

We assume that the intrinsic axes of the stellar and dark matter
component are aligned. For the axisymmetric models this means that the
minor and major axes of the surface mass densities of
the stars and dark matter are aligned as well. However, in case of a
triaxial dark matter component the projected axes are misaligned with
respect to those of the intrinsically rounder stellar component.

%%%FIG
\begin{figure*}
\begin{center}
\includegraphics[width=0.75\textwidth]{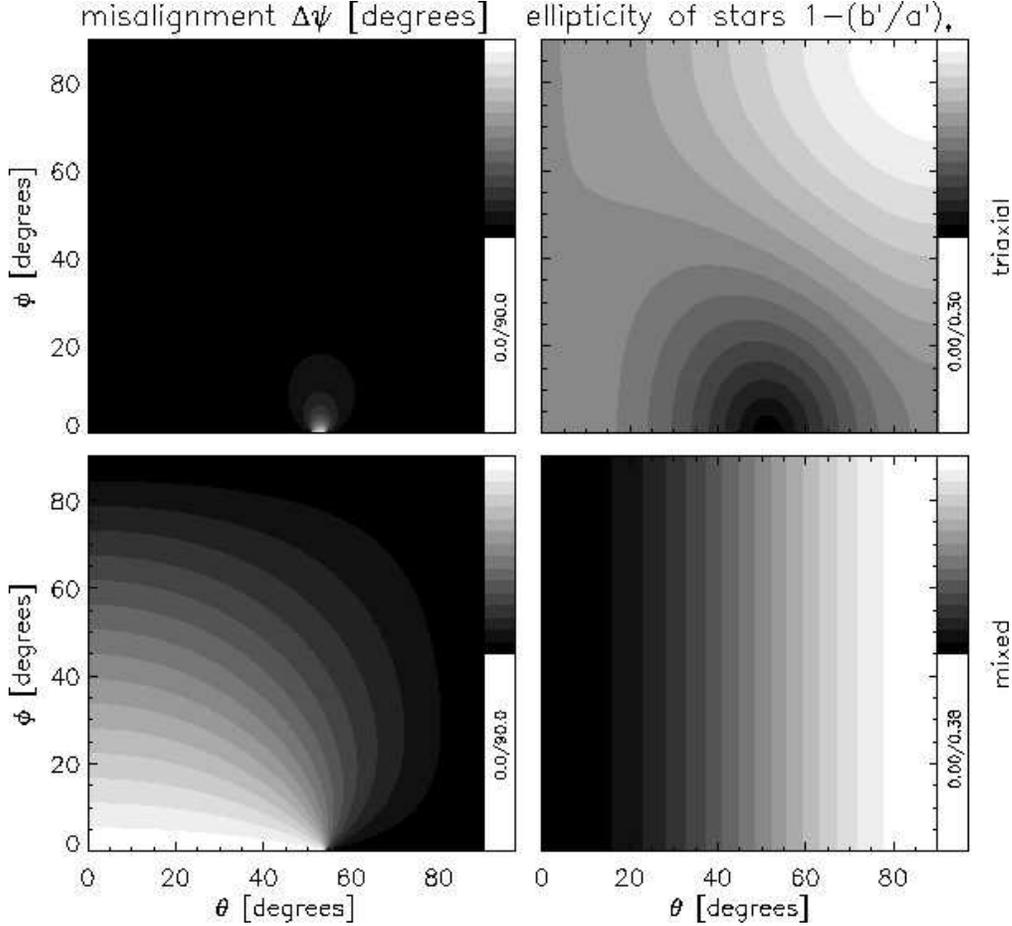}
\caption{\label{F:misalignment} The left panels show the misalignment
  $\Delta\psi$ between the projected axes of the surface mass density
  of the stellar and dark matter component of the triaxial (top) and
  mixed (bottom) shape models, as a function of the polar viewing
  angle $\vartheta$ and azimuthal viewing angle $\varphi$. In
  practice, $\Delta\psi$ is estimated from modelling of the lensing
  images and the position angle of the stars on the sky. The latter
  becomes very uncertain if the stars appear too round on the plane of
  the sky, i.e., in case of a very small ellipticity $1-(b'/a')_\star$
  shown in the right panels.  Note that these are not equal-area plots
  since the horizontal axis is $\theta$ rather than $\cos{\theta}$.}
\end{center}
\end{figure*}
%%%FIG
%Therefore, we show in the middle column the ellipticity
%$1-(b'/a')_\star$ of the stars, which we combine in the right column
%with $\Delta\psi$ to arrive at an 'effective' misalignment
%$[1-(b'/a')^2]^{1/2} \sin\Delta\psi$ (see Section~\ref{SSS:misalignment}
%for details).  

The misalignment $\Delta\psi = | \psi_\mathrm{dm} - \psi_\star |$
follows from equation~\eqref{eq:misalignment_psi} for the different
intrinsic axis ratios of the dark matter and stellar component
(Table~\ref{T:axisratios}) as a function of the polar and azimuthal
viewing angles $(\vartheta,\varphi)$. The left panels of
Fig.~\ref{F:misalignment} show the resulting $\Delta\psi$ for the
triaxial (top) and mixed (bottom) shape model. Depending on the
viewing direction, the projected axis ratios of the stars and dark
matter can become completely orthogonal ($\Delta\psi = 90$\dgr).  At
the same time, however, the projected ellipticity of the stars
$1-(b'/a')_\star$ goes to zero, as shown in the right panels of
Fig.~\ref{F:misalignment}. This means that the stars appear round on
the plane of the sky, so that their position angle on the sky is very
hard to establish.

Even ignoring the stellar ellipticity, the probability of a
significant misalignment is rather small in the case of random viewing
directions\footnote{We show in Paper~II that there is a selection bias
  in orientation for non-spherical lens galaxies.}. For the triaxial
shape model only $\sim 1.2$\% of the area on the viewing sphere leads
to misalignments $\Delta\psi \ge 10$\,\dgr. This number increases to
$\sim 50.8$\% for the mixed shape model, but at the same time the area
on the viewing sphere that leads to a round appearance of the stars is
also larger: $\sim 8.3$\% of the area has $(b'/a')_\star \ge 0.95$,
while this is only $\sim 2.1$\% for the triaxial shape model.

Our use of models that lead to almost no situations with significant
projected ellipticity {\em and} misalignment of DM and stellar
components is consistent with studies of relatively isolated lens
systems, for 
which the position angles of the light and of the projected total mass
agree to within $\lesssim 10^{\circ}$ \citep{1998ApJ...509..561K,
  2002sgdh.conf...62K, 2006ApJ...649..599K, 2008ApJ...684..248B}.
Moreover, both the observed surface brightness and stellar kinematics
of early-type galaxies are well-fitted by dynamical models with the
intrinsic density parametrised by multiple (Gaussian) components that
have different shapes but are all co-axial
\citep[e.g.][]{1994A&A...285..723E, 2006MNRAS.366.1126C}. This
includes the photometric and kinematic twists observed in particularly
the giant ellipticals (\citeauthor{2008MNRAS.385..647V}
\citeyear{2008MNRAS.385..647V}, \citeyear{2008PhDT.........5V}).

The above is all empirical evidence in favour of an initial guess
involving a high degree of alignment.  While there are many reasons
why these assumptions are not perfect, here we do not explore the
effects of intrinsic misalignment between the shapes, because of the
enormous amount of parameter space involved. Future work should
explore this issue further, as well as the issue of warps in the
intrinsic and/or projected profiles.

%---------------------------------------------------------------------
\subsection{Deprojected S\'ersic simulations}
\label{SS:sersicsim}
%---------------------------------------------------------------------

In addition to the simulations with cusped density profiles, we also
create a set of simulations with deprojected S\'ersic density profiles
(Section~\ref{SS:sersicdens}), which also provide a good description of
both stars and dark matter over a range of mass scales
(Section~\ref{SS:densprof}). As before, we choose the different parameters
for the dark matter and stellar density to create observationally
realistic galaxy models.

\subsubsection{Density parameters}
\label{SSS:sersicdensityparameters}

For the total mass of the stellar component we use the same values
given in Table~\ref{T:massscales} for the lower and higher mass
scales. Even though the total mass of the dark matter component is
also finite in this case, we adopt for consistency the same virial
mass scale by setting the mass within the virial radius $\rvdm$ equal
to $\Mvdm$ as given in Table~\ref{T:massscales}. In spherical
geometry, the scale radius of the S\'ersic density is the half-mass
radius, which for the stellar component is $\Rhs$ given in
Table~\ref{T:massscales}.  This, in turn, is equal to the effective
radius $R_e$ in the case of a constant stellar mass-to-light ratio.
For the dark matter component we numerically derive the radius $\Rhdm$
that encloses a projected mass that is half of the virial mass $\Mvdm$
of the spherical NFW dark matter halo, and use the same half-mass
radius for the deprojected S\'ersic profiles. This procedure results
in $\Rhdm = 2.2699 \, \rsdm \simeq 112$\,kpc and $\Rhdm = 1.6882 \,
\rsdm \simeq 339$\,kpc for the lower and higher mass scale,
respectively

The only parameter left to choose is the S\'ersic index $n$. To derive
the S\'ersic indices $\ns$ of the stellar component, we use the
relation $\log\ns = -0.10 \, (M_B + 18) + 0.39$ for Virgo early-type
galaxies as derived by \cite{2006ApJS..164..334F}. A luminosity of
$L_*$ corresponds to an absolute Johnson $B$-band and SDSS
$r$-band magnitude of about $M_{B*} = -19.4 + 5\log{(h=0.72)} = -20.1$
\citep{2007ApJ...665..265F} and $M_{r*} = -20.28 + 5\log{(h=0.72)} =
-21.0$ \citep{2003ApJ...592..819B}, respectively. With $B-r = 1.32$ as
a typical colour for elliptical galaxies \citep{1995PASP..107..945F},
this means that our choice for the lower and higher mass scales of $\sim 2
L_*$ and $\sim 7 L_*$ in the $r$-band translate to $\sim 1.3 L_*$ and
$\sim 4.7 L_*$ in the $B$-band, or about $M_B = -20.4$ and $M_B =
-21.8$. We thus obtain S\'ersic indices $\ns = 4.29$ and $\ns = 5.87$
for the stellar component of the lower and higher mass scale,
respectively. For the dark matter component, we use the S\'ersic fits
to simulated $\Lambda$CDM halo profiles presented in
\cite{2005ApJ...624L..85M}.  Taking into account the difference in
virial mass definition, we obtain $\ndm = 2.96$ and $\ndm = 2.65$ for
the lower and higher mass scales, respectively.

In addition to these fiducial S\'ersic indices, we also consider
values consistent with the observed scatter. For a given absolute
luminosity $M_B$, the scatter in the observed $\log\ns$ in
\cite{2006ApJS..164..334F} is approximately $\pm 0.1$\,dex. This
scatter 
results in a range in $\ns$ of $[3.41,5.40]$ and $[4.66,7.39]$ for the
lower and higher mass scales, respectively. Similarly, for a given
virial mass $\Mv$, the range of fitted $\log\ndm$ in
\cite{2005ApJ...624L..85M} is approximately covered by $\pm
0.05$\,dex. Since these findings are based on a low number (19) of
simulations, we consider indices $\pm 0.10$\,dex around the fiducial
values. This scatter implies a range in $\ndm$ of $[2.35,3.73]$ and
$[2.10,3.33]$ for the lower and higher mass scales, respectively.

\subsubsection{Normalisation}
\label{SSS:sersicnormalization}

For the cusped density profiles in Section~\ref{SSS:normalization},
changing the dark matter inner slope from the fiducial value $\gdm=1$
only affects the inner parts of the density profile. As a result, the
corresponding change in virial mass $\Mvdm$ (about $-20$\% and $+30$\%
when changing to respectively $\gdm=0.5$ and $\gdm=1.5$), could easily
be compensated by changing the density amplitude $\rho_0$ and/or the
scale radius $r_s$. For the deprojected S\'ersic density profiles,
changing the index $n$ affects the whole density profile rather than
just the inner parts, so that the corresponding changes in the mass
are much larger. Changing the fiducial dark matter index $\ndm$ to the
limits of the adopted range (while keeping $\ns$ fixed), results in a
factor $\sim 3$ change in the virial mass $\Mvdm$ for both mass
scales.  Changing the fiducial stellar index $\ns$ to the limits of
the adopted range, results in a factor $>100$ change in the total
stellar mass $\Mts$. Preserving the mass thus requires very large
variations in the central density $\rho_0$ and/or the half-mass radius
$\Rh$.

Certainly for the stellar component, changing $\Rh$ is not an option,
since early-type galaxies obey a size-luminosity relation
\citep[e.g.][]{2003MNRAS.343..978S, 2007AJ....133.1741B}.  Although
there is some freedom due to scatter in this relation and the stellar
mass-to-light conversion, this relation excludes larger variations in
$\Rhs$ when preserving $\Mts$. A similar conclusion is reached based
on the \cite{1977ApJ...218..333K} relation $\Ie \propto R_e^{-0.4g}$,
where $\Ie = L_\mathrm{tot}/(\pi R_e^2)$ is the mean surface
brightness within the effective radius $R_e$. In case of spherical
symmetry and a constant mass-to-light ratio $\Rhs = R_e$ and
$L_\mathrm{tot} \propto \rho_0 R_e^3 f(\ns)$, where the latter
dependence on the S\'ersic index follows from
equation~\eqref{eq:masssersic}. Therefore, obeying the Kormendy relation
means $\rho_0 \, R_h^{(1+0.4g)} \, f(\ns) = \mathrm{cnst}$, while
preserving the total stellar mass implies $\rho_0 \, R_e^3 \, f(\ns) =
\mathrm{cnst}$. Since $g \simeq 3$ even at high(er) redshift
\citep[e.g.][]{2003ApJ...595..127L, 2005A&A...442..125D}, these two
relations together imply that $R_e$ remains to be approximately
constant, while $\rho_0$ can be varied to compensate for the change in
$\ns$.

A similar scaling relation seems to hold for NFW dark matter halos
between their concentration $\cdm$ (or scale radius in case of a
constant virial radius), and virial mass $\Mvdm$ of the form $\cdm
\propto \Mvdm^{-0.13}$ \citep{2001MNRAS.321..559B}. Since the
dependence on $\Mvdm$ is weak and the scatter in the relation is
rather large, we could not use it before in Section~\ref{SSS:normalization}
when changing the inner slope $\gdm$ of the cusped density profile to
break the degeneracy between the different normalisation options
(which we showed in case of preserving $\Mvdm$ was not necessary,
since the effect of the different normalisations on the lensing
cross-sections is similar). \cite{2007ApJ...666..181S} show that when
an Einasto profile is used instead of a NFW profile, a similar
concentration-mass relation exists for dark matter halos. Because
deprojected S\'ersic profiles fit simulated dark matter halos equally
well as Einasto profiles (Section~\ref{SS:densprof}), we expect also in
that case a correlation between $\cdm$ and $\Mvdm$. As mentioned
above, for the S\'ersic density the effect of changing $\ndm$ is much
stronger than changing $\gdm$ for the cusped density, so that varying
$\cdm$ (while fixing $\rho_0$) to preserve $\Mvdm$ would violate this
concentration-mass relation. As a result, for both the stellar and
dark matter component we vary $\rho_0$ to preserve the (total and
virial) mass, while we keep the half-mass radius $\Rh$ fixed.

\subsubsection{Density profiles}
\label{SSS:sersicdensityprofiles}

The profile of the spherical deprojected S\'ersic mass density
$\rho(r)$ is shown in Fig.~\ref{F:dens} in the middle row and
compared to the cusped mass density in the bottom row, for the lower
and higher mass scales. The thin curves indicate the stellar and dark
matter profiles which combined yield the thick curves. The solid
curves are based on the fiducial S\'ersic indices for both components,
whereas the two other curves indicate the variation in the mass density
when adopting the above limits in $\ndm$ and $\ns$. In a similar way,
we show in Fig.~\ref{F:3dlogslope} the corresponding negative
logarithmic slope $\gamma(r)$, in Fig.~\ref{F:surf} the S\'ersic
surface mass density $\Sigma(R')$, in Fig.~\ref{F:2dlogslope} the
corresponding negative logarithmic projected slope $\gamma'(R')$, in
Fig.~\ref{F:2dfdm} the projected dark matter fraction $\fpdm(<R')$,
and finally in Fig.~\ref{F:alpha} the deflection angle.

As for the cusped density in Section~\ref{SSS:projectedprofiles}, we find
that both the slope of the composite surface mass density and the
deflection angle (Fig.~\ref{F:2dlogslope} and Fig.~\ref{F:alpha}) are
consistent with an isothermal density around the Einstein radius.
Especially for the lower-mass scale, the projected slope for all cases
converges to $\gamma'=1$ around $\Rein$, and the curves of the
deflection angle are extremely flat at and beyond $\Rein$. As for
cusped models, this finding implies that constraints from strong
lensing cannot trivially be extrapolated to smaller and larger radii.

%---------------------------------------------------------------------
\subsection{Computation of surface mass density maps}
\label{SS:compsurfmaps}
%---------------------------------------------------------------------

The lensing properties of a galaxy follow from its surface mass
density $\Sigma(x',y')$. For the above galaxy models this is the sum
of the surface mass density of the stellar and dark matter components,
each of which can be computed as follows.
For a given intrinsic shape $(a,b,c)$ seen from a viewing direction
$(\vartheta,\varphi)$, the projected semi-axis lengths $(a',b')$ are
given by equations~\eqref{eq:apbpellipse}--\eqref{eq:defB}. At each
position $(x',y')$ on the plane of the sky,
equation~\eqref{eq:ellipse} then yields the elliptic radius $m'$.  The
surface mass density follows by evaluation of
equation~\eqref{eq:surfmassell} for a cusped density $\rho(m)$ given
in equation~\eqref{eq:denscusp}, or in the case of a S\'ersic density
directly as $\Sigma(m')$ given in equation~\eqref{eq:surfsersic}.

However, rather than doing the numerical evaluation for each position
and viewing direction, we instead compute $\Sigma(m')$ on a fine
one-dimensional grid in $m'$, \emph{once} for each intrinsic density
profile and shape. We then derive $\Sigma(x',y')$ by linearly
interpolating onto this grid at the $m'$ value that (according to
equations~\ref{eq:ellipse}--\ref{eq:defB}) corresponds to the position
$(x',y')$ and chosen viewing direction $(\vartheta,\varphi)$.  In
particular, in the case of the cusped density, this procedure means
that we can avoid doing the numerical evaluation of
equation~\eqref{eq:surfmassell} on a two-dimensional grid repeatedly
for each viewing direction.

Along with the construction of the cusped and S\'ersic galaxy models
described above, we implemented and tested this efficient computation
of the surface mass density in
\idl.\footnote{http://www.ittvis.com/ProductServices/IDL.aspx} We use
a linear grid in $m'$, except in the inner parts where we sample
logarithmically in $m'$ to accurately resolve a central peak in the
density. Even so, when we compute the surface mass density on a
regular square grid in $(x',y')$ with a given dimension (box size) and
sampling (resolution), the resulting fixed pixel size might not be
small enough to account for a strong central cusp. This is not the
result of the interpolation but is due to the discreteness of the
map-based approach. The effects of finite resolution and box size on
the surface mass density and corresponding lensing properties are
further tested and discussed in Sections~\ref{SS:resolution}
and~\ref{SS:boxsize}.

We choose the grid in $(x',y')$ such that the $x'$-axis is aligned
with the major axis of the surface mass density of the dark matter
component. Even though we assume that the intrinsic axes of the
stellar and dark matter component are aligned, in case of a triaxial
dark matter component the \emph{projected} axes are misaligned with
respect to those of the intrinsically rounder stellar component. As
discussed in Section~\ref{SSS:misalignment} above, the resulting
misalignment angle $\Delta\psi = | \psi_\mathrm{dm} - \psi_\star |$
follows from equation~\eqref{eq:misalignment_psi}, and depends on the
viewing direction $(\vartheta,\varphi)$.
To take this misalignment of the stellar component with respect to the
dark matter component into account, we first rotate $(x',y')$ over the
misalignment angle $\Delta\psi$
\begin{eqnarray}
  %\label{eq:rotatedeltapsi_xp}
  \nonumber
  x_p & = & x' \cos\Delta\psi + y'\sin\Delta\psi,
  \\
  %\label{eq:rotatedeltapsi_yp}
  \nonumber
  y_p & = & -x'\sin\Delta\psi + y'\cos\Delta\psi,
\end{eqnarray}
and instead of equation~\eqref{eq:ellipse}, we then use $m'^2 =
(x_p/a')^2+(y_p/b')^2$ to define the elliptic radius $m'$. In case of
the axisymmetric models $\Delta\psi=0$, so that the stellar and dark
matter component are always aligned. Finally, we add the maps of the
stellar and dark matter components together to arrive at the total
surface mass density $\Sigma(x',y')$ of the galaxy model.

%%%FIG
\begin{figure*}
\begin{center}
\includegraphics[width=0.9\textwidth]{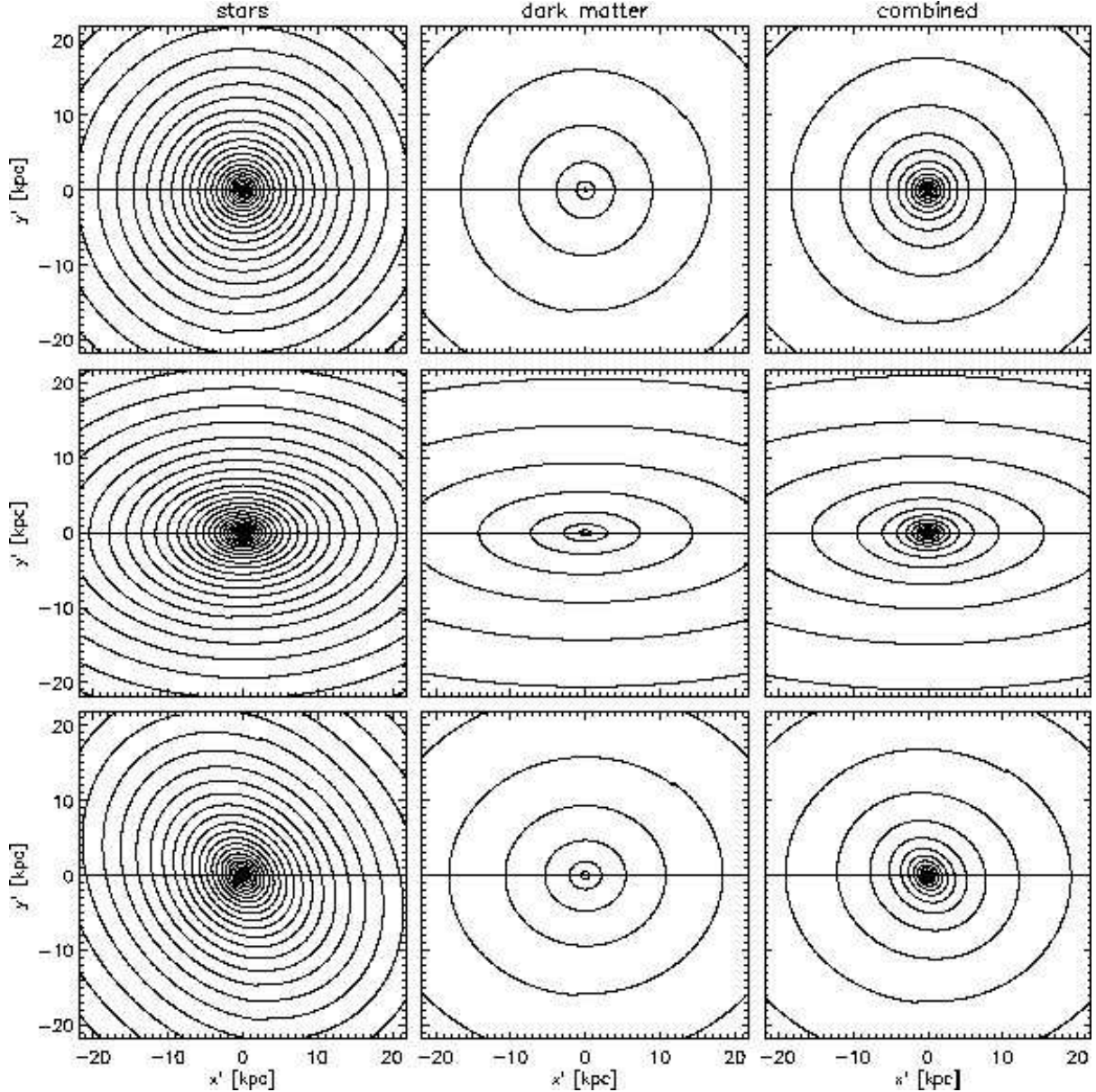}
\caption{\label{F:surfmaps} Examples of surface mass density maps
  $\Sigma$ as a function of position $(x',y')$ on the sky-plane in
  case of lower, galaxy-scale mass models with cusped density
  profiles. From top to bottom, the very oblate mass model viewed
  close to face-on with inclination $i \simeq 14$\dgr\ and close to
  edge-on with $i \simeq 81$\dgr, and the mixed shape model viewed at
  a polar angle $\vartheta \simeq 50$\dgr\ and azimuthal angle
  $\varphi \simeq 13$\dgr. The latter is close to the viewing
  direction that results in the maximum combination of ellipticity of
  the stars (left column) and misalignment with respect to the dark
  matter (middle column), resulting in a twist in the combined surface
  mass density (right column). The contour levels correspond to a
  factor of ten decrease in $\Sigma$ going outwards, clearly showing
  that the stars are more concentrated than the dark matter.}
\end{center}
\end{figure*}
%%%FIG

In Fig.~\ref{F:surfmaps}, we show examples of derived surface mass
density maps for lower, galaxy-scale mass models with cusped density
profiles. The figure includes, from top to bottom, the very oblate
mass model viewed close to face-on and close to edge-on, and the mixed
shape model. The latter is seen close to the viewing direction that
results in the maximum combination of ellipticity of the stars (left
column) and misalignment with respect to the dark matter (middle
column), resulting in a twist in the combined surface mass density
(right column). In Section~\ref{SS:projangle} we address the technical
issue of how many viewing angles we need to sample in order to
accurately characterise how the lensing behaviour changes with viewing
direction.

%=====================================================================
\section{Lensing calculations}
\label{S:lensingcalc}
%=====================================================================

In this section, we describe in detail how we do the lensing
calculations necessary for our analysis.  We present numerical methods
designed to balance efficiency with accuracy
(Sections~\ref{SS:kapmap}--\ref{SS:imsepdistr}), together with
extensive tests to validate those methods
(Sections~\ref{SS:validation}--\ref{SS:projangle}). All of the
calculations described here can be done with \gravlens\ version 1.99k,
which is a version of the original software by
\citet{2001astro.ph..2340K} that has been extended for this
project\footnote{The features that are new in \gravlens\ v1.99k are
  identified in the text. All the new features have been tested, and
  have instructions available through the \gravlens\ \texttt{help}
  command, but are not fully documented in the manual.  Version 1.99k
  can be accessed using the link for latest updates at
  \texttt{http://redfive.rutgers.edu/\~{}keeton/gravlens}.}.

%---------------------------------------------------------------------
\subsection{Lens potential, deflection, and magnification}
\label{SS:kapmap}
%---------------------------------------------------------------------

Given the two-dimensional convergence distributions for our models,
we need to solve the Poisson equation \eqref{eq:Poisson} to find
the lens potential.  In the present work, we actually need the first
and second derivatives of the lens potential (for the deflection
and magnification, respectively), and do not work with the potential
directly.  We use different approaches to obtain the derivatives,
depending on the model properties.  Note that we apply the methods
discussed here to the stellar and dark matter components separately,
and construct the composite lens model from a simple linear
superposition of the mass components.

If the model has circular symmetry and the convergence is known
analytically, the deflection can be obtained from the one-dimensional
integral over $\kappa(\theta)$ given in equation~\eqref{eq:circdef}.  Often
this integral can be evaluated analytically, so the lens model is
fully analytic. This procedure is used for circular deprojected
S\'ersic models. (For more detail about lensing by circular S\'ersic
models, see \citealt{2004A&A...415..839C} and
\citealt{2009JCAP...01..015B}.)

If the model has elliptical symmetry and the convergence is known
analytically, the first and second derivatives of the lens potential
can be obtained from a set of one-dimensional integrals over
$\kappa(m')$ and its derivative $\rmd\kappa/\rmd m'$
\citep{1990A&A...231...19S,2001astro.ph..2341K}.  For elliptical
deprojected S\'ersic models, we evaluate the integrals
numerically,\footnote{We note that the steep central profile of
  S\'ersic models can lead to numerical precision requirements that
  are more stringent than usual.  With inadequate precision, there is
  a systematic and cumulative error that effectively changes the
  density profile.  We achieve the necessary precision by decreasing
  the \gravlens\ integration tolerance parameter \texttt{inttol} from
  its default value of $10^{-6}$ down to $10^{-10}$.}  yielding what
we call numerical lens models. Circular and elliptical S\'ersic models
are new features in \gravlens\ that have been added for this project.

For some models, not even the convergence is known analytically, so
the projection integral in equation~\eqref{eq:surfmassell} must be
evaluated numerically. In such cases we compute the convergence
$\kappa(\vec{R}')$ on a map and then solve the Poisson equation using
Fourier methods.  Here, $\vec{R}'=(x',y')$ is a two-dimensional
angular position on the sky in the lens plane, with $\vec{k}'$ its
counterpart in Fourier space.  We use FFTs to find the Fourier
transform $K(\vec{k}')$ of the convergence. We then rewrite the
real-space Poisson equation~\eqref{eq:Poisson} in Fourier space as
\begin{equation} \label{eq:PoissonF} 
  -k'^2 F(\vec{k}') = 2 K(\vec{k}')\,,
\end{equation}
which is readily solved for the Fourier transform $F(\vec{k}')$ of the
lens potential $\phi(\vec{R}')$.  Derivatives of the lens potential
are easily computed in Fourier space: for example, the Fourier
transform of $\vec{\nabla}\phi(\vec{R}')$ is $i \vec{k}' F(\vec{k}')$.
We can then use inverse FFTs to convert back to real space.  All of
this analysis can be done with the \gravlens\ routine
\texttt{kap2lens} (which has been added and tested for this project).
We refer to lens models computed this way as map-based lens models,
and we use them for our cusped models.\footnote{Hernquist and NFW
  models could be treated analytically in the spherical case and
  numerically in the elliptical case, since the convergence is known
  analytically.  However, we must treat generalised NFW models with
  $\gamma \ne 1$ with the map-based approach, since the convergence is
  not analytic.  We opt to handle all the cusped models in the same
  way for uniformity.}

There are four important technical details to consider in the Fourier
analysis. The first issue is avoiding ``edge effects'' that may arise
because we are using a rectangular box with an elliptical density map,
and because FFTs naturally impose periodic boundary conditions.  We
embed the $\kappa$-map in a grid at least twice as large as the
desired box in each direction (and then round up to the next power of
two, as required for FFTs). We take advantage of the fact that our
density models have circular or elliptical symmetry, and we know (from
Newton's third law) that the gravitational force inside an elliptical
shell due to that shell is zero.  We can therefore set the density to
zero outside of a reference ellipse that is larger than the region
where we wish to study lensing, without affecting the lensing
deflections or magnifications (or differential time delays, although
we do not study those). Specifically, if the box side length is $L$,
and the angle of the projected mass distribution relative to the $x'$
axis is $\psi$ (see also Section~\ref{SS:ellipsoidalshapes} and
equation~\ref{eq:misalignment_psi} in particular), we first define the
axes of the projected mass distribution
\begin{eqnarray}
  x_p & = & x' \cos\psi + y'\sin\psi 
  \nonumber \\
  y_p & = & -x'\sin\psi + y'\cos\psi 
  \nonumber
\end{eqnarray}
and then set $\kappa$ to zero when
\begin{equation}
  \label{eq:zerokappa}
  (x_p)^2 + \left(\frac{y_p}{b'/a'}\right)^2 \; > \; \left(\frac{L}{2}\right)^2.
\end{equation}

The second issue arises because equation~\eqref{eq:PoissonF} is
ill-behaved at $\vec{k}' = 0$.  We avoid this problem by setting all
Fourier transforms to zero at $\vec{k}' = 0$.  Since $K(0) \propto
\int \kappa(\vec{R}')\,\rmd\vec{R}'$, putting $K(0) = 0$ amounts to
requiring that the total mass in the FFT box vanish.  That, in turn,
means the code effectively adds a negative, uniform mass sheet to
offset the positive mass in our model $\kappa$-map.  We can compensate
by explicitly inserting a positive mass sheet to compensate for the
effective negative mass sheet. We have carefully tested this approach
to verify that it yields accurate lensing calculations (see
Section~\ref{SS:validation}).
 
As a third issue, we need to make sure our results are not affected by
the resolution of the grid on which we compute $\kappa$. Finally, we
must necessarily work in a finite box, and we need to check that our
results are not affected by the size of the box. We examine the latter
two issues carefully below in Sections~\ref{SS:resolution}
and~\ref{SS:boxsize}, respectively.

%---------------------------------------------------------------------
\subsection{Image configurations}
\label{SS:imgconfig}
%---------------------------------------------------------------------

The \gravlens\ software features a general ``tiling'' algorithm to
solve the lens equation and find the images of a given source (see
\citealt{2001astro.ph..2340K} for details).  We can check the
recovered images using general considerations from lens theory.
First, we classify the images as minima, maxima, and saddle points as
discussed in connection with equation~\eqref{eq:parities} in
Section~\ref{SS:lenstheory}.  We can then categorise the image
configuration as follows:
\begin{itemize}

\item A 2-image lens has one minimum and one saddle point (plus
one maximum that we do not analyse).

\item A 3-image lens has two minima and one saddle point.

\item A 4-image lens has two minima and two saddle points (plus
one maximum that we do not analyse).

\end{itemize}
Note that all three cases satisfy the image number relation in
equation~\eqref{eq:imgnum}.  Recall from Section~\ref{SS:lenstheory}
that we find but do not analyse maximum images because they are
rarely observed and do not play a major role in most strong lensing
applications.

According to lens theory, these are the only viable multiply-imaged
configurations that can be produced by an isolated galaxy whose
central surface density is shallower than $\Sigma \propto R'^{-1}$
\citep[see][]{1981ApJ...244L...1B,1992grle.book.....S}.  Therefore,
any set of recovered images that does not fall into one of these
three categories must represent a numerical error.

In general, the error rate in \gravlens\ is very low. The few errors
that do occur arise from two causes, which are related to the tiling
algorithm.  First, when the source lies very close to the caustic, two
of the images can lie so close together (straddling the critical
curve) that they are difficult for the tiling algorithm to resolve.
Second, the tiling algorithm has adaptive resolution, which is
valuable but presents certain challenges in regions where the
resolution changes \citep[cf.\ Fig.\ 4 of][]{2001astro.ph..2340K}.  In
most cases checking the image configuration identifies the rare
errors.  There is one error that can slip past this check: if the code
fails to find one saddle point of a 4-image lens, it will misclassify
the system as a 3-image lens.  We shall see below
(Section~\ref{SS:projangle}) that this error does occur, but at a low
rate that does not corrupt our results at a significant level.

%---------------------------------------------------------------------
\subsection{Point source cross-sections}
\label{SS:csec}
%---------------------------------------------------------------------

Our initial focus will be on understanding intrinsic selection biases
in strong lensing surveys (see Paper II).  We therefore study lensing
cross-sections with the idea that any physical effect that enhances
the cross-section will be more likely to appear in lens samples,
while any effect that reduces the cross-section will be disfavoured.
Historically, most lenses found in systematic surveys had point-like
images (quasars or radio sources).  That is changing with the advent
of surveys like SLACS \citep{2006ApJ...638..703B, 2008ApJ...682..964B},
but the statistics of galaxy/galaxy strong lensing are much more
complicated than the statistics of point source lensing
\citep{2008ApJ...685...57D}, so we currently focus on the point
source case.

At the most basic level, the lensing cross-section is just the
area of the source plane that leads to multiple imaging.  We call
this the total cross-section (\st), and subdivide it into the
cross-sections for the regions that produce 2, 3, or 4 images
(\sig{2}, \sig{3}, and \sig{4}).  In real surveys, we also need
to account for the fact that brighter lensed images are easier
to observe; this ``magnification bias'' effect is discussed in
Section~\ref{SS:magbias}.  The cross-sections have dimensions of area
on the sky, so the natural unit is arcsec$^2$.

We compute the cross-sections using Monte Carlo techniques.  We place
a source at a random position behind the lens, and solve the lens
equation numerically to find all the images of this mock lens.  We
repeat this process many times, tabulate the number of mock 2-image,
3-image, and 4-image lenses, and use the input number density of
sources to compute the associated cross-sections.  We find that using
5000 sources provides a good balance between run time and statistical
uncertainties.

%%%FIG
\begin{figure*}
\begin{center}
\includegraphics[width=3.0in,angle=0]{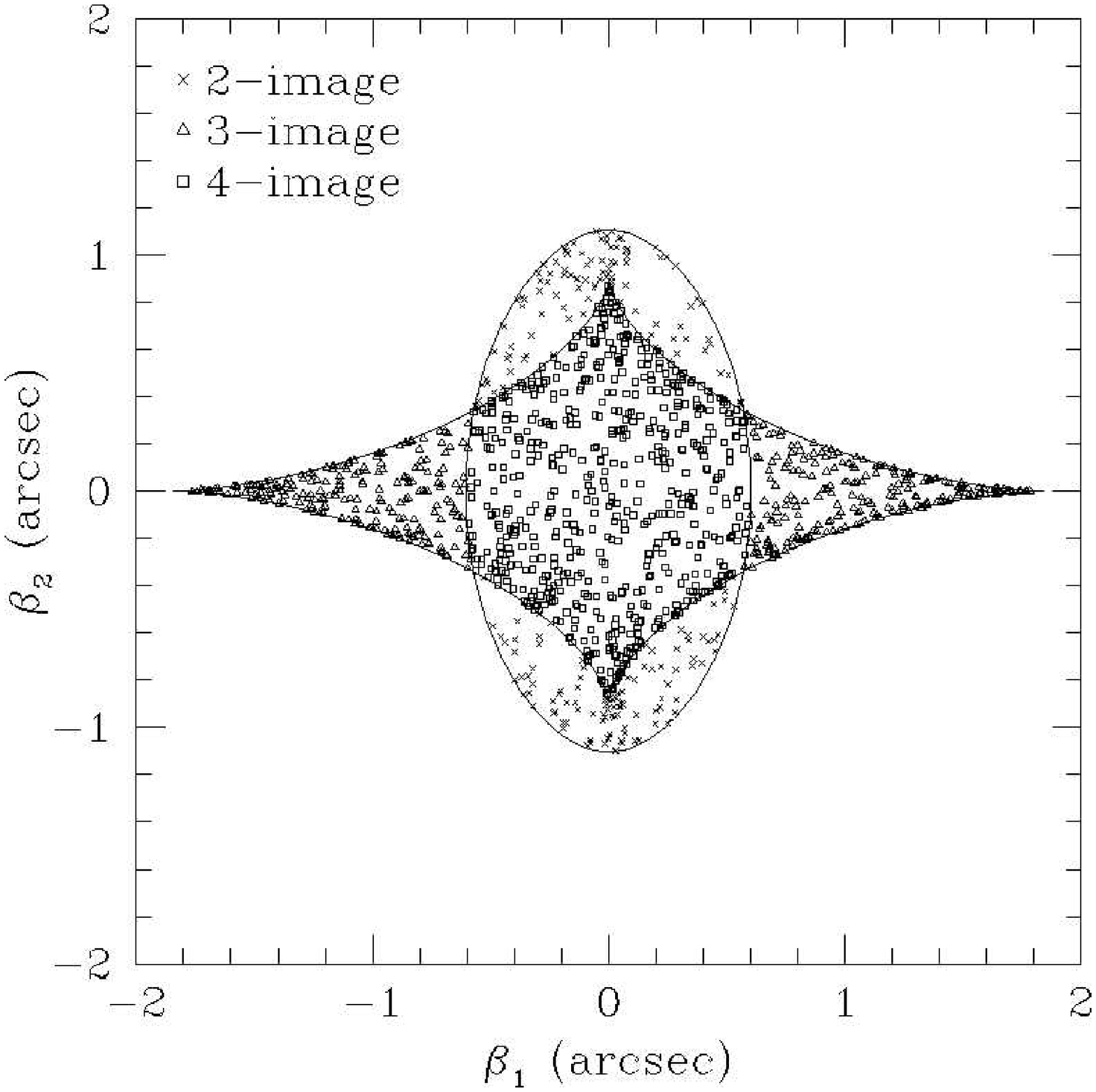}
\includegraphics[width=3.0in,angle=0]{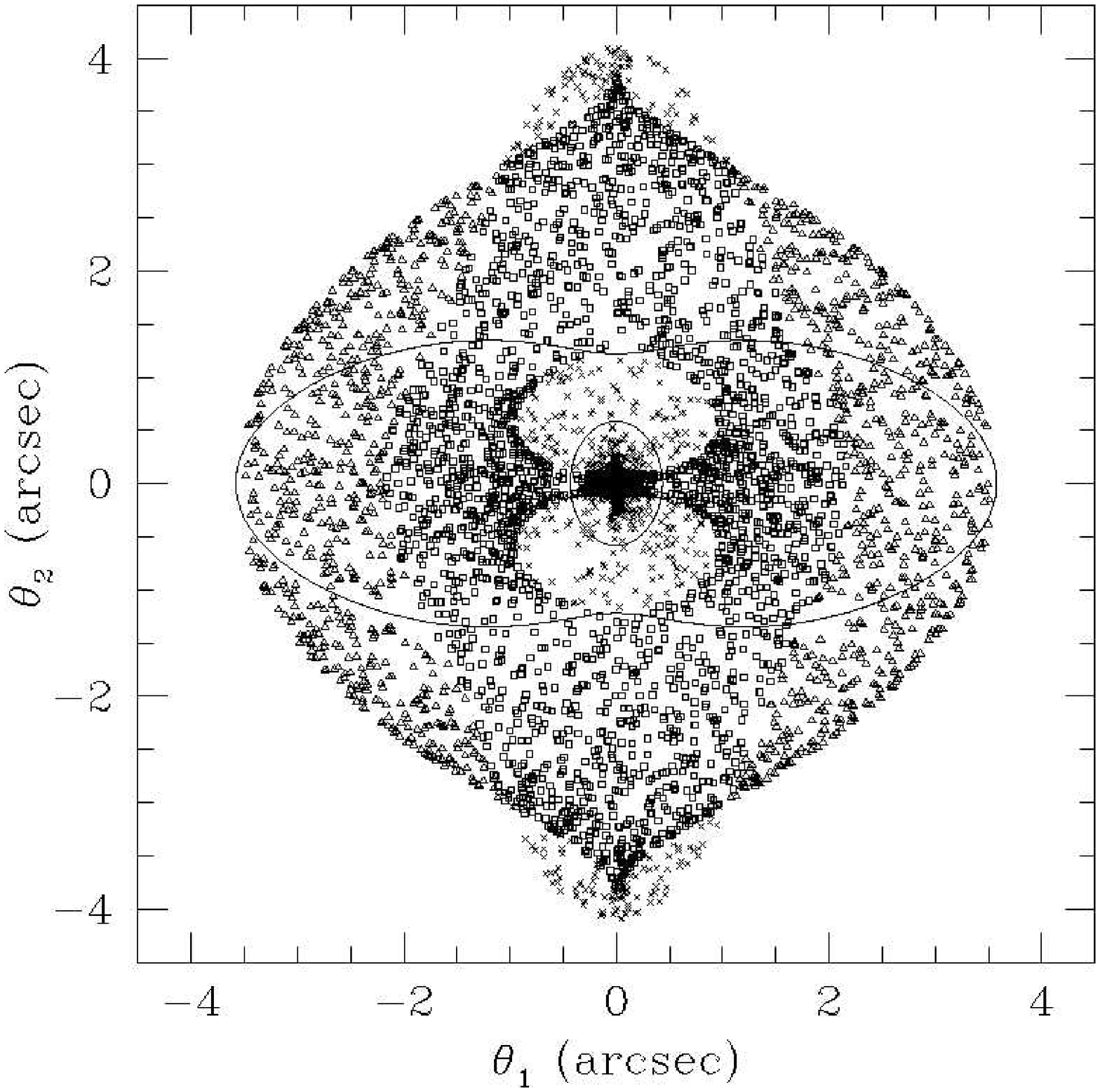}
\end{center}
\caption{\label{F:mockfig} Sample mock lenses for our very oblate
  higher mass, group scale model viewed edge-on. The left panel shows
  the caustics in the source plane, along with random source positions
  selected with magnification weighting (see Section~\ref{SS:magbias}). The
  right panel shows the critical curves in the lens plane, along with
  the images of the random sources. The horizontal axis is aligned
  with the major axis of the surface mass density of the dark matter
  component (which in turn is aligned with the stellar component in
  this axisymmetric case). The point type denotes the number of images,
  as indicated in the legend.  The left panel clearly shows the
  division of the source plane into 2, 3, and 4-image regions bounded
  by the caustics.  The right panel shows that the lens plane is also
  divided into distinct regions, although they are not bounded by
  easy-to-find curves \citep[cf.\ Fig.~1 of][]{2002ApJ...577...51F}.
  This figure includes only 2000 sources for clarity, compared with
  5000 used in our full analysis.  }
\end{figure*}
%%%FIG

The most obvious approach is to distribute sources randomly and
uniformly in the smallest circle enclosing the lensing caustics
in the source plane.  However, this approach yields poor sampling of
high-magnification systems that are rare but important, especially
for magnification bias (see Section~\ref{SS:magbias}).  For an alternate
approach, \citet{2004ApJ...612..660K} pointed out that picking
positions randomly and uniformly in the \emph{image} plane, and
then mapping them to the source plane, yields a set of source
positions that are random but weighted by the total magnification
of all images (see Fig.~\ref{F:mockfig}).  We must account for the
magnification weighting in order to compute the cross-section and
magnification bias properly, but that is a straightforward task.  Compared
with uniform source sampling, magnification-weighted source sampling
is no more complicated in concept or algorithm, and yields superior
results for Monte Carlo calculations of lensing cross-sections and
magnification bias.

We determine the statistical errors on the cross-sections using
bootstrap resampling.  We create 500 resampled sets of mock lenses,
and use the variance among them as the statistical uncertainty.  All
of this analysis (including bootstrap resampling) can be done with the
\gravlens\ routine \texttt{mockcsec} (new in version 1.99k).

%---------------------------------------------------------------------
\subsection{Magnification bias}
\label{SS:magbias}
%---------------------------------------------------------------------

Most lensing surveys have a flux limit that influences the
statistical distribution of lenses.  The fact that brighter images
are easier to observe leads to a lensing ``magnification bias''
\citep[e.g.,][]{1984ApJ...284....1T}.  We account for this fact by
computing not just the lensing cross-section itself, but the
combination of cross-section and magnification bias, which we call
the biased cross-section.

Consider a survey in which the number of sources brighter than flux
$S$ is $N_\mathrm{src}(>\!S)$, while the survey flux limit is $S_0$.
The survey is sensitive to systems in which the intrinsic flux $S$ and
the lensing magnification $\mu$ combine such that $\mu S > S_0$.
Thus, the total number of lenses found in the survey has the form of
an integral over the source plane, 
\begin{equation}
  N_\mathrm{lens} \propto \int N_\mathrm{src}(>\!S_0/\mu)\ \rmd\vec{\beta}.
\end{equation}
This number is  divided by the total number of sources to obtain 
the lensing probability, leading to the following definition for
the biased cross-section
\citep{2004ApJ...612..660K,2005ApJ...622...81M,2005ApJ...624...34H}:
\begin{equation} \label{eq:magbias}
  \sigma_n = \int_{n} \frac{N_\mathrm{src}(>\!S_0/\mu)}{N_\mathrm{src}(>\!S_0)}\ 
    \rmd\vec{\beta}
\end{equation}
The subscript $n$ indicates that we can compute the biased
cross-section for lenses with $n$ images by restricting the
integral to the appropriate region in the source plane (cf.\ 
Fig.~\ref{F:mockfig}).  (Note that we do not distinguish between
the symbols for unbiased and biased cross-sections, because we have
tried to be explicit in the subsequent text about which we are
using.)  The cross-section can be combined with the number density
of sources to obtain the observed number of lens systems.

While magnification bias depends explicitly on the flux limit of a
survey, it depends implicitly on the resolution as well.  For a survey
that detects lenses but does not resolve the images (assuming they are
resolved in follow-up observations), the initial detection depends on
the combined flux of all images, so it is the total magnification that
enters equation~\eqref{eq:magbias}.  For a survey that can resolve the
images at the outset, the discovery of lensing depends on the flux of
the second-brightest image,\footnote{A 2-image system in which one
  image is fainter than the flux limit will not be identified as a
  lens.  A system with more than two images will be identified as a
  lens if at least two images are above the flux limit.  (If some
  images are below the flux limit, the true number of images may not
  be known at first, but the system will still be identified as a
  lens.)}  so it is the second-ranked magnification that is used in
equation~\eqref{eq:magbias}.  To keep our analysis as general as
possible, we present results for both cases.

To compute biased cross-sections we must specify the source counts,
$N_\mathrm{src}(>\!S)$, or equivalently the differential counts,
$\rmd N_\mathrm{src}/\rmd S$, near the survey flux limit.  We consider
optical quasar surveys to determine a reasonable model.  For optical
surveys, $\rmd N/\rmd S$ is typically modelled as a broken power-law,
so we consider the two extremes.  Based on the results for SDSS
and 2QZ in the $g$- and
$i$-bands \citep{2004MNRAS.349.1397C,2006AJ....131.2766R}, we find
that $\rmd \log{N_\mathrm{src}}/\rmd \mbox{mag}= \alpha$ where
$\alpha \sim 1$ at the bright end and $\sim 0.1$ at the faint end.
This corresponds to
\begin{equation}
  \label{eq:diffsourcecount}
  \frac{\rmd N_\mathrm{src}}{\rmd S} \propto S^{-\nu},
\end{equation}
where $\nu=2.5\alpha+1$. Thus, our two values of $\nu$ are $3.5$ and
$1.25$. (A faint-end slope of $\nu=1.25$ is consistent with the
spectroscopic survey of faint quasars by
\citealt{2006AJ....131.2788J}.) To account for the break in the
power-law, we implement a functional form commonly used for quasar
luminosity functions,
\begin{equation}
  \label{eq:qsolumfunc}
  \Phi(L) \propto \frac{1}{(L/L_*)^{\nu_1} + (L/L_*)^{\nu_2}}
\end{equation}
where we use $\nu_1$ and $\nu_2$ of $3.5$ and $1.25$.  For simplicity,
we consider limiting luminosities rather than limiting fluxes, but we
examine a broad range of values: $0.04L_*$, $0.4L_*$, and $4L_*$
(chosen deliberately to sample the full range of effective slopes with
even spacing in magnitude).  These correspond to effective slopes
$\nu_\mathrm{eff}$ in equation~\eqref{eq:diffsourcecount} of $1.25$,
$1.5$, and $3.4$, although we caution that the luminosity function is
far from a pure power-law (especially near the intermediate limiting
luminosity), so magnification bias depends not just on
$\nu_\mathrm{eff}$ but on the full shape of the luminosity function.

% \crk{Rachel quoted the ratios of number densities, $4.9:1:1/82$,
% but it's not clear to me that they are relevant.  The biased
% cross-section is normalised such that it is independent of the
% number of sources.  So the normalization of the luminosity function
% doesn't matter; only the shape does.} \rsm{My thinking had been that
% the reader could use the ratios of number densities in combination
% with the biased cross-sections to estimate
% ratios of lensing systems.}

In summary, we track $7$ values of cross-section: the unbiased one,
and $6$ biased ones ($=2$ weighting schemes $\times$ $3$ limiting
magnitudes).  The \gravlens\ routine \texttt{mockLF} (new in version
1.99k) can be used to define a double power-law source luminosity
function that \texttt{mockcsec} uses to compute biased cross-sections.

%---------------------------------------------------------------------
\subsection{Image separation distributions}
\label{SS:imsepdistr}
%---------------------------------------------------------------------

In addition to the lensing cross-sections, we also consider the
distribution of image separations $\Delta\theta$.  Here, we define the
image separation for a given lensing system as the maximum distance
between any two images in the system.  This definition is simple and
well-defined even for image systems with arbitrary numbers of images,
and it has been used before \citep[e.g.,][]{2003Natur.426..810I,
2004ApJ...610..663O,2005ApJ...624...34H}. 

In principle, especially for systems with $>2$ images, there may be
more sophisticated ways to characterise image separations that
incorporate information about the full configuration of images.
While \citet{2002ApJ...578...25K} and \citet{2007ApJ...660....1O}
have taken first steps in this direction, there has been no
systematic exploration of various image separation statistics.
We defer consideration of these issues to future work. 

%---------------------------------------------------------------------
\subsection{Validation}
\label{SS:validation}
%---------------------------------------------------------------------

To validate our Monte Carlo methods for computing lensing cross-sections,
we present a simple test case with a singular isothermal sphere lens,
for which cross-sections can be computed analytically.  For concreteness
we set the Einstein radius to be $\Rein = 1\arcsec$ and we consider
a simple power-law source luminosity function with $\nu=1.25$, as
appropriate for the faint end of our fiducial broken power-law model.
(The biased cross-section cannot be computed analytically for the
full broken power-law luminosity function.) The ratio of the numerical
to analytic cross-section is $0.988\pm0.003$ for the unbiased
cross-section, as well as for the biased cross-section using the total
magnification. (The statistical error bar is from bootstrap
resampling.)  Thus, the numerical cross-sections are accurate to about
1 per cent. 

To test our Fourier methods for handling map-based lens models, we
present an example using our higher mass scale deprojected S\'ersic
model.  We compute the cross-sections with an analytic circular lens
model, now using our fiducial broken power-law source luminosity
function and tracking 7 cross-sections (one raw and six biased).  We
then create a $\kappa$-map that is $30\arcsec$ on a side and has
$800^2$ pixels,\footnote{The box size is the fiducial value determined
  from the box size tests discussed in Section~\ref{SS:boxsize}, and
  the resolution is twice the fiducial resolution discussed in
  Section~\ref{SS:resolution}.  While we clearly show in that
  subsection that the fiducial resolution is sufficient for cusped
  models, we have found that twice better resolution is necessary to
  resolve the very steep inner parts of the higher mass, deprojected
  S\'ersic model (stellar component).} and recompute the
cross-sections using a map-based lens model.  The ratio of the
cross-section from the map-based model to that from the analytic model
is between 0.960 and 1.041 for all 7 cross-sections.  The statistical
uncertainties depend on the bias mode, ranging from $0.001$ (for the
unbiased cross-section) to $0.023$ (for the biased cross-section using
the second-brightest magnification and a limiting luminosity of $4
L_*$).  The ratios become even closer to unity as the resolution
increases, but we defer further discussion of the resolution to
Section~\ref{SS:resolution}.  The main conclusions we draw from this
test are that the Fourier methods are valid and that map-based lens
models yield accurate cross-sections.

%---------------------------------------------------------------------
\subsection{Resolution}
\label{SS:resolution}
%---------------------------------------------------------------------

As noted in Section~\ref{SS:kapmap}, we need to ensure that the
lensing properties of our model galaxies are not affected by the
resolution of the grid on which we solve the Poisson equation.  Here
we present convergence tests to determine acceptable pixel sizes.
Before designing these tests, we must determine precisely how the
pixel size can cause problems with resolution.  There are two separate
issues: (i) In cases where the Einstein radius is quite small, the
finite pixel size may cause \gravlens\ to calculate a polygonal rather
than elliptical inner critical curve (in the lens plane), and
consequently a polygonal outer source plane caustic as well. (ii) In
cases where the inner slope of the density profile is very steep, a
large pixel size may mean that its slope, and therefore the location
of the inner critical curve, are poorly determined.  We consider the
two cases separately below.

\subsubsection{Obtaining a smooth inner critical curve}
\label{SSS:smoothcritcurve}

For these tests, we use the mass models with $\gdm=0.5$, since they
have the smallest Einstein radii and cross-sections, and therefore
yield the most stringent resolution requirements from a standpoint of
having a smooth inner critical curve.

We begin with a pixel size of $0.025''$ and $0.075''$ for the galaxy
and group scale models, respectively (with $10''$ and $30''$ boxes;
see Section~\ref{SS:boxsize}).  We vary the pixel size to a factor of $2$
smaller, and a factor of $2$ and $4$ larger in the hopes (a) of seeing
that the lensing results have converged by the smallest pixel sizes,
and (b) of determining the largest pixel sizes that yield reliable
results. We do these tests using a limited set of models: a spherical
model, followed by the very oblate, very prolate and triaxial models
seen from the most extreme 2, 2, and 3 viewing directions along the
principal axes.

The figures of merit for the convergence tests are the lensing
cross-sections ($\sig{2}$, $\sig{3}$, $\sig{4}$ and $\st$ separately),
although we also examine the critical curves and caustics. We are
interested in cross-sections both with and without magnification bias
(according to the various magnification schemes discussed in
Section~\ref{SS:magbias}).  For the purpose of this test, we require
convergence of the cross-sections at the 5 per cent level.

%%%FIG
\begin{figure}
\begin{center}
$\begin{array}{c@{\hspace{0.2in}}c}
\includegraphics[width=1.4in,angle=0,trim=0.4in 0 0 0]{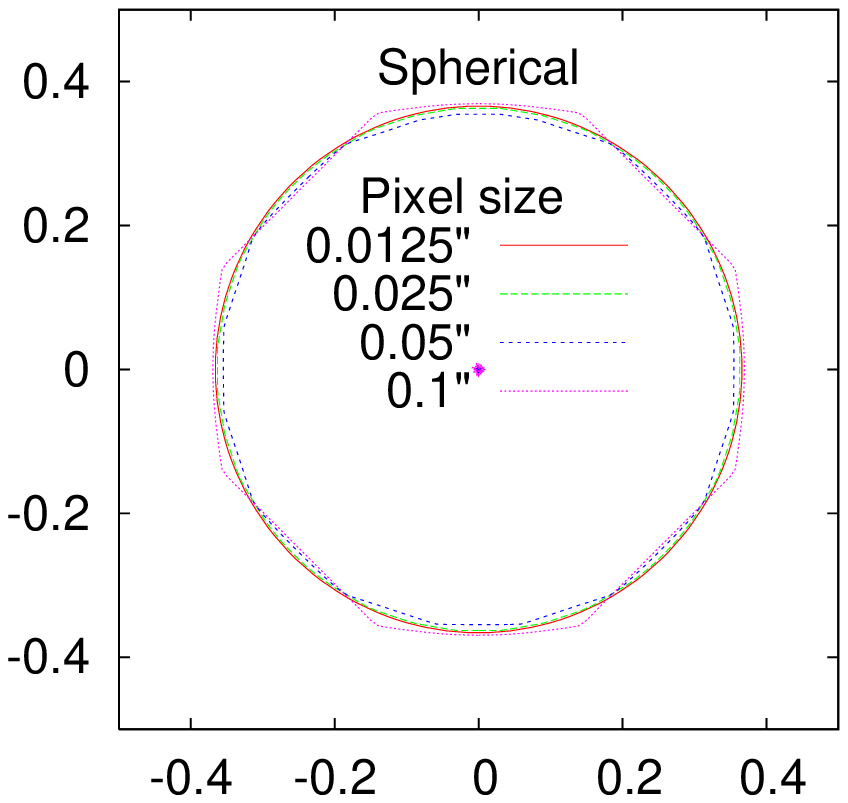} &
\includegraphics[width=1.4in,angle=0,trim=0.4in 0 0 0]{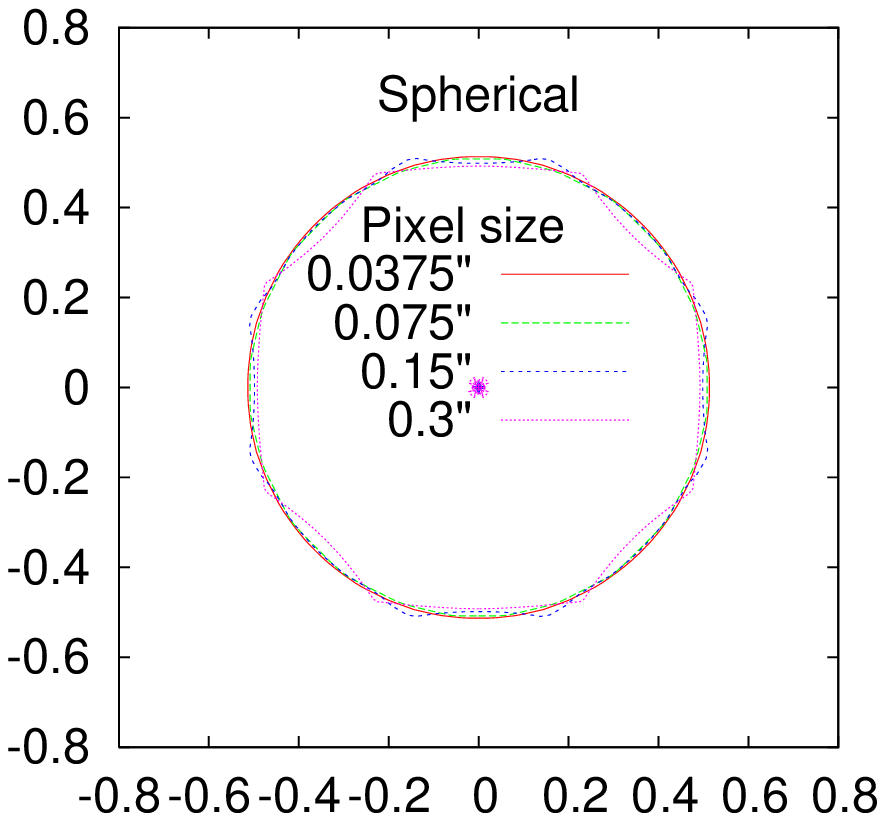} \\
\includegraphics[width=1.4in,angle=0,trim=0.4in 0 0 0]{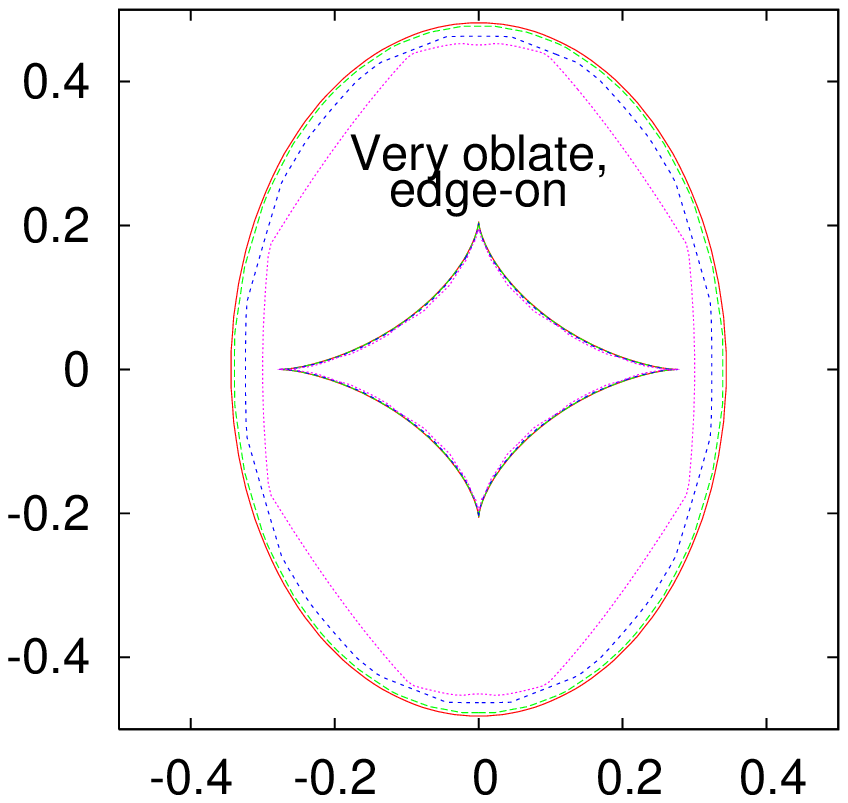} &
\includegraphics[width=1.4in,angle=0,trim=0.4in 0 0 0]{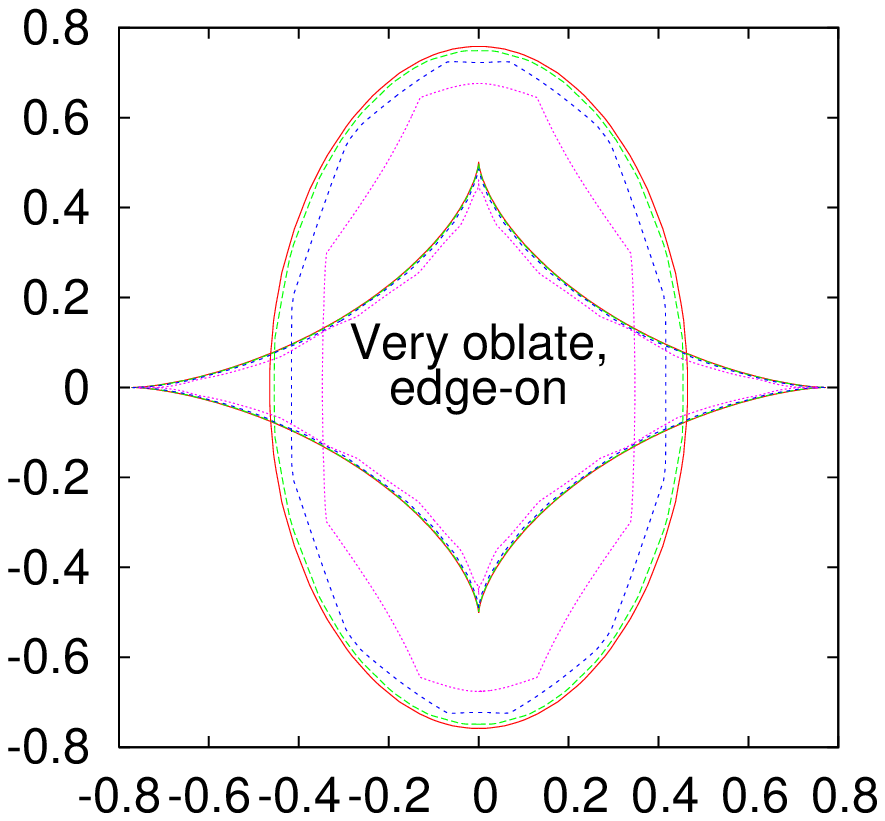} \\
\includegraphics[width=1.4in,angle=0,trim=0.4in 0 0 0]{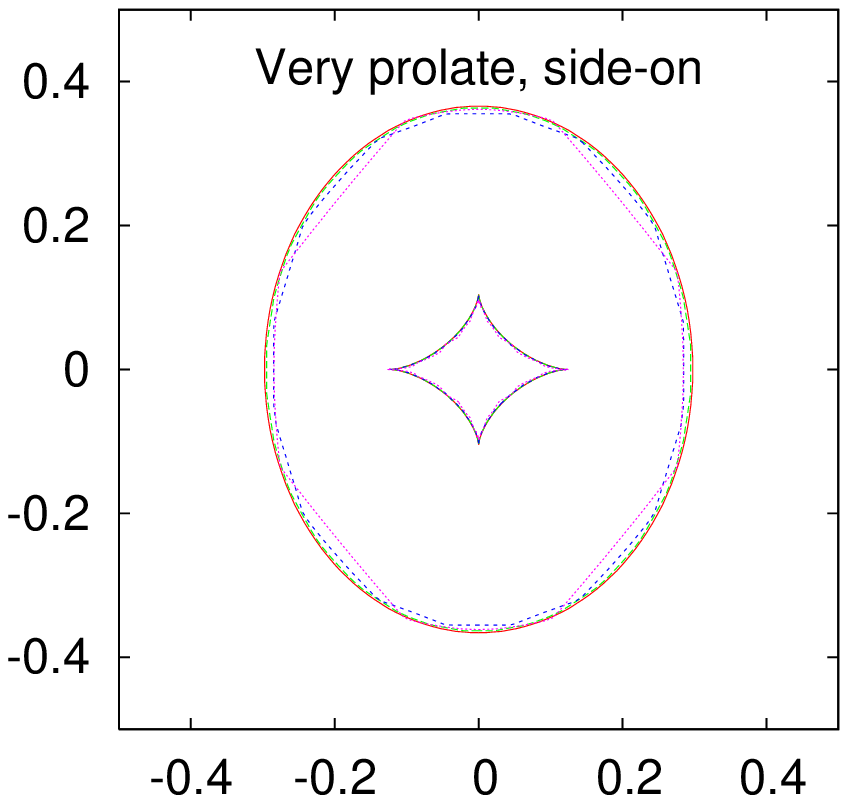} &
\includegraphics[width=1.4in,angle=0,trim=0.4in 0 0 0]{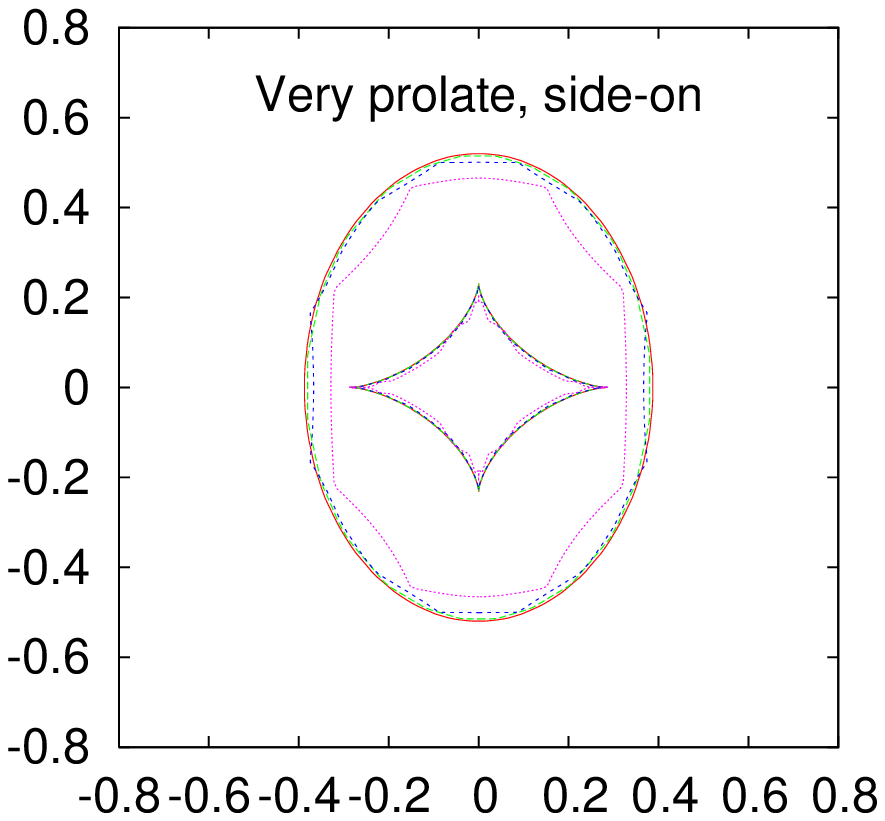} \\
\end{array}$
\caption{\label{F:causticres} Source plane caustics for the lower
  (left) and higher (right) mass models with three of the eight
  extreme lens configurations used for the $\gdm=0.5$ resolution
  tests: spherical, oblate (edge-on), and prolate (side-on) from top
  to bottom. The pixel scales are as labelled on the top plot for each
  mass scale. The axes are in the plane of the sky (in units of
  arcsec) with the horizontal axis aligned with the major axis of the
  surface mass density of the dark matter component.}
\end{center}
\end{figure}
%%%FIG

The effects of resolution are seen qualitatively in the caustics,
which are shown in Fig.~\ref{F:causticres} for three of the eight
lens configurations.  (Note that the mass density distributions are
always elongated along the $x'$ axis, but as usual the caustics are
elongated in the perpendicular direction.)  The quantitative results
of the resolution tests can be summarised as follows:
\begin{itemize}
\item For the lower mass scale, the most stringent requirements on the
  resolution come from the very oblate, edge-on model, for which we
  require $0.025\arcsec$ pixels. Even for this model, the resolution
  seems to have converged, i.e., the difference between $0.025\arcsec$
  and $0.0125\arcsec$ pixels is much less than our 5 per cent
  convergence criterion.  For the very prolate,
  side-on model, $0.025\arcsec$ pixels are formally insufficient for
  the 4-image cross-section.  However, the 4-image cross-section is
  only $\sim 5$ per cent of the total cross-section, and with such a
  small value the statistical uncertainties are large, so the 
  failure is actually marginal and not a significant concern.
\item For the higher mass scale, the most stringent resolution test
  also comes from the very oblate, edge-on case.  The second-best
  resolution ($0.075\arcsec$ pixels) marginally satisfies our
  convergence criterion for the 4-image case, and marginally fails for
  the 3-image case. However, this marginal failure is unlikely to be
  too crucial, since the 3-image cross-section is such a small
  fraction of the total cross-section.  In all other cases, the
  second-best resolution ($0.075\arcsec$ pixels) is clearly sufficient
  at the 5 per cent level.
\item In general, as we have predicted, the radial caustic is much
  more strongly affected by pixelisation effects than the tangential
  caustic, since the radial caustic maps to the inner critical curve
  in the lens plane.  The radial caustic tends to become more
  polygonal rather than elliptical as the resolution is degraded.  In
  Fig.~\ref{F:causticres}, it is clear that the caustics are nearly
  identical for both the best and second-best resolutions, and that
  the two poorer resolutions are insufficient.
% \item There are 3-image lenses for all scenarios in which
%   the tangential caustic sticks out of the radial caustic.  In
%   practice, this only occurs for the higher mass scale, for which the
%   more highly non-spherical dark matter component dominates relative to
%   the rounder stellar component.
\end{itemize}
In conclusion, our second-best resolutions for both mass scales ,
$0.025\arcsec$ ($\simeq 0.11$\,kpc) and $0.075\arcsec$ ($\simeq
0.33$\,kpc), appear to be safe choices from the standpoint of
obtaining smooth critical curves and caustics, and the results have
converged in comparison with higher resolution simulations at the $5$
per cent level.  In dimensionless units, these resolutions correspond
to $\sim 25$ pixels lengths per $\Rein$.

\subsubsection{Resolving a steep inner slope}
\label{SSS:resolvinginnerslope}

The other resolution issue, the ability to resolve the slope of the
density profile in the inner regions, is most easily tested using our
steepest cusped model, $\gdm=1.5$.  In this case, we repeat the
previous tests but only for the spherical case.  

Our results suggest that while this effect may be significant for the
poorest resolution we consider, the results have converged at better
than the 5 per cent level for our fiducial resolution.  For example,
in the previous subsection (spherical $\gdm=0.5$ case), the fiducial
resolution gave cross-sections that were 1.4 and 2.7 per cent lower
than the best resolution (lower and higher mass models, respectively);
the poorest resolution led to a highly significant 20 and 35 per cent
reduction of the cross-section.  For $\gdm=1.5$, those numbers are 1.1
and 0.9 per cent reduction in \st\ for the fiducial resolution, or 15
and 11 per cent reduction for the poorest resolution.  In all cases,
these results are for the unbiased cross-section, for which it is
clear that the fiducial resolution is sufficient to solve both
possible resolution problems.  For the biased cross-section, the
resolution requirements appear to be slightly less stringent.

While we do not use a map-based approach for the deprojected S\'ersic
profile simulations, similar convergence tests suggest that such an
approach would lead to more stringent resolution requirements (at
least a factor of two better), due to
the necessity of resolving a very steep inner slope in the innermost
regions. The reason these models may be particularly problematic is
that they do not converge to a single inner slope, unlike the cusped
models. 

%---------------------------------------------------------------------
\subsection{Box size}
\label{SS:boxsize}
%---------------------------------------------------------------------

We also need to check that the size of the box in which we solve
the Poisson equation does not affect our lensing results.  We have again
run convergence tests, but now using mass models with $\gdm=1.5$
since these have the largest Einstein radii and hence the greatest
sensitivity to the box size.  We include non-spherical models since
the elongation of the mass distribution may create extra demands
on the size of the box in the direction of the major axis.  As in the resolution tests, we use only
the extreme non-spherical configurations, as a worst-case scenario.

%%%FIG
\begin{figure}
\begin{center}
  $\begin{array}{c@{\hspace{0.4in}}c}
    \includegraphics[width=1.4in,angle=0,trim=0.5in 0 0 0]{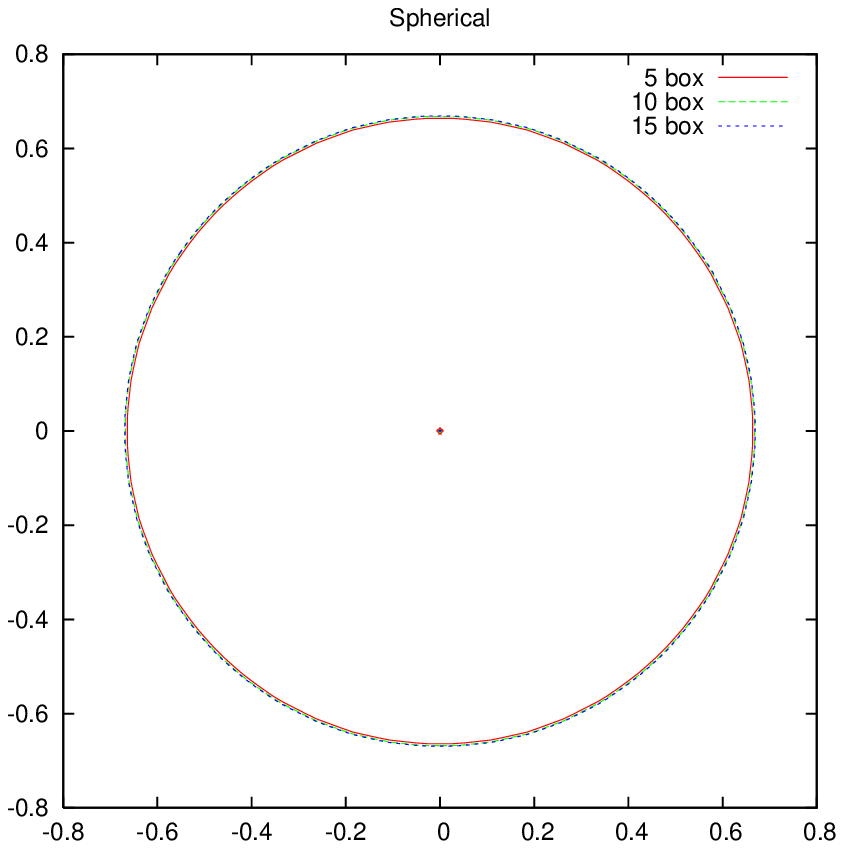} &
    \includegraphics[width=1.4in,angle=0,trim=0.5in 0 0 0]{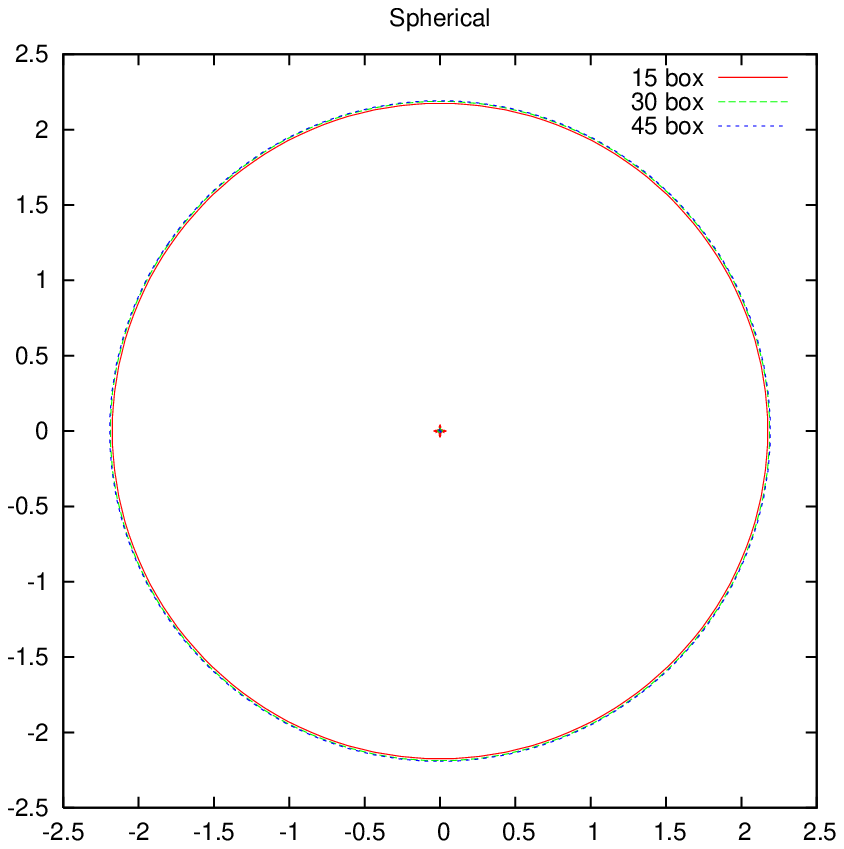} \\
    \includegraphics[width=1.4in,angle=0,trim=0.5in 0 0 0]{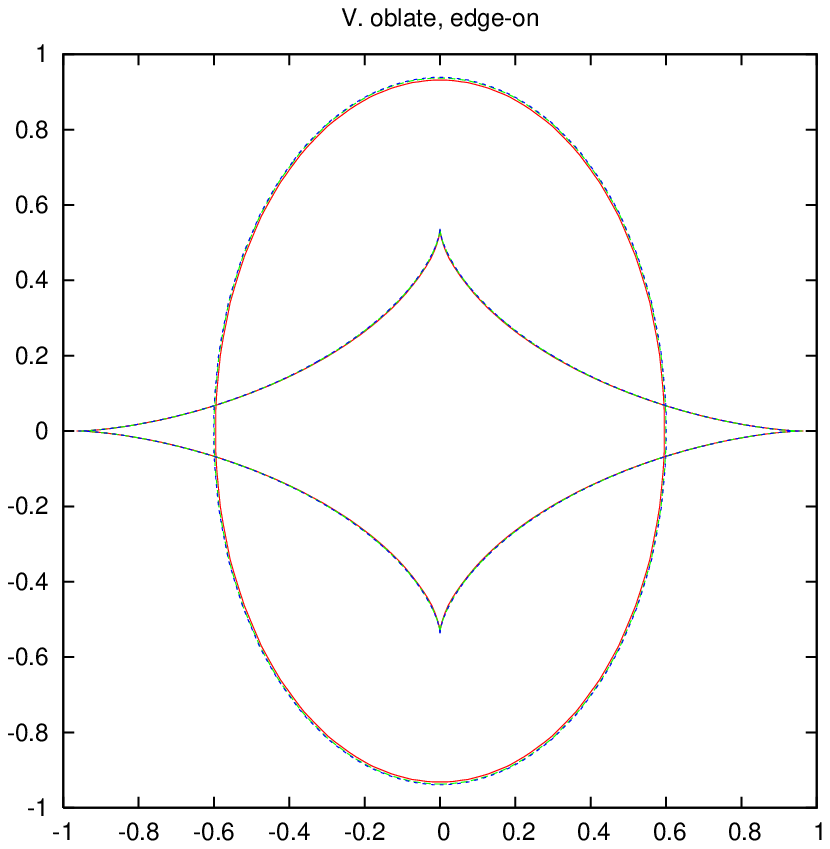} & 
    \includegraphics[width=1.4in,angle=0,trim=0.5in 0 0 0]{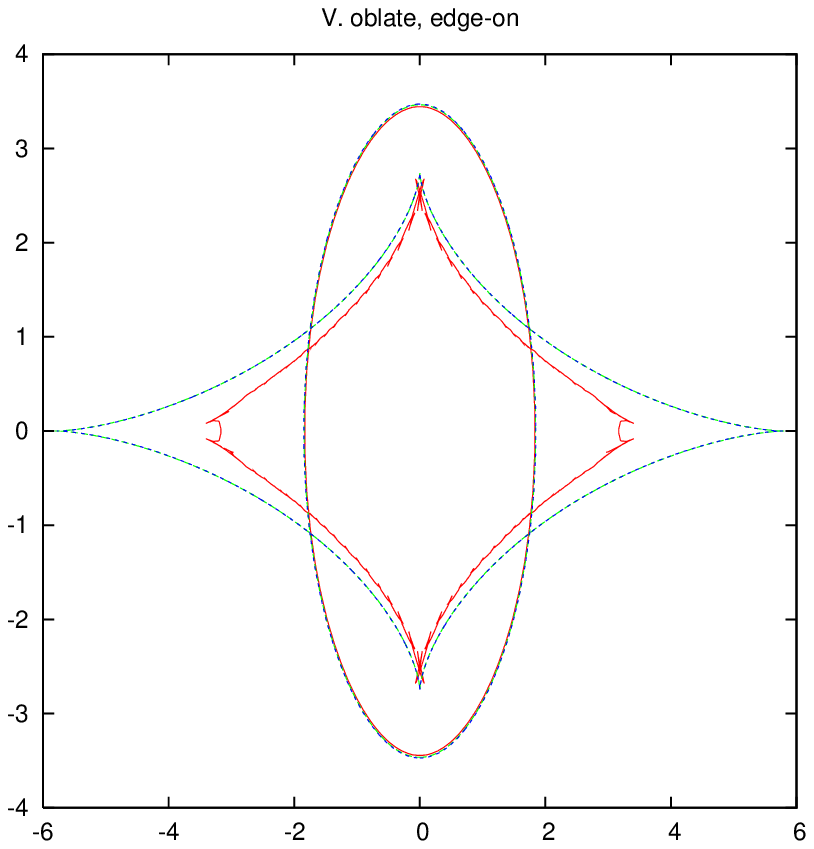} \\
    \includegraphics[width=1.4in,angle=0,trim=0.5in 0 0 0]{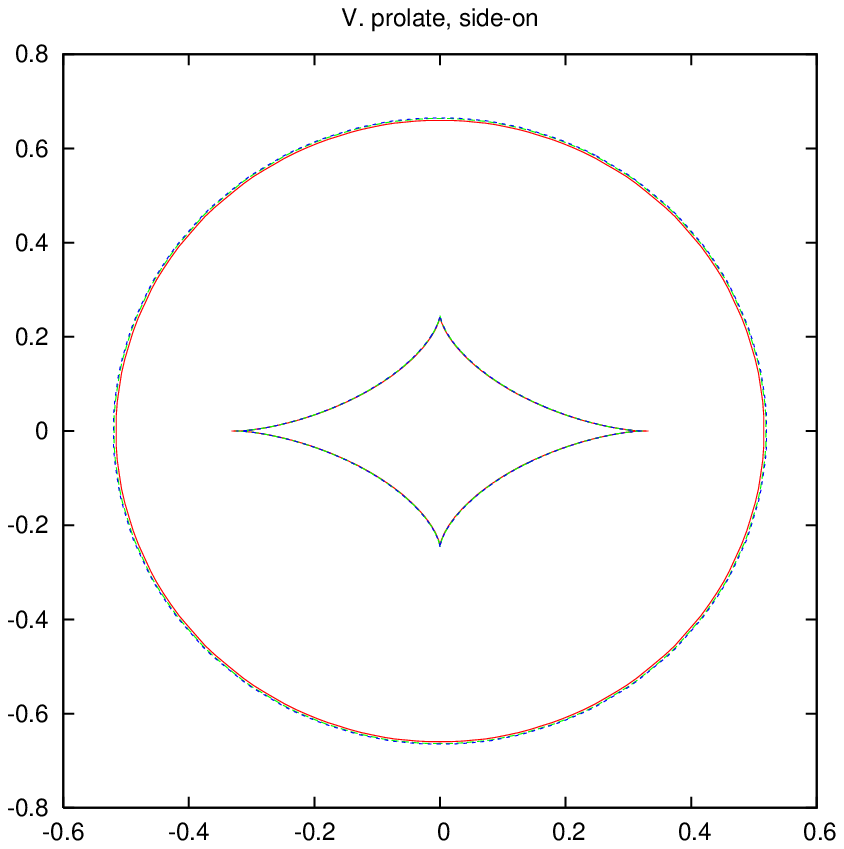} &
    \includegraphics[width=1.4in,angle=0,trim=0.5in 0 0 0]{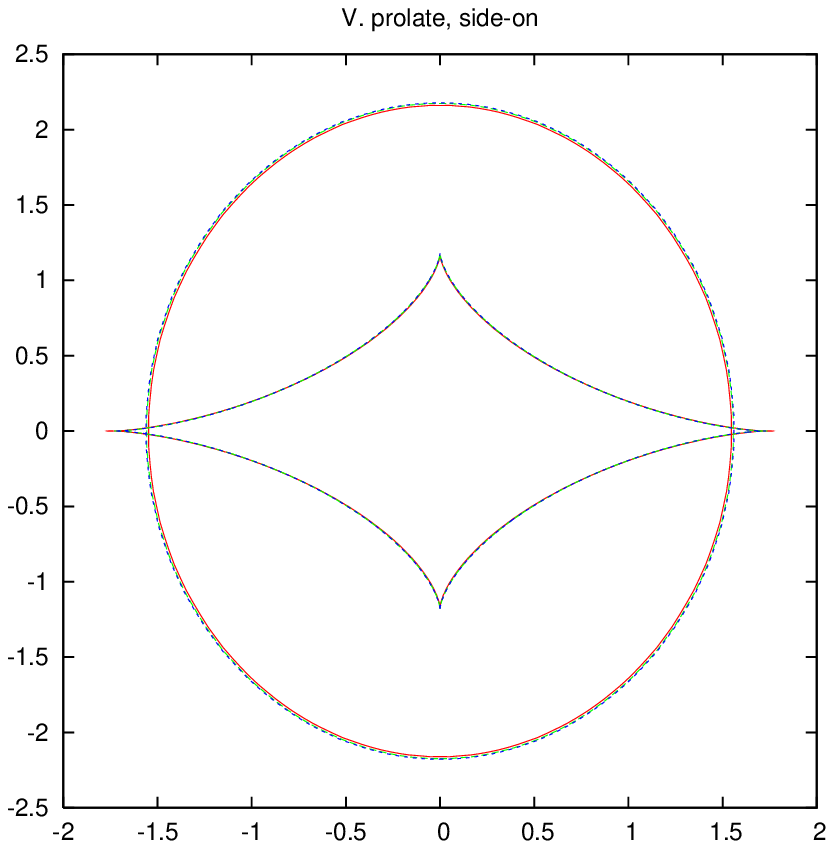} \\
  \end{array}$
  \caption{\label{F:causticbox} Source plane caustics for the lower
    (left) and higher (right) mass models, with the three of the eight
    extreme lens configurations used for the box size tests:
    spherical, edge-on oblate, and side-on prolate (top to bottom).
    Box sizes are labelled on the top plot for each mass scale. The
    meaning of the axes is the same as in Fig.~\ref{F:causticres}.}
\end{center}
\end{figure}
%%%FIG

To begin, we use $10 \arcsec \times 10 \arcsec$ and $30 \arcsec \times
30 \arcsec$ boxes for the galaxy and group scale models, respectively
(with our fiducial $0.025\arcsec$ and $0.075\arcsec$ pixels; see
Section~\ref{SS:resolution}). We then examine box sizes of $5\arcsec$,
$10\arcsec$, and $15\arcsec$ for galaxy scale models, and $15\arcsec$,
$30\arcsec$, and $45\arcsec$ for group scale models.
Fig.~\ref{F:causticbox} shows the effects of the box size on the
caustics, for three of the eight extreme lens configurations.  A
summary of the quantitative results for all eight lens configurations
are as follows:
\begin{itemize}
\item For the lower mass scale, the default box size of $10\arcsec$ is
  sufficient.  There is one marginal case---the 3-image cross-section
  for the very oblate, edge-on model---but this cross-section is a
  small fraction ($<1$ per cent) of the total cross-section for that
  model, and the statistical errors in determining it are large.  The
  smallest ($5\arcsec$) box typically yields unbiased cross-sections
  that pass our 5 per cent convergence criterion, but the biased
  cross-sections can fail dramatically (up to 25 per cent reduction
  of \st\ relative to the largest box), which eliminates the possibility
  of using such a small box.
\item For the higher mass scale, the default box size of $30\arcsec$
  is typically sufficient to achieve $5$ per cent accuracy compared to
  the larger box size, though occasionally the results are marginal,
  particularly for highly biased cross-sections.  Visually, the
  caustics appear to be nearly identical for the $30\arcsec$
  (fiducial) and $45\arcsec$ (largest) boxes
  (Fig.~\ref{F:causticbox}), with rather bizarre and very significant
  distortion for the $15\arcsec$ box for some viewing directions, for
  which the cross-sections are also significantly reduced.
\item Some caustics appear smoother than for the corresponding cases
  in the resolution tests in Section~\ref{SS:resolution}.  The difference
  comes from the fact that here we use $\gdm=1.5$ models, which have
  Einstein radii that are significantly larger than for the $\gdm=0.5$
  models used for the resolution tests.
\item The effect of insufficient box size is to shrink both the radial
  and tangential caustics, and to alter the shape of the tangential
  caustic.
\end{itemize}
The caustics and cross-sections together imply that even with the
largest possible Einstein radius at each mass scale, the medium-sized
boxes, $10\arcsec$ ($\simeq 43.6$\,kpc) and $30\arcsec$ ($\simeq
131$\,kpc), are sufficient.  The box size requirement in dimensionless
units is $\sim 16\,\Rein$.

%---------------------------------------------------------------------
\subsection{Sampling viewing angles}
\label{SS:projangle}
%---------------------------------------------------------------------

We are interested in the lensing properties of galaxies not only for
specific viewing angles, but also when averaged over all viewing
angles.  We must choose how to sample the possible viewing directions
in order to obtain accurate averages with as little computational
effort as possible.  We are guided by two principles.  First, we
expect the lensing quantities to vary with viewing direction in some
smooth and, in the oblate and prolate cases, monotonic way.  Second,
this variation is more directly related to the shape of the projected
surface mass density than to the viewing angles per se.

\subsubsection{Sampling in projected axis ratios}
\label{SSS:samplingprojaxisratios}

Instead of directly sampling angles relative to the intrinsic axes
$\vartheta$ and $\varphi$, we sample linearly in the ratios $a'/a$ and
$b'/a$, i.e., the projected major and minor semi-axis length $a'$ and
$b'$ as given in equation~\eqref{eq:apbpellipse}, normalised by the
intrinsic length scale $a$. Since this choice does not correspond to
an equal-area sampling on the viewing sphere (i.e., uniform in $-1 \le
\cos\vartheta \le 1$ and $0 \le \varphi \le 2\pi$), we need to include
a weighting factor to recover accurate averages.  The orientation-averaged
value of some quantity $f$ (which in our case will be a lensing
cross-section) can be written as
\begin{eqnarray}
  \label{eq:angleavg}
  \langle f \rangle 
  & = & 
  \frac{2}{\pi} \int_0^{\pi/2} \int_0^{\pi/2} 
  f(\vartheta,\varphi) \; \sin\vartheta \; 
  \rmd\vartheta \; \rmd\varphi, \\
  \label{eq:abavg}
  & = &
  \frac{2}{\pi} \int_c^b \int_b^a 
  f(\vartheta,\varphi) \; \sin\vartheta \; 
  J(a',b') \; \rmd a' \; \rmd b',
\end{eqnarray}
where the expression for the Jacobian $J(a',b')$ is given in
Appendix~\ref{A:samplingviewing}, as is the inversion of
equation~\eqref{eq:apbpellipse} to find $(\vartheta,\varphi)$ in terms of
$(a',b')$.  In the oblate and prolate cases, one of the two integrals
is irrelevant since the lensing depends on only one of the angles, but
in the triaxial case both integrals are important.

By construction, the individual mass components we use have axis
ratios that are constant as a function of radius.  However, since
the dark matter and stellar components have different intrinsic
shapes (Table~\ref{T:axisratios}), the combined, total density
has intrinsic and projected axis ratios that vary with radius (see
also Fig.~\ref{F:surfmaps}). Fortunately, the precise choice of
axis ratios to use for the sampling is not terribly important,
as long as it is treated properly according to equation~\eqref{eq:abavg}.
We choose to sample in the axis ratios of the dark matter
component, i.e., in what follows $a'/a$ and $b'/a$ always refer to
the dark matter component.

There are two important issues related to the sampling:
\begin{enumerate}
\item We must first test our intuition that cross-sections vary
  smoothly with $a'/a$ and $b'/a$.  We can do this by using a large
  number $N_s$ of samplings (we use $N_S = 47$ for the oblate case,
  $N_S = 31$ for the prolate case, and $N_S = 23 \times 15 = 345$ for
  the triaxial case).
\item We must determine how few samplings can be used to obtain
  orientation-averaged cross-sections that are accurate at the 5 per cent
  level.  We can do this by undersampling the original number of
  samplings and seeing how few samplings we can use while still
  recovering the average cross-sections to within 5 per cent.
\end{enumerate}

For the viewing angle sampling tests, we use cusped density profiles
with dark matter slope $\gdm=1.5$ and the extreme oblate and prolate
shapes, since these will have the strongest variation of cross-section
with inclination. We also consider the triaxial case, for which we
test the sampling in both $a'/a$ and $b'/a$. In addition to the total
strong lensing cross-section, we require accurate recovery of the
2, 3, and 4-image cross-sections separately.

\subsubsection{Variation of cross-section with projected axis ratios}
\label{SSS:crosssectprojaxisratios}

%%%FIG
\begin{figure*}
\begin{center}
$\begin{array}{c@{\hspace{0.1in}}c}
\includegraphics[width=\columnwidth,angle=0,trim=0.3in 0 0.3in 0]{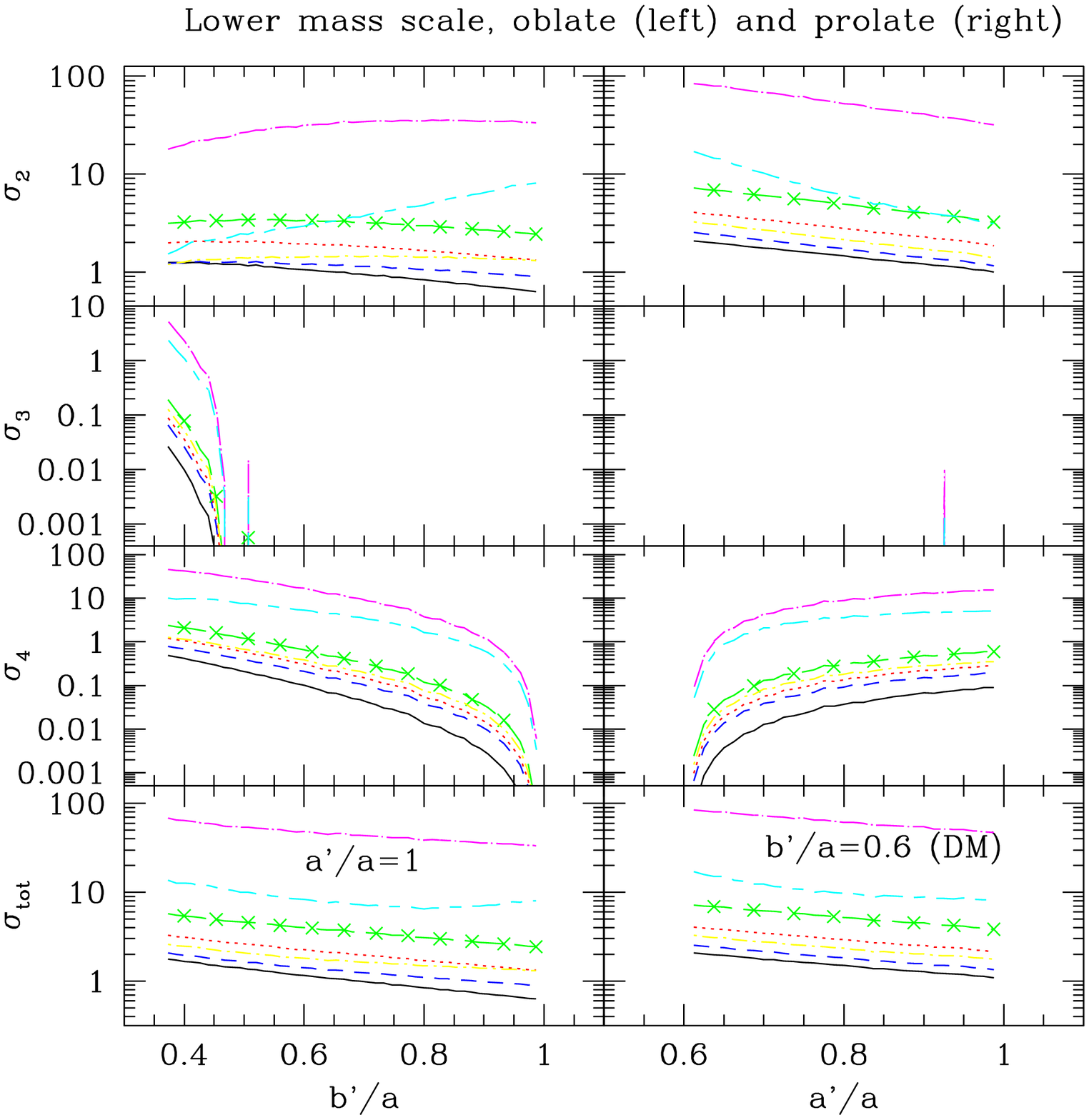} &
\includegraphics[width=\columnwidth,angle=0,trim=0.0in 0 0.6in 0]{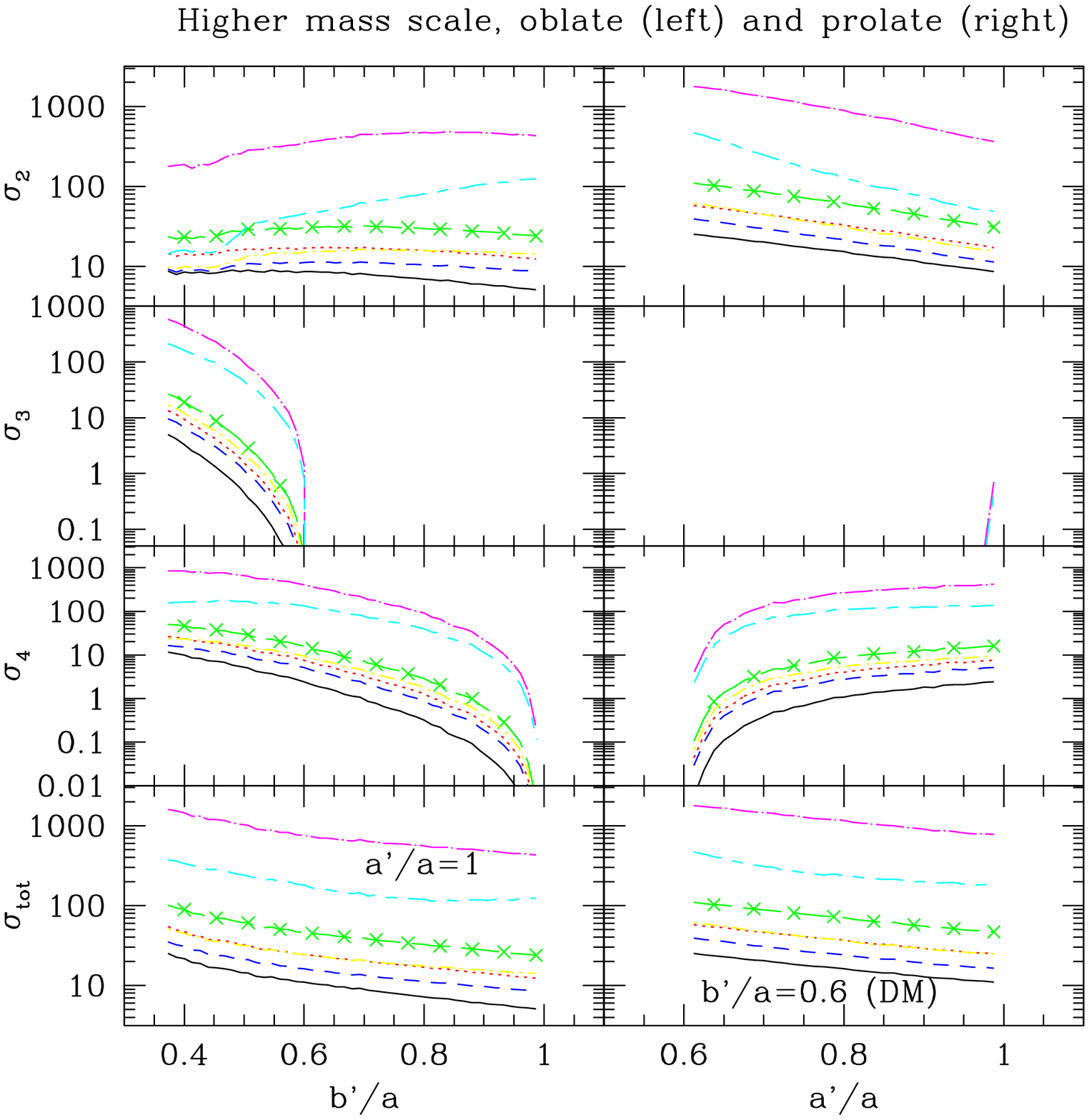} \\
\end{array}$
\caption{\label{F:incboth} Lensing cross-sections for the lower (left
  plot) and higher (right plot) mass scale, for oblate and prolate
  models, for 2, 3, 4-image systems separately as labelled, and the
  total cross-section at the bottom.  The lines are as follows: black
  solid $=$ unbiased, red dotted and blue dashed $=$ weighted by total
  and second-brightest magnification respectively for $0.04L_*$
  limiting magnitude, green long-dashed and yellow dot-dashed $=$ the
  same for $0.4L_*$ limiting magnitude, and magenta dot-long dashed
  and cyan short-long dashed $=$ the same for $4L_*$ limiting
  magnitude.  The lines show the curves resulting from our fiducial
  sampling in $a'$ and $b'$; crosses show our eventual choice of
  sampling as described in Section~\ref{SSS:undersample}.  The 3-image
  cross-section should be identically zero for $b'/a$ or $a'/a$ above
  some threshold; the occasional spikes at small but finite values
  indicate small numerical errors due to misclassification of images.}
\end{center}
\end{figure*}
%%%FIG

We show the variation of the cross-sections with viewing angle in
Fig.~\ref{F:incboth} for both mass models, using the very oblate and
very prolate shapes.  The symmetry of the oblate case means we only
need to vary the polar viewing angle $\vartheta$, which we accomplish
by taking equal steps in $a'/a$.  Similarly, for the prolate case we
only need to vary the azimuthal viewing angle $\varphi$, again taking
equal steps in $a'/a$.  For this plots, we use our fiducial number of
samplings (47 for the oblate case, and 23 for the prolate case) to
generate the cross-section curves, and we show the undersampled by
four case with crosses.  We defer discussion of the physical intuition
behind the viewing angle dependence to Paper II, and focus here on the
technical issues of smoothness and sampling.  It is clear that, as
expected, the variation of the individual and total cross-sections is
a smooth function of the projected axis ratios for prolate and oblate
dark matter plus stellar density shapes.

Note that the 3-image cross-section should be identically zero for
any lens configuration for which the tangential caustic lies entirely within
the radial caustic.  This means that $\sigma_3$ should be zero for our
oblate models when $b'/a$ is above some threshold, and for our
prolate models throughout the full range of $a'/a$.  Our numerical
3-image cross-sections are not strictly zero at a very few values of
the projected axis ratio due to extremely rare errors
in finding and classifying images (see Section~\ref{SS:imgconfig}).
However, the spurious 3-image cross-sections are well below $0.1$ per cent of 
the total cross-sections, so they are a negligible source of error.

% %%%FIG
% \begin{figure*}
% \begin{center}
% $\begin{array}{c@{\hspace{0.00in}}c}
% \includegraphics[width=0.45\textwidth,angle=0]{lower_raw_csec.ps} &
% \includegraphics[width=0.45\textwidth,angle=0]{higher_raw_csec.ps} \\
% \end{array}$
% \caption{\label{F:bothmtriax} Base-10 logarithm of the unbiased
%   lensing cross-sections for the lower (left) and higher (right) mass
%   models for the triaxial shape, as a function of $a'/a$ and $b'/a$.
%   Note the different scales for the contours on each plot.  $\sig{3}$
%   is not shown because it is negligible.  The crosses indicate our
%   final choice of sampling, as described in Section~\ref{SSS:undersample}.}
% \end{center}
% \end{figure*}
% %%%FIG

%%%FIG
\begin{figure}
  \begin{center}
    \includegraphics[width=1.0\columnwidth]{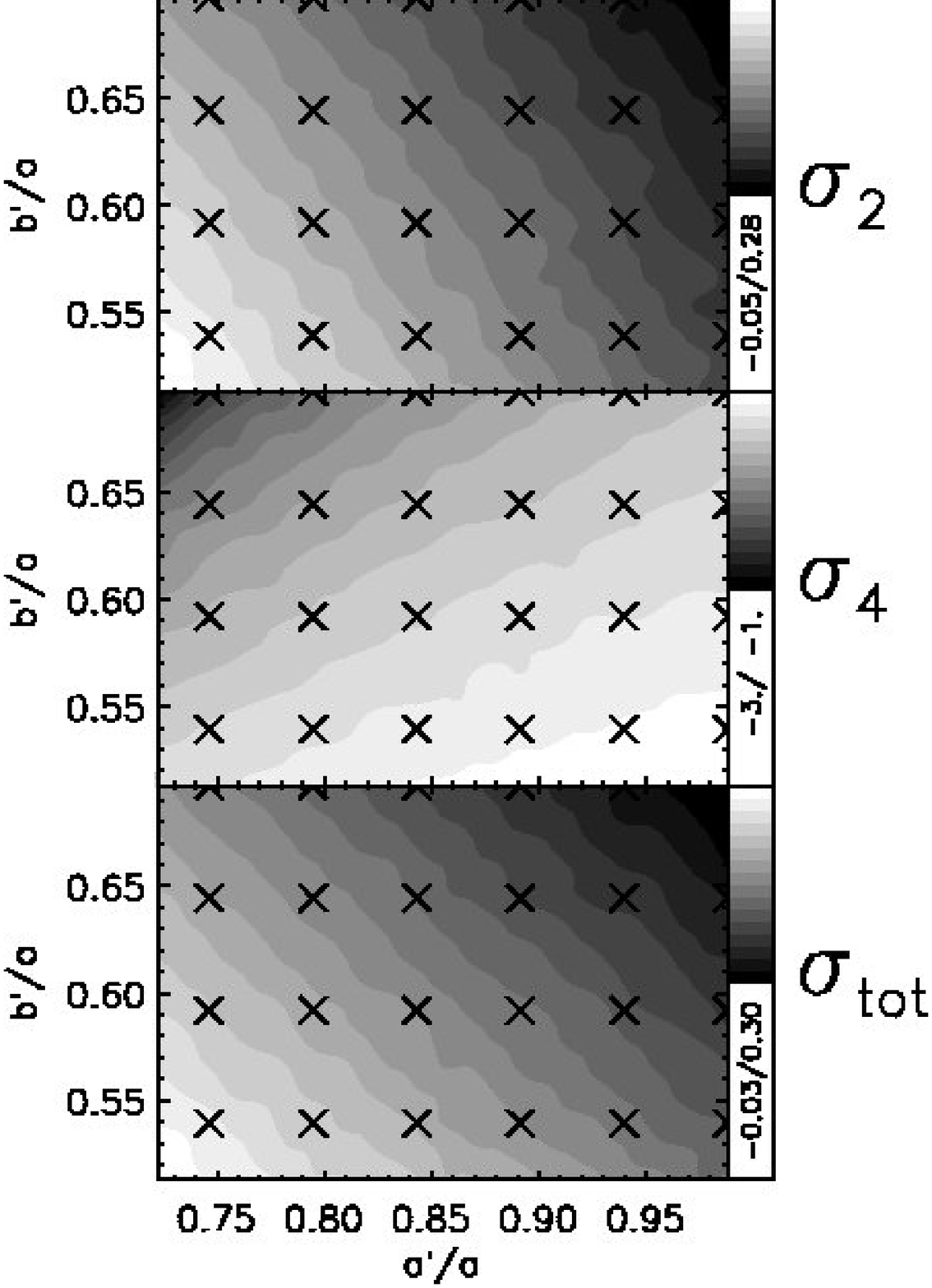}
    \caption{\label{F:bothmtriax} Base-10 logarithm of the unbiased
      lensing cross-sections for the lower mass, triaxial shape model,
      as a function of $a'/a$ and $b'/a$. Note that the contour scales
      are different on each plot, and $\sig{3}$ is not shown because
      it is negligible. The crosses indicate our final choice of
      sampling, as described in Section~\ref{SSS:undersample}.}
  \end{center}
\end{figure}
%%%FIG

%% For the lower and higher mass triaxial models, the variation of the
%% unbiased cross-sections with $a'/a$ and $b'/a$ is shown in
%% Fig.~\ref{F:bothmtriax}, while the changes for the biased
%% cross-sections are described qualitatively. 

In Fig.~\ref{F:bothmtriax}, we show the variation of the unbiased
cross-sections with $a'/a$ and $b'/a$ for the lower mass triaxial
model.  (The results are very similar for the higher mass triaxial
model, except for an overall difference in amplitude.)  The changes
for the biased cross-sections are described qualitatively.  We again
defer discussion of the physical intuition behind the
orientation-dependence of the cross-section to Paper II.  Here we
merely observe that for the triaxial case, the unbiased cross-sections
$\sig{2}$, $\sig{4}$, and \st\ are smooth functions of $a'/a$ and
$b'/a$, without multiple local minima/maxima or other features that
would require dense sampling. The biased cross-sections also vary
smoothly, regardless of which magnification bias mode we use. One
difference is that the contours of the 2-image biased cross-section
tend to lie at nearly constant $a'/a$, not as tilted as in the
unbiased case (see Paper~II).

\subsubsection{Minimum number of samplings}
\label{SSS:undersample}

Next we consider how few samplings we can get away with and still
recover accurate results for cross-sections averaged over viewing
directions.  To do so, we start with our fiducial number of samplings,
and compare the total average cross-sections (properly weighting each
sampling using the Jacobian) when undersampling by a factor of 2, 4,
and 8. In the triaxial case, we always consider undersampling by the
same factor in both $a'/a$ and $b'/a$.  Generally, given $N_s$
samplings in $a'/a$ or $b'/a$, we have $(1+N_s)/u$ viewing angles when
undersampling by a factor of $u=2^v$ for $v \ge 1$.

We find that for all mass scales, magnification bias modes, and
numbers of images, undersampling by a factor of 4 in both $a'/a$ and
$b'/a$ allows us to recover orientation-averaged cross-sections to within
5 per cent accuracy when compared with the full sampling.  In general,
3-image systems are the most difficult to sample properly due to their
strong variation with inclination (as in Fig.~\ref{F:incboth}).
This level of undersampling corresponds to only $12$ samplings in
$b'/a$ in the oblate case, $8$ samplings in $a'/a$ in the prolate
case, and $6$ (in $a'/a$) $\times$ $4$ (in $b'/a$) $=24$ samplings
total in the triaxial case.  For reference, in Fig.~\ref{F:incboth}
and~\ref{F:bothmtriax} we indicate our chosen level of undersampling
with crosses.

%=====================================================================
\section{Conclusions}
\label{S:conclusions}
%=====================================================================

% IDL: Start with spherical reference virial mass, virial radius,
% scale radius; reset $\rho_0$ (or $r_s$) given slope parameters
% $(\gamma,n)$; reset scale length $a$ given axis ratios $(p.q)$;
% yields final $\rho(m)$. The compute $\Sigma(m')$ on fine grid in
% $m'$; given box size and resolution, create grid $(x',y')$; given
% viewing direction $(\vartheta,\varphi)$ compute corresponding $m'$
% values and interpolate; yields maps $\Sigma(x',y')$.
%
% Scripts: create sets of maps; coordinate gravlens evaluation and
% results handling
%
% Gravlens: updated version to efficiently do many, many lensing
% calcuations

% We have presented a flexible simulation pipeline for coherent
% investigations of selection and modelling biases in strong lensing
% surveys. We use the pipeline in paper II to assess whether the distribution of
% physical parameters, such as the profile and shape of the (projected)
% density, in lens galaxies truly reflects the distribution among the
% galaxy population in general.  In that paper we also quantify in what
% ways deviation of the lens galaxy population from the typical galaxy
% population are most dominant. This pipeline also enables us to
% simulate realistic strong lensing data, which we will use to study the
% effects of modelling assumptions.

We have presented a flexible simulation pipeline for coherent
investigations of selection and modelling biases in strong lensing
surveys. We have focused on point-source lensing by two-component
galaxy models meant to emulate realistic early-type, central galaxies
at two different mass scales: a lower, $\sim 2L_*$ galaxy mass scale
and a higher, $\sim 7 L_*$ group mass scale (with $L_*$ in the
$r$-band).
Below is a list of our main conclusions regarding the construction of
the simulation pipeline for lensing by realistic galaxy models:
\begin{itemize}
\item We include both cusped and deprojected S\'ersic density
  profiles, with observationally-motivated choices of masses and scale
  lengths, and a range in density profile parameters, separately for
  the stellar and dark matter component. We use seven different models
  for the galaxy shapes with the stellar component rounder than the
  dark matter component, but with the axes intrinsically aligned.
  [Sections~\ref{SS:massscales}--\ref{SS:sersicsim}]
\item When we change the density profile parameters away from the
  adopted fiducial values, we preserve the total and virial mass of
  the stellar and dark matter components (respectively) by changing
  the density amplitude. We keep the scale radius fixed to comply
  with observed size-luminosity relations for early-type galaxies and
  concentration-mass relations for dark matter halos. Similarly, for the
  non-spherical shapes we change the scale length to preserve the mass
  but keep the scale radius (and hence the concentration) the same as
  for the fiducial spherical case. [Sections~\ref{SSS:normalization},
  \ref{SSS:scalelengthnonspherical}, \ref{SSS:sersicnormalization}]
\item We choose fiducial lens and source redshifts of $0.3$ and $2.0$,
  resulting in Einstein radii $\Rein$ at about one effective radius
  $R_e$. [Section~\ref{SS:redshifts}]
\item We require multiple ways of handling magnification bias to
  sample different parts of the observed quasar luminosity function
  and to allow for application to different survey limitations in
  image resolution and flux. [Section~\ref{SS:magbias}]
  %modes (those that resolve images from the outset, and those that do not).
\item When the surface mass density is known analytically (as for our
  deprojected S\'ersic models), we compute the corresponding lensing
  deflection and magnification analytically (for circular symmetry) or
  with numerical integrals (for elliptical symmetry).  When the
  surface mass density cannot be computed analytically, as for the
  cusped models, we construct a surface density map and use Fourier
  methods to compute the lensing deflection and magnification.  We
  validate and use Monte Carlo methods to calculate lensing
  statistics, including unbiased and biased cross-sections and image
  separations. [Sections~\ref{SS:kapmap}--\ref{SS:validation}]
\item For our map-based lensing calculations, tests for convergence at
  the 5 per cent level suggest that the required map resolution is
  $\sim 25$ pixel lengths per $\Rein$, corresponding to $0.025\arcsec$
  ($\simeq 0.11$\,kpc) and $0.075\arcsec$ ($\simeq 0.33$\,kpc) for the
  lower and higher mass scale, respectively. The resolution needs to
  be fine enough to simultaneously resolve the steepness of the inner
  slope of the density and recover the smoothness of the shape of the
  inner critical curve. Similarly, convergence tests suggest that the
  required box size is $\sim 16\,\Rein$, corresponding to $10\arcsec$
  ($\simeq 43.6$\,kpc) and $30\arcsec$ ($\simeq 131$\,kpc) for the
  lower and higher mass scale, respectively.
  [Sections~\ref{SS:resolution} and~\ref{SS:boxsize}]
\item Lensing cross-sections for fairly general triaxial models vary
  in a smooth and monotonic way with viewing direction, which can be
  efficiently sampled through projected axis ratios (rather than
  viewing angles) of the surface density. Surprisingly few samplings
  are necessary: 10, 8, and 24 viewing directions can effectively
  determine the orientation-averaged cross-sections for oblate, prolate,
  and triaxial models (respectively) with 5 per cent precision.
  [Section~\ref{SS:projangle}]
\end{itemize}
We have implemented the construction of the realistic galaxy models
and efficient computation of the corresponding surface mass density
maps in \idl\ (see also Section~\ref{SS:compsurfmaps}). All subsequent
strong lensing calculations are done with an updated and extended
version of \gravlens. We use scripts to coordinate and analyse the
extensive data flow to and from these two parts, and to make it into a
coherent simulation pipeline.

Our investigations have revealed a number of useful, general points
related to the interpretation of observational strong lensing results:
\begin{itemize}
\item In the vicinity of the Einstein radius, the intrinsic and
  projected logarithmic slopes of the total density are close to the
  values $\gamma=2$ and $\gamma'=1$ of an isothermal profile, for all
  of our galaxy models (see Figs.~\ref{F:3dlogslope}
  and~\ref{F:2dlogslope}, and the text of
  Sec.~\ref{SSS:intrinsicprofiles}). Consequently, a measurement of
  the total  
  density profile near the Einstein radius cannot (alone) be used to
  determine the dark matter inner slope for a two-component model.
  This is a non-trivial conclusion given the variety of intrinsic
  non-isothermal density profiles used for both the stellar and dark
  matter component, including profiles with a true central cusp and
  those that do not asymptote to a particular inner slope at any
  scale.
\item We also find that the lensing deflection curves are nearly flat
  around (and even beyond) $\Rein$, similar to a flat deflection curve
  for an isothermal model (see Fig.~\ref{F:alpha}), particularly for
  the lower-mass galaxy scale. This again implies that constraints
  from strong lensing cannot be extrapolated to radii much smaller or
  larger than the Einstein radius.
  We emphasize that the parameters for the galaxy models were
  \emph{not} a priori chosen to mimic this effect, but that it truly
  seems to be an ``isothermal conspiracy''
  \citep[e.g.,][]{2003ApJ...595...29R}.
  % end section 3.2. of Guangfei & Kochanek (2007):
  % http://adsabs.harvard.edu/abs/2007ApJ...671.1568J
  % based on Hernquist + power-law (=isothermal around effective
  % radius) models by Rusin, Kochanek & Keeton (2003)
  % http://adsabs.harvard.edu/abs/2003ApJ...595...29R
\item Because of our assumption of alignment between the intrinsic
  axes of the stellar and dark matter components, only the two shape
  models with a triaxial dark matter component and a rounder (triaxial
  or oblate) stellar component can have a significant misalignment
  between their projected axes.  Even then, significant misalignment
  occurs only for a limited number of viewing directions, several of
  which lead to a relatively round projected stellar component such
  that the misalignment is not very important in practice (the
  position angle of a round stellar component is on the sky, after
  all, difficult to establish). The small misalignments inferred for
  relatively isolated lens systems, as well as detailed dynamical
  modelling of early-type galaxies, support near intrinsic alignment
  between stars and dark matter (see Section~\ref{SSS:misalignment}).
\end{itemize}

In Paper~II, we use the flexibility of the pipeline to study selection
biases related to the galaxy mass, shape, orientation, and various
parameters of the dark matter and stellar profiles. In subsequent
work, we will investigate modelling biases by analysing the mock lens
systems produced by the pipeline using the lens modelling tools within
the \gravlens\ package. We have therefore created a lensing simulation
pipeline that will be crucial for deriving inferences about galaxy
density profiles and shapes, $H_0$, and other parameters from the
thousands of lenses that will be discovered in the coming years in
large photometric surveys such as Pan-STARRS, LSST, SNAP, and SKA.

Furthermore, following a similar approach as in
\citeauthor{2008MNRAS.385..614V} (2008a),
%\cite{2008MNRAS.385..614V}, 
we plan to extend the pipeline to produce projected kinematics of our
galaxy models that mimic the two-dimensional kinematic observations of
early-type galaxies. We will then investigate how well we can expect
to recover the intrinsic density profile and shape of early-type
galaxies, based solely on dynamical models fitted to these simulated
kinematics, or on a combination of kinematics and strong lensing
(using techniques similar to \citeauthor{2008arXiv0807.4175V} 2008b).
%\citep[using techniques similar to][]{2008arXiv0807.4175V}.  
This work will be valuable for understanding the modelling biases in
analyses of the kinematic data that are becoming available for
hundreds of galaxies at increasing redshifts, and for understanding
both modelling and selection biases in joint lensing+kinematics
studies.

%=====================================================================
% ACKNOWLEDGMENTS
%=====================================================================

\section*{Acknowledgments}
\label{S:acknowledgments}

We are thankful to Chung-Pei Ma, Michael Kuhlen and Scott Tremaine for
stimulating discussions on related topics. 
We thank the referees for constructive comments and suggestions.
This work has made use of the public \textsc{contra} software package
provided by Oleg Gnedin to perform adiabatic contraction calculations.

GvdV and RM acknowledge support provided by NASA through Hubble
Fellowship grants HST-HF-01202.01-A and HST-HF-01199.02-A,
respectively, awarded by the Space Telescope Science Institute, which
is operated by the Association of Universities for Research in
Astronomy, Inc., for NASA, under contract NAS 5-26555. CRK
acknowledges support from NSF through grant AST-0747311, and from NASA
through grant HST-AR-11270.01-A from the Space Telescope Science
Institute, which is operated by the Association of Universities for
Research in Astronomy, Inc., under NASA contract NAS 5-26555.

%=====================================================================
% REFERENCES
%=====================================================================

%\bibliographystyle{mn2e}
%\bibliography{../allpapers} 

%=====================================================================
% APPENDICES
%=====================================================================

\appendix

%=====================================================================
\section{Sampling of viewing direction via projected axis ratios}
\label{A:samplingviewing}
%=====================================================================

The smooth variation in lensing cross-section with viewing direction
is more directly related to the (elliptic) shape of the surface mass
density than the viewing angles. Hence, instead of sampling
directly $\vartheta$ and $\varphi$, we sample linearly in the
projected axis ratios $a'$ and $b'$ given in
equation~\eqref{eq:apbpellipse}. Inversion of the latter relation
yields
\begin{eqnarray}
  \label{eq:thvsapbp}
  \cos^2\vartheta & = &  
  \frac{({a'}^2-c^2)({b'}^2-c^2)}
  {(a^2-c^2)(b^2-c^2)},
  \\
  \label{eq:phvsapbp}
  \tan^2\varphi & = & 
  \frac{({a'}^2-b^2)(b^2-{b'}^2)(a^2-c^2)}
  {(a^2-{a'}^2)(a^2-{b'}^2)(b^2-c^2)}.
\end{eqnarray}
The Jacobian $J(a',b')$ of the transformation between
$(\vartheta,\varphi)$ and $(a',b')$ is given by
\begin{equation}
  \label{eq:Jacobianthph2apbp}
  \sin\vartheta \, J(a',b') = 
  \frac{{a'}{b'}({a'}^2-{b'}^2)(b^2-c^2)}
  {({a'}^2-b^2)(b^2-{b'}^2)({a'}^2-c^2)({b'}^2-c^2)}.
\end{equation}
In the axisymmetric limit one of the viewing angles becomes redundant.
In the oblate case ($a=b>c$), we set, without loss of generality,
$\varphi=\pi/2$ so that $a'=a$, and
\begin{equation}
  \label{eq:thphvsapbp_obl}
  \cos^2\vartheta = \frac{{b'}^2-c^2}{a^2-c^2},
  \quad
  \sin\vartheta \, \rmd\vartheta = 
  \frac{{b' \, \rmd b'}}{\sqrt{(a^2-c^2)({b'}^2-c^2)}}.
\end{equation}
Similarly, in the prolate case ($a>b=c$) we take $\vartheta=\pi/2$ so
that $b'=c$, and
\begin{equation}
  \label{eq:thphvsapbp_pro}
  \tan^2\varphi = \frac{{a'}^2-c^2}{a^2-{a'}^2},
  \quad
  \rmd\varphi = 
  \frac{{a' \, \rmd a'}}{\sqrt{(a^2-{a'}^2)({a'}^2-c^2)}}.
\end{equation}

\end{document}